\documentclass[a4paper,fleqn,usenatbib,useAMS, twocolumn]{mn2e}
\usepackage{color, xcolor, journals}
\usepackage{natbib}
\usepackage{graphicx}   
\usepackage{amsmath}    
\usepackage{amssymb}    
\usepackage{multicol}        
\usepackage{bm}     
\usepackage{pdflscape}  

\newcommand\nn         {\nonumber}

\newcommand{\be}{\begin{equation}}
\newcommand{\ee}{\end{equation}}
\newcommand{\ba}{\begin{eqnarray}}
\newcommand{\ea}{\end{eqnarray}}

\newcommand{\gt}{\geq}
\newcommand{\mx}{\mbox}

\newcommand{\map}{M_{\rm ap}}

\def\gt{\geq}
\def\SN{{\mathcal S}/{\mathcal N}}

\def\btheta{{\mx {\bm $\theta$}}}

\def\map{{M_{\rm ap}}}

\def\deg2{\rm deg^2}
\def\arcmin2{\rm arcmin^2}

\def\nn{{\nonumber}}
\def\om{{\Omega_{\rm m}}}
\def\s8{{\sigma_8}}

\usepackage{mathabx}
\usepackage{hyperref}

\newcommand{\ngmix}{\textsc{ngmix}}
\newcommand{\imshape}{\textsc{im3shape}}
\newcommand{\skynet}{\textsc{skynet}}
\newcommand{\balrog}{\textsc{Balrog}}
\newcommand{\snr}{\ensuremath{\mathcal S / \mathcal N}}
\newcommand{\apertsize}{\ensuremath{\theta_{\rm{max}}}}
\newcommand{\SE}{\ensuremath{S_8}}
\newcommand{\review}{}

\newcommand{\psfex}{\textsc{PSFEx}}
\newcommand{\sex}{\textsc{SExtractor}}
\newcommand{\swarp}{\textsc{SWarp}}

\defcitealias{DESCS2015}{DES15}
\defcitealias{TheDarkEnergySurveyCollaboration2015}{DES15}

\usepackage{eso-pic}

\AddToShipoutPictureBG*{%
  \AtPageUpperLeft{%
    \hspace{0.75\paperwidth}%
    \raisebox{-3.5\baselineskip}{%
      \makebox[0pt][l]{\textnormal{DES 2015-0113}}
}}}%

\AddToShipoutPictureBG*{%
  \AtPageUpperLeft{%
    \hspace{0.75\paperwidth}%
    \raisebox{-4.5\baselineskip}{%
      \makebox[0pt][l]{\textnormal{FERMILAB-PUB-15-566-AE}}
}}}%

\title[Cosmology from peak statistics in DES SV]{Cosmology constraints from shear peak statistics in Dark Energy Survey Science Verification data}

\author[]{
\parbox{\textwidth}{
\Large
T.~Kacprzak$^{1}$,\thanks{Corresponding author: \texttt{tomasz.kacprzak@phys.ethz.ch}}
D.~Kirk$^{2}$,
O.~Friedrich$^{3,4}$,
A.~Amara$^{1}$,
A.~Refregier$^{1}$,
L.~Marian$^{5}$,
J.~P.~Dietrich$^{6,7}$,
E.~Suchyta$^{8}$,
J.~Aleksi\'c$^{9}$,
D.~Bacon$^{10}$,
M.~R.~Becker$^{11,12}$,
C.~Bonnett$^{9}$,
S.~L.~Bridle$^{13}$,
C.~Chang$^{1}$,
T.~F.~Eifler$^{8,14}$,
W.~Hartley$^{1}$,
E.M.~Huff$^{15,16}$,
E.~Krause$^{12,8}$,
N.~MacCrann$^{13}$,
P.~Melchior$^{18}$,
A.~Nicola$^{1}$,
S.~Samuroff$^{13}$,
E.~Sheldon$^{11}$,
M.~A.~Troxel$^{13}$,
J.~Weller$^{7,6,4}$,
J.~Zuntz$^{13}$,
T. M. C.~Abbott$^{1}$,
F.~B.~Abdalla$^{2,3}$,
R.~Armstrong$^{4}$,
A.~Benoit-L{\'e}vy$^{5,2,6}$,
R.~A.~Bernstein$^{7}$,
E.~Bertin$^{5,6}$,
D.~Brooks$^{2}$,
D.~L.~Burke$^{8,9}$,
A.~Carnero~Rosell$^{10,11}$,
M.~Carrasco~Kind$^{12,13}$,
J.~Carretero$^{14,15}$,
F.~J.~Castander$^{14}$,
M.~Crocce$^{14}$,
C.~B.~D'Andrea$^{16,17}$,
L.~N.~da Costa$^{10,11}$,
S.~Desai$^{18,19}$,
H.~T.~Diehl$^{20}$,
A.~E.~Evrard$^{21,22}$,
A.~Fausti Neto$^{10}$,
B.~Flaugher$^{20}$,
P.~Fosalba$^{14}$,
J.~Frieman$^{20,23}$,
D.~W.~Gerdes$^{22}$,
D.~A.~Goldstein$^{24,25}$,
D.~Gruen$^{8,9}$,
R.~A.~Gruendl$^{12,13}$,
G.~Gutierrez$^{20}$,
K.~Honscheid$^{26,27}$,
D.~J.~James$^{1}$,
K.~Kuehn$^{28}$,
N.~Kuropatkin$^{20}$,
O.~Lahav$^{2}$,
M.~Lima$^{29,10}$,
M.~March$^{30}$,
J.~L.~Marshall$^{31}$,
P.~Martini$^{26,32}$,
C.~J.~Miller$^{21,22}$,
R.~Miquel$^{33,15}$,
J.~J.~Mohr$^{18,19,34}$,
R.~C.~Nichol$^{16}$,
B.~Nord$^{20}$,
A.~A.~Plazas$^{35}$,
A.~K.~Romer$^{36}$,
A.~Roodman$^{12,26}$
E.~S.~Rykoff$^{8,9}$,
E.~Sanchez$^{37}$,
V.~Scarpine$^{20}$,
M.~Schubnell$^{22}$,
I.~Sevilla-Noarbe$^{37,12}$,
R.~C.~Smith$^{1}$,
M.~Soares-Santos$^{20}$,
F.~Sobreira$^{10}$,
M.~E.~C.~Swanson$^{13}$,
G.~Tarle$^{22}$,
D.~Thomas$^{16}$,
V.~Vikram$^{38}$,
A.~R.~Walker$^{1}$,
Y.~Zhang$^{22}$
\vspace{0.2cm} (The DES Collaboration) \\
\emph{(Affiliations are listed at the end of paper)} 
}
}

\begin{document}
\maketitle
\begin{abstract}
Shear peak statistics has gained a lot of attention recently as a practical alternative to the two point statistics for constraining cosmological parameters.
We perform a shear peak statistics analysis of
the Dark Energy Survey (DES) Science Verification (SV) data, using weak gravitational lensing
measurements from a 139 deg$^2$ field.
We measure the abundance of peaks identified in aperture mass maps,
as a function of their signal-to-noise ratio, in the signal-to-noise range $0<\snr<4$.
To predict the peak counts as a function of cosmological parameters we use a suite of $N$-body simulations spanning 158 models
with varying $\Omega_{\rm m}$ and $\sigma_8$, fixing $w = -1$,
$\Omega_{\rm b} = 0.04$, $h = 0.7$ and $n_s=1$, to which we have applied the DES SV mask and redshift distribution.
In our fiducial analysis we measure $\sigma_{8}(\Omega_{\rm m}/0.3)^{0.6}=0.77 \pm 0.07$,
after marginalising over the shear multiplicative bias and the error on the mean redshift of the galaxy sample.
We introduce models of intrinsic alignments, blending, and source contamination by cluster members.
These models indicate that peaks with $\snr>4$ would require significant corrections, which is why we do not include them in our analysis.
We compare our results to the cosmological constraints from the two point analysis on the SV
field and find them to be in good agreement in both the central value and its uncertainty.
We discuss prospects for future peak statistics analysis with upcoming DES data.

\end{abstract}

\section{Introduction}

Weak gravitational lensing (WL) is a promising and powerful
probe for constraining cosmology because of its ability to map
the 3D matter distribution of the Universe in an unbiased way.
The effects of WL are observable through small, but spatially coherent, distortions of galaxy shapes.
This technique was successfully used to constrain cosmological parameters by several lensing surveys, most recently by:
Canada-France-Hawaii Telescope Lensing Survey (CFHTLenS) \citep{Heymansetal2013,Kilbinger2013},
COSMOS \citep{Schrabback2010},
and Sloan Digital Sky Survey (SDSS) \citep{Huff2011}.
Most recently, the first weak lensing cosmological results from the Dark Energy Survey (DES)
were presented by \citet{DESCS2015} (hereafter DES15).

Among WL observables, the shear two point (2-pt) correlation function has so far
received the most attention from the WL community
\citep{Jarvisetal2003, Hoekstraetal2006, Sembolonietal2006,
  Hetterscheidtetal2007, Kilbingeretal2013, Heymansetal2013}.
This statistic is a powerful tool for constraining
cosmology and the impact of systematic and measurement errors on it
have been extensively studied \citep[see][for a review]{Kilbinger2015}.
It has also been used to plan and forecast
coming missions such as {\it Euclid} \citep{Refregier2010, Euclid2011} and LSST
\citep{lsst2009}.

To optimally exploit the power of WL surveys to constrain
cosmological models, it is commonly believed that using one type of statistic alone will
not suffice \citep{Petri2014,Osato2015a};
this is because different probes are generally affected by systematics in a different
way, and combining and comparing them will help test and understand and calibrate them better.
Moreover, alternative statistics can capture additional information from the non-Gaussian features
in the matter distribution.

Shear peak statistics is one of these alternative probes of WL.
It aims to extract the cosmological information from the `peaks',
i.e. regions of the map high signal-to-noise ($\SN$), produced by
overdense regions of the matter density field projected along the line
of sight. Massive clusters imprint peaks in WL maps, which can
be used to detect and measure cluster masses, as first pointed out by
the pioneering papers of \citet{KaiserSquires1993}, \citet{Tyson1990} and \citet{Miralda-Escude1991}.
Many of the peaks with lower $\SN$ are produced, not by single clusters, but by the projection of many halos along the line-of-sight \citep{YKW+11}.
Random noise can also produce spurious ``peaks'' in maps made from data.

With the introduction
of the aperture mass by \cite{Schneider1996} the idea of detecting
clusters as points of high $\SN$ in WL maps really took wing. A series
of studies investigating optimal aperture filters, projection effects
on cluster mass determination, forecasts for future WL surveys, and
detections in available WL data followed \citep{Hamanaetal2004,
  Cloweetal2004, Wangetal2004, Maturietal2005, HennawiSpergel2005,
  TangFan2005, Dahle2006, MarianBernstein2006, Schirmeretal2007,
  Maturietal2007, Berge2008, Abateetal2009, Marianetal2010}. Indeed,
for a long time, shear peaks were mainly regarded as means for WL
cluster detection, before being considered as a WL
probe in its own right. This last idea became popular when studies doing `blind' peak
detection in WL maps, generated from $N$-body simulations, showed that
the peak abundance scales with cosmological parameters in the same way
as the halo mass function \citep{Reblinskyetal1999, Marianetal2009,
  Marianetal2010}, and therefore can be used to constrain the
cosmological model \citep{DietrichHartlap2010, Kratochviletal2010,
  Marianetal2012, Bardetal2013}.
The shear-peak abundance can also
constrain primordial non-Gaussianity of the local type, being one of
the most effective WL probes for this purpose \citep{Marianetal2011,
  Maturietal2011, Hilbertetal2012}. Further analysis of simulated WL
maps showed that peak profiles and peak correlation functions can
significantly improve the constraints on cosmology relative to the
peak abundance alone \citep{Marianetal2013}.

The shear peak abundance has the advantage, relative to the
cluster mass function, that it does not depend on a mass-observable
relation (the shear signal can be used directly to constrain
cosmology, without having to be converted into a virial mass).
Another advantage of this method is that it is sensitive to
non-Gaussian features in the mass distribution \citep{Berge2010, Pires2012}.
However, it has the disadvantage that the analytical predictions are relatively complicated
\citep{Shan2013, Lin2015, Maturietal2010, Reischke2015}.
Nonetheless, the consensus so far among peak studies is that, as long as real data
maps are compared to simulated maps that have been imprinted with the
same characteristics -- survey masks, source distribution etc. -- and
the same analysis is applied to both, the lack of reliable analytical
predictions can be circumvented.

Several recent studies have made measurements of the WL peak
abundance from data, in particular from the Canada-France-Hawaii
Telescope (CFHT). \cite{Liuetal2015Z} have used the CFHT Lensing
Survey (CFHTLenS) \citep{heymans2012} shape catalogues \citep{Milleretal2013} to obtain
convergence maps which they smoothed with Gaussian filters of various
sizes to identify peaks as local maxima. The measured peak
abundance -- which included also the smallest peaks, and even regions
of negative convergence -- was then compared to results from simulated
maps corresponding to cosmologies with varying $\om, \s8, w$, and thus
constraints on the cosmological model were obtained. The latter were
found to be similar to those yielded by the convergence power
spectrum, while combining the two probes tightened the constraints by
a factor of $\approx$ 2. \cite{Liuetal2015W} used the CFHT Stripe 82
survey to also create Gaussian-smoothed convergence maps, where peaks
were detected as points of local maxima, this time applying a more
conservative detection threshold of $\SN>3$. Using covariance matrices
measured from the data, the authors derived constraints on $\om,
\s8$. Finally, \cite{Hamanaetal2015} used Subaru/SuprimeCam data
\citep{Miyazakietal2002} to detect WL peaks in an area of $\sim 11\,
\deg2 $. This was also done in convergence maps, but only high $\SN
(\gt 5)$ peaks were selected. These were shown to correspond to
optically confirmed clusters.

In this paper we present a measurement of the WL peak abundance from
another data set, the Dark Energy Survey Science Verification (SV)
data. Unlike the previous studies, we measure peaks using the
aperture mass maps, not convergence, though we point out that convergence maps of
the SV data have been presented in \cite{Vikrametal2015,
  Changetal2015}. We use simulated WL maps with cosmologies spanning
the $\{\om, \s8\}$ plane to derive cosmological constraints from our
measurements.
We extensively explore the possible systematics affecting the peak
statistics measurement. In our analysis, we model and marginalise the
shear multiplicative bias and the error in the mean of the redshift
distribution.
Additionally, we explore (i) the impact of the contamination of the
source galaxy sample with cluster galaxies, (ii) loss of
background galaxies due to enhanced blending at the positions of clusters
and (iii) the impact of intrinsic alignment of shapes of galaxies with
respect to the centres of the peaks.

Finally, we present constraints on the $\om$ and $\s8$ parameters,
when other cosmological parameters are fixed to $\Omega_{\rm b} = 0.04$, $h = 0.7$ and $n_s=1$.
We make a comparison between the results from our analysis
and the WL 2-pt presented in \citetalias{TheDarkEnergySurveyCollaboration2015}.
This allows us to check the consistency of the results between these two
methods, which may respond to different systematics in
a different way. We discuss the impact of systematics and
prospects for future peak statistics analyses with DES.

The paper is structured as follows: in Section \ref{sec:data}, we describe
the shear catalogue, the photo-$z$ catalogue, and the numerical simulations used
for this work. In Section \ref{sec:map_making} we outline our filtering method
and how we find and define the shear peaks. In Section
\ref{sec:peak-function-measurements},  we present our measurements of the peak
function, then discuss the systematic
effects in Section \ref{sec:systematics}. Section \ref{sec:inference} contains the details of construction of the
likelihood function. In Section \ref{sec:cosmological_constraints}, we present
the cosmological constraints and finally, in Section \ref{sec:conclusions},
we draw our conclusions.

A number of appendices give further details on aspects of our work: Appendix \ref{app:systematics_model} describes the modelling of the effect of the multiplicative bias and redshift error nuisance parameters on the peak counts. Details of our interpolation schemes are given in Appendix \ref{app:peakfun_modelling}. The calculation of boost factors is described in Appendix \ref{app:boost_factors} and the modelling of intrinsic alignments in Appendix \ref{app:intrinsic_alignments}. Appendix \ref{app:balrog} summarises the \balrog\ catalogues used in this work and Appendix \ref{app:des_mask} describes how we used an interpolation scheme to apply the DES $n(z)$ and mask to the simulations.

\section{The data}
\label{sec:data}

The Dark Energy Survey is a five-year optical and near-infrared ($grizY$) survey of $5000$ deg$^2$,
to limiting magnitude $i_{AB} \lesssim 24$, using the Blanco 4-m Telescope at the
Cerro Tololo Inter-American Observatory site in Chile \citep{TheDarkEnergySurveyCollaboration2005}.
The survey instrument, the Dark Energy Camera \citep[DECam,][]{Flaugher2015},
is a wide-field (2.2~deg in diameter), thick-CCD camera,
which was commissioned in fall 2012.
During the Science Verification (SV) period, which lasted from November 2012 to February 2013,
data were taken in a way mimicking the full survey on a relatively small sky area,
approaching full-survey depth in some areas.

We use the \ngmix\ SV shear catalogue\footnote{\url{http://des.ncsa.illinois.edu/releases/sva1}} described by \citet{Jarvis2015}, which covers an area of 139~deg$^2$.
We provide an overview of salient features for this work in section~\ref{sec:shear-cat}
and refer readers to \citet{Jarvis2015} for full details.
The dataset has been used in other studies, for example
to constrain cosmology using 2-pt functions \citep[\citetalias{TheDarkEnergySurveyCollaboration2015}]{Becker2015}
and to make weak lensing mass maps \citep{Vikrametal2015,Changetal2015}.

As one of the goals of this paper is to make a comparison between the
shear peak statistics and the two point statistics methods, we will
take as the fiducial configuration the same data as in the 2-pt
cosmological analysis of \citetalias{TheDarkEnergySurveyCollaboration2015}.
We use the identical source galaxy sample and corresponding shear
measurements, as well as the source redshift distribution $n(z)$ relying
on photometric redshifts for the non-tomographic configuration.

To make an empirical prediction of the peak abundance as a function of
cosmology, we use a suite of $N$-body simulations, described in the
Section \ref{sec:simulations}.  These simulations were taken from
\citet{Dietrich2009} and span 158 cosmological models in the
$\Omega_{\rm m}$ and $\sigma_8$ plane.

We did not attempt to calculate the peak functions analytically, as was demonstrated by \citet{Shan2013,Maturietal2010,Reischke2015}.
We decided to use a fully computational approach, which has the advantage of allowing us to incorporate the exact DES mask and shape noise in an easy way.

\subsection{Shear catalogue description}
\label{sec:shear-cat}

Two lensing catalogues were created from the DES SV observations, using the
\ngmix\ \citep{Sheldon2014} and \imshape\ \citep{Zuntz2013} shape measurement methods,
which contain 3.44 million and 2.12 million galaxies respectively.
The point spread function (PSF) modelling was done with the
\psfex\ software \citep{Bertin2011}.   Each galaxy comes with
a two-component shear estimate, a corresponding sensitivity
correction and a statistical weight.  These catalogues were
thoroughly tested for systematics in \citet{Jarvis2015} and \citet{Becker2015},
and {\review show B-modes and PSF leakage consistent with zero}.
In this work we employ only the
\ngmix\ catalogue because of its higher source density, a decision we share with \citet{Becker2015}, DES15 and \citet{Kirketal2015}.
 The raw number density of that catalogue is 6.9
galaxies/$\arcmin2$, and the effective number density is 5.7 galaxies/$\arcmin2$, after
weighting by the signal-to-noise of galaxies.
Tests on simulations have shown possible sources of multiplicative
systematics related to model bias \citep{Voigt2010, Bernstein2010,
  Kacprzak2014}. \citet{Jarvis2015} recommended the use of a Gaussian prior of width $\sigma_m = 0.05$ on the
multiplicative correction factor.
\subsection{Photometric redshifts catalogue description}

The photometric redshift solutions for objects in the DES SV shear
catalogues were subjected to a series of validation tests,
described in \citet{Bonnett2015}. In that work, four of the
best-performing algorithms were examined, finding good agreement
between them. For this
analysis we use the redshift results obtained from running the
\skynet\ code \citep{Graff2013}, which was also taken as the
fiducial set of solutions in the shear analysis of
\citetalias{TheDarkEnergySurveyCollaboration2015}. For further details of
the implementation of \skynet\ and performance we refer the reader to
\citet{Bonnett2015}. The resulting catalogue was trimmed to
$0.3<z<1.3$, based on the
mean redshift of the \skynet\ probability distribution function. These cuts exclude the least certain redshifts whilst
having minimal impact on the lensing measurements.
In \citetalias{TheDarkEnergySurveyCollaboration2015} the uncertainty
on the mean redshift was set to $\Delta z=0.05$ and the marginalisation
with that prior included an independent parameter for each redshift bin.
For the case of the non-tomographic measurements, a single prior with the
same width was adapted.

\subsection{Simulations}
\label{sec:simulations}

\begin{figure}
\begin{center}
\includegraphics[width=\columnwidth]{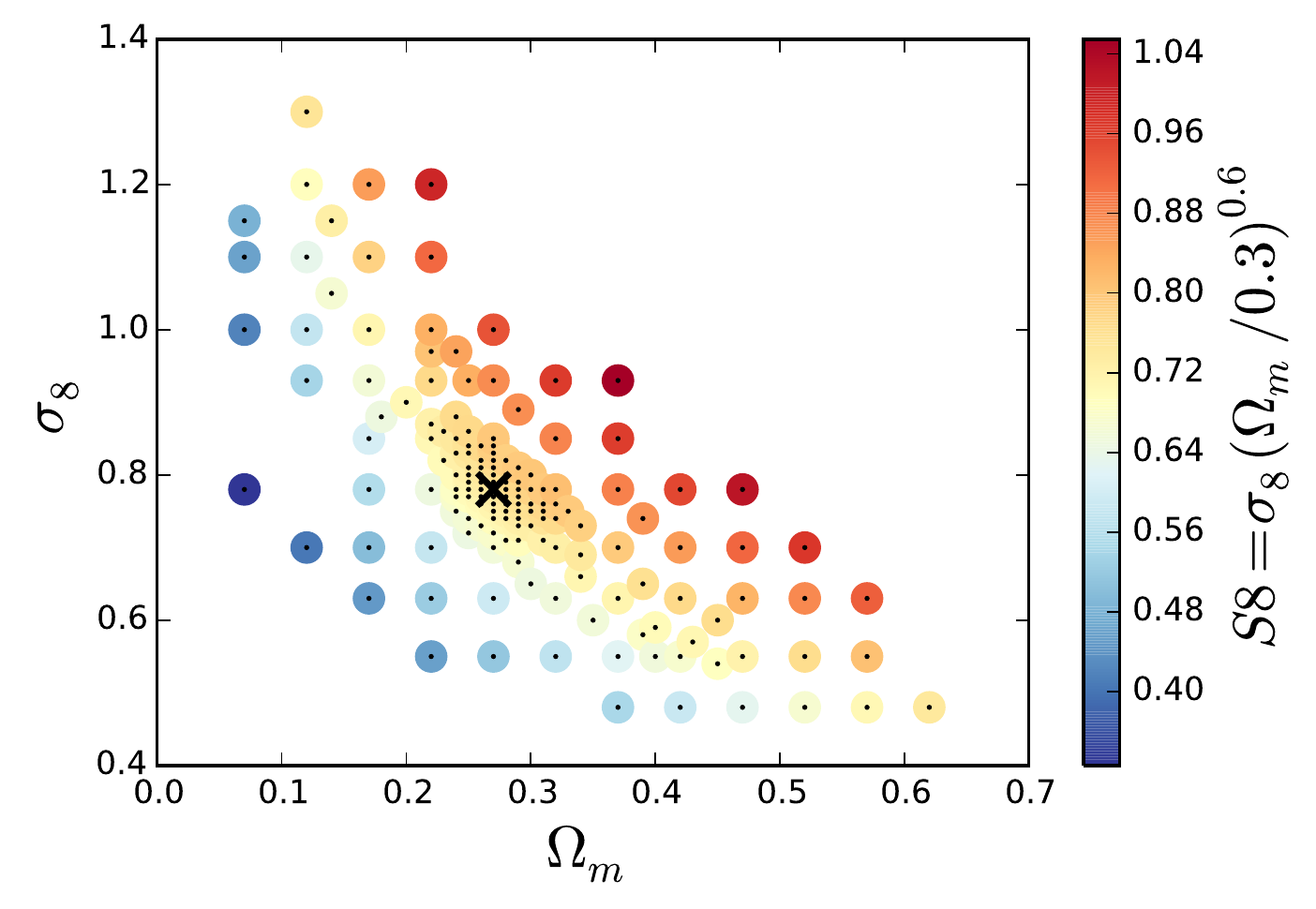}
\end{center}
\caption[]{The grid of cosmological models in the simulations from \citet{Dietrich2009}.
Colour corresponds to the value of $\SE=\sigma_8 (\Omega_{\rm m}/0.3)^{0.5}$ parameter.
The central cosmological model is marked by black cross.
Other cosmological parameters for these simulations were: $h_{70} = 1$, $n_s=1$ and $\Omega_b=0.04$.
\label{fig:sim_grid}}
\end{figure}

We use a set of $N$-body simulations from \citet{Dietrich2009},
created with the publicly available TreePM code GADGET-2
\citep{Springel2005}.  They use the $\Lambda CDM$ model, with initial
conditions set by the transfer function of \citet{Eisenstein1998}.  The
simulation space spans two cosmological parameters: $\sigma_8$,
$\Omega_{\rm m}$. The curvature is fixed at $\Omega_{\rm k}=0$, causing
$\Omega_\Lambda$ to vary accordingly.
The grid of cosmological models has 158 unique parameters pairs and is shown in Figure \ref{fig:sim_grid}.
Other cosmological parameters
were set to fixed values: $\Omega_{\rm b} = 0.04$, $n_s=1$, $h_{70}=1$.
All simulations used $256^3$ dark matter particles in a box with 200
$h^{-1}_{70} \rm{Mpc}$ side length.  Particle masses vary with cosmology and
range from $m_p=9.3 \times 10^{9} M_{\odot}$ for $\Omega_{\rm m}=0.07$ to
$m_p=8.2 \times 10^{10} M_{\odot}$ for $\Omega_{\rm m}=0.62$.
The particle mass for
our fiducial, non-tomographic cosmology is $m_p=3.6 \times 10^{10}
M_{\odot}$. The force softening in these simulations was
  set to $\epsilon = 25\,h^{-1}_{70} \rm{kpc}$.
Propagation of light rays through the simulated matter distribution is
done using the multiple lens-plane algorithm
\citep[for example][]{Hilbert2009,Blandford1986}; for more details about
ray-tracing used here, see \citet{Dietrich2009}. The central
cosmological model was simulated 35 times. The values for the central-cosmology
parameters are $\om=0.27$, $\Omega_\Lambda=0.73$, \SE=0.78.

{\review Each cosmological model in the simulations, except the central, has 5 patches of $6 \times  6$ deg,
corresponding to five projections of the simulated N-body boxes, which gives 180 deg$^2$.}
The central model has 35 such patches.
Each of this patches comes with a catalogue of galaxies with density of 25 galaxies/$\arcmin2$, positioned uniformly in ra and dec.
The redshift distribution in these catalogues is deeper than the SV
survey (see Appendix \ref{app:des_mask} for details).
We do not use these catalogues directly in our analysis, as they do not have the proper DES mask and redshift distribution.
Instead, using these catalogues, we produce new catalogues with applied DES mask and $n(z)$.

For the peak statistics, it is crucial to ensure that the survey mask
is correctly included in the simulations.  \citet{Bard2014} studied
the impact of masked regions and designed a forward-modelling approach
to tackle the problem.  We also find our measurement to be highly
impacted by the survey mask. To assess the impact of the mask, we
compare the number of shear peaks found in the randomised shear
maps in both the simulations and DES data.  After trying several
schemes, we decided to create simulations which have exactly the
same positions of galaxies as the DES data.
{\review To do this, for each patch, we take the positions of DES galaxies
and assign shear values, according to the simulations.
We also kept the shape noise, weight and multiplicative calibration factor
of the DES galaxy.}  This way we
produced simulated catalogues which differed only by the shear signal,
which was taken from simulations.  To assign a shear value from
simulations at a position of a DES galaxy, we used an interpolation
method, described in Appendix \ref{app:des_mask}.  This assignment is
done only using galaxy positions, and ignoring the corresponding DES
galaxy redshift; the relation between position and redshift
is broken in the simulations.  The redshifts are drawn from the DES
$n(z)$ for the \skynet\ photo-$z$ catalogue, described in
\cite{Bonnett2015}, and is the same as in
\citetalias{TheDarkEnergySurveyCollaboration2015}.  This approach is similar
to the one taken by \citet{Liuetal2015W}, with the difference that we do not
use each individual galaxy photo-$z$.  In Appendix \ref{app:des_mask} we
test this interpolation method and find it to perform very well on
noise-free data. See Section \ref{sec:map_making} and Appendix \ref{app:des_mask}
for details regarding making flat shear field cutouts and applying the survey mask.

As the peaks caused by random noise fluctuations (from both shape and
measurement noise) dominate our signal,
we have to include them in the analysis of simulations. Additionally,
to get the empirical prediction for the number of peaks as a function
of cosmology, we have to make sure that the uncertainty on the number
of peaks in simulations caused by shape and measurement noise fluctuations is small.
Therefore, for every cosmological
model, we run 300 noise realisations of the DES footprint.
These noise realisations are done by rotating each ellipticity
by a random angle, while keeping the ellipticity magnitude fixed.
The sensitivity correction and the statistical weights also remain the same.
{\review This configuration can be considered as a realistic draw from
the ellipticity noise distribution, as only the position angle is changed; the
ellipticity modulus, which can depend on the observing conditions,
such as, for example, PSF and sky background level, is preserved.}
{\review This allows us to get the uncertainty on the number of peaks in a \snr\ bin for all
cosmological models to be close to $5\%$ of the uncertainty on the number of peaks in the DES
measurement for that bin.
This way the uncertainty on the number of peaks predicted from simulations does not decrease the quality of constraints significantly.
}
Additionally, the interpolation of the peak counts as
a function of cosmology reduces the uncertainty
caused by shape noise and cosmic variance, as we expect the peak counts
to vary smoothly with cosmology.
{\review Interpolation of likelihood should also benefit from that feature.}
(see Section \ref{sec:interpolation} for details on interpolation schemes).
We found that the cosmological constraints
do not change when the number of noise realisations is changed from 300 to 200,
and we conclude that adding further noise realisations would not
change the constraints. There are 35 simulations of the central cosmology, which
results in 10500 total noise realisations of this model.

\section{Map making}
\label{sec:map_making}

We create aperture mass maps from the DES SV shape catalogues.  The full survey
area is divided into 20 patches, each of size $3 \times 3\,\deg2$.
This procedure makes the maps easier to create and also enables us to
apply the DES mask to simulations, as the simulation tiles have a size
of $6 \times 6\,\deg2$, which is easy to divide into four $3 \times 3$ patches.
{\review We found that using smaller patches would cause larger loss in the area covered due to edge effects.
Also, using larger patches of size $6 \times 6\,\deg2$ allows us to cover the
complicated DES SV footprint without losing a large fraction of the area.}
By dividing the footprint into these patches we will exclude a
small fraction of galaxies which do not belong to any patch.

After applying the DES mask and noise, the
simulations are analysed exactly in the same way as the DES data.
An aperture mass map is calculated on a grid covering the $3 \times 3\,\deg2$ patches,
 at a resolution 30 arcsec per pixel side.
{\review We verified that an increase of the resolution to 20 arcsec
per pixel did not affect the shape and uncertainty on the peak
function for the central cosmological model.}
Each patch is then used to create the mass map, using the
aperture mass filter technique \citep{Schneider1996}.

\subsection{The aperture mass filter}
\label{sec:aperture_mass_filter}

The aperture mass method consists of the smoothing of the field with a
filter function obeying certain properties. In the case of the
tangential shear, the filter $Q$ must have finite support, i.e. it
goes to 0 after a certain radius, which defines the aperture radius.
For each point on the map grid, the estimator for the aperture mass
is
\be
\map(\btheta_0) = \frac{1}{n_{\rm g}} \sum_i Q_i \,e^{\rm t}_i, \
\ee
where $\btheta_0$ is the pixel centre position, the index $i$ runs over $n_{\rm g}$
source galaxies within the aperture radius $\theta_{\rm max}$, $Q_i \equiv
Q(\btheta_i-\btheta_0)$ is the value of the filter at radius
$\btheta_i$ relative to $\btheta_0$, and $e^{\rm t}_i$ is the value of the
tangential shear of galaxy $i$ with respect to position $\btheta_0$, such that
\begin{equation}
e^{\rm t}_i = -\Re( e_i \exp[{-2j \phi_i}]),
\end{equation}
where $\phi_i$ is the angular position of galaxy $i$ about the centre of the pixel.
The $\SN$ of this estimator is
\be
\SN(\btheta_0) = \sqrt{2} \frac{ \sum_i Q_i e^{\rm t}_i }{\sqrt{ \sum_i Q_i^2 (e_{1,i}^2 + e_{2,i}^2) } }\ ,
\ee
where $e_{1,i}^2$ and $e_{2,i}^2$ corresponds to two components of the ellipticity of galaxy $i$.
We used these equations, modified to include shear sensitivity
correction and statistical weights,
\ba
\map(\btheta_0) & = & \frac{ \sum_i Q_i e^{\rm t}_i w_i } { \sum_i w_i s_i} \ ; \nn \\
\SN(\btheta_0) & = & \sqrt{2} \frac{ \sum_i Q_i e^{\rm t}_i w_i }
{ \sqrt{ \sum_i  Q_i^2 (e_{1,i}^2 + e_{2,i}^2) w_i^2} } \frac{\sum_i w_i} {\sum_i w_i s_i}\ ,
\ea
where $w_i$ is the statistical weight and $s_i$ is the shear
sensitivity correction as described in \citet{Jarvis2015}.
{\review
As the multiplicative correction affects the
variance of the noise, we found it necessary to include it in the processing of the
simulated data too.
We include it in the following way:
first we multiply the shear in the simulation corresponding sensitivity,
 then we process the simulation data the same way as the DES data, including the sensitivity correction.
By using this procedure we recover the correct shear in the analysis of the simulations.
This is possible because we do not expect the sensitivity at a position of a DES galaxy to be correlated with
the shear of simulated peak.
}

We consider an aperture mass filter with a shape matching the lensing
signal of NFW halos \citep{Navarro1997}. This filter is expected to
be optimal for the detection of clusters, as discussed in many
previous studies \citep{HennawiSpergel2005, MarianBernstein2006,
  Marianetal2010, Marianetal2012}. In particular we follow the work of
\citet{Schirmer2007, Dietrich2007, Dietrich2009} and adopt the
following expression for our filter:
\be
Q(\theta) = \frac{\tanh(\theta/\theta_c)}{\theta/\theta_c}
\frac{1}{1+\exp(6-150\theta)+\exp(-47+50\theta)}  \ ,
\ee
where $\theta$ [deg] is the distance relative to the aperture centre, scaled
by the aperture radius, and $\theta_c$ is a free parameter for which
we adopt $\theta_c=0.15$, the same value as \citet{Hetterscheidt2005} and \citet{Dietrich2009}. Regarding the aperture radius, we consider three
values, $\theta_{\rm max} = \{12, 20, 28\}$ arcmin.
We use $\theta_{\rm max} = 20$ as our fiducial measurement, and present
results from other filter scales.
We found that $\theta_{\rm max} = 20$ and $\theta_{\rm max} = 12$ give constraints of similar width, and slightly better than that from $\theta_{\rm max} = 28$.
We processed the maps with fixed filter size values, without using an adaptive scheme \citep{Marian2012}.
We did not attempt to combine the results from many filter sizes, as
it was done by \citet{Liu2014}.
This choice of filter function heavily downweights the central
  portion of peaks where the profile shape of real halos could deviate
  from those in the simulations due to baryonic effects and force
  softening.

\subsection{Mass maps and peak identification}

We process both the DES and simulated catalogues with the aperture mass filter and use these processed maps to identify the shear peaks.
To identify a peak, we select map pixels which have higher mass intensity than all their eight closest neighbours.
This approach is similar to others used in peak statistics \citep{Liuetal2015W, Liu2014, Marian2012}, although variations on this scheme have been proposed \citep[][for example]{Dietrich2009}.
To remove measurements from very low density areas and patch edges we additionally require a peak to have no less than 0.5 galaxies per arcmin$^2$ {\review within the area inside the aperture}.
In total, in DES SV, we identified 969 peaks above $3\sigma$ threshold.
An average number of peaks for the randomised maps was 676.4.
Number of peaks above $\snr > 0$ was 20165 and 20904.9 for DES and random peaks, respectively.

\begin{figure}
\begin{center}
\includegraphics[width=\columnwidth]{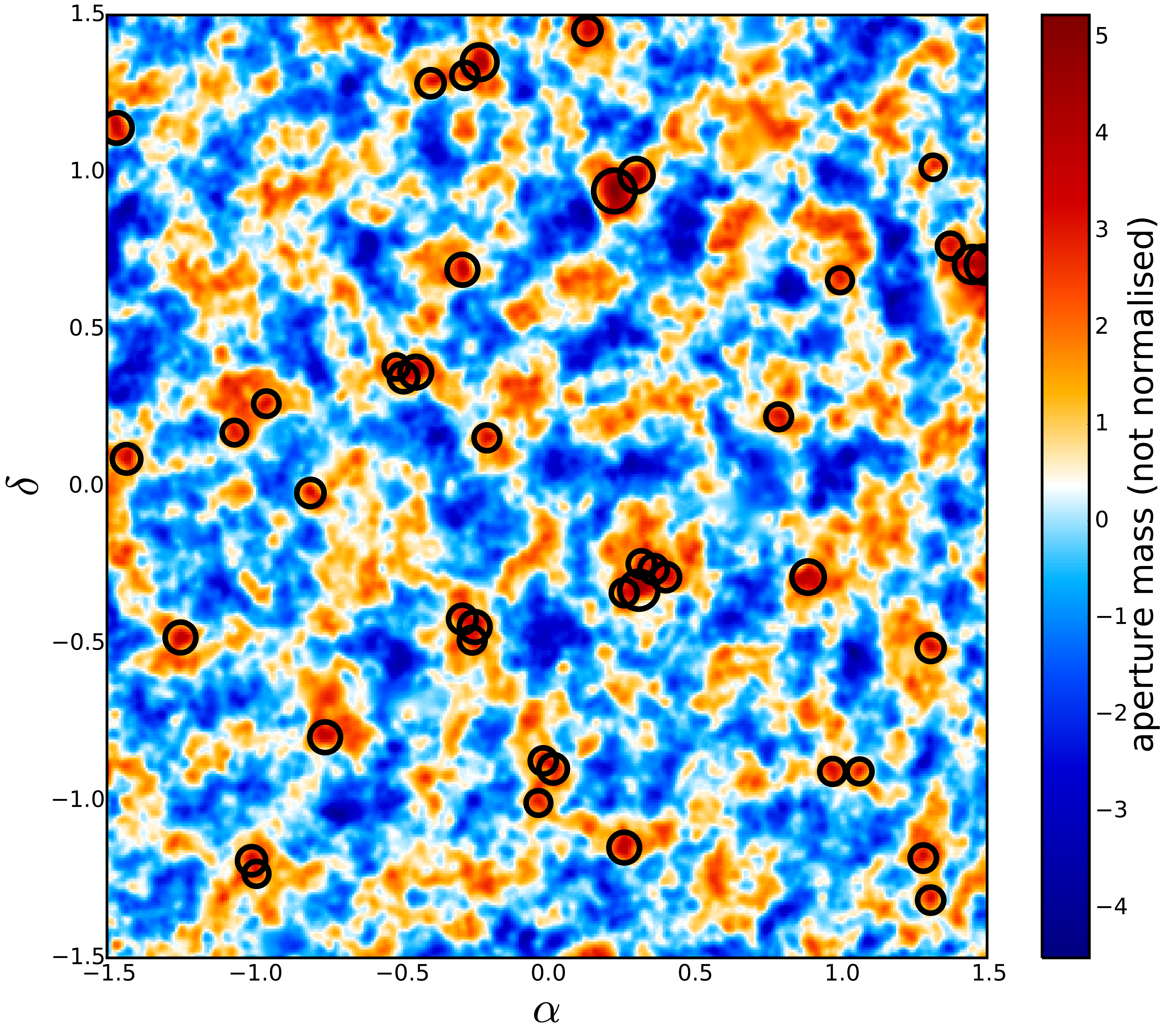}
\end{center}
\caption[]{Example aperture mass map from a 3 x 3 deg patch of DES SV data.
The centre of this patch is ra=79.0, dec=-59.5 [deg]. Black circles identify peaks detected above a $\snr > 3$ threshold, and their size changes with their \snr.
Not all identified peaks are lenses, most of them are actually random noise fluctuations.
The colour bar shows the value of the aperture mass.
\label{fig:example_map}}
\end{figure}

An example {\review DES} map of size $3 \times 3$ deg is presented in Figure \ref{fig:example_map}.
In this map, there were 44 peaks identified above the $\snr > 3$ threshold, marked in black circles.
Not all of these peaks correspond to real clusters, as some of them are created by random noise fluctuations.

\section{Peak functions}
\label{sec:peak-function-measurements}

We construct peak functions from the aperture mass maps.
In our work, we define a peak function to be a count of the number of peaks in bins of their \snr.
Previous works often used binning in \snr, although using the actual values of the $\kappa$ map is also an option \citep{Liuetal2015W}.
The most common choice of the \snr\ range is to focus on the high \snr\ peaks, although \citet{Bard2014} and  \citet{Liuetal2015W}
demonstrated that peaks with very low and even negative \snr\ carry a large proportion of the cosmological information \citep{DietrichHartlap2010,KLW+12}.
Many of the peaks are projections of many halos along the line of sight \citep{Yang2011a, Marian2009}.
Here, we decided to focus on low ($\snr \in [0,2]$) and medium ($\snr \in [2,4]$) \snr\ peaks.
We next detail the considerations that we took into account when determining the number of \snr\ bins and their upper limit.

\subsection{Signal-to-noise range}
\label{sec:signal-to-noise-range}

Given the limited number of realisations of the central cosmology which are used to create the covariance matrix, we focus on using as few \snr\ bins as possible, without significant loss of information.
We use 13 equally-spaced bins, since a larger number does not strengthen cosmological constraints.
{\review We did not consider \snr\ bins that did not have an equal width.
For example, \citet{Dietrich2009} and \citet{Liuetal2015W} used roughly logarithmic bin widths.
}
We leave this sort of binning optimization to future work.
We also verified that our estimated covariance matrix is accurate enough for this length of data vector (see Section \ref{sec:covariances}).

We considered two arguments for deciding upon the value of upper limit of the \snr\ range.
Firstly, the high mass end of the peak function corresponds to big clusters and can carry significant cosmological information \citep[see for example][]{Reischke2015}.
Cluster science also aims to extract that information \citep[see for example][]{Rozo2010, Allen2011}.
However, accurate measurement of cluster mass with weak lensing is a difficult task.
Accurate ``boost factors'' have to be calculated to account for extra galaxies found in the cluster and the decrease in the number of lensing source galaxies due to blending; both these effects cause a decrease in the signal of a peak.
Additionally, intrinsic alignments can significantly change the estimated \snr\ of the peak, especially in the case of non-tomographic analysis.
Both \citet{Applegate2012} and \citet{Melchior2014} used boost correction factors to calibrate cluster masses, and these corrections were on the order of $10\%$.
In this work, we also make an estimate of the impact of the boost factor and intrinsic alignments on the peak \snr, and find that the highest peaks with $\snr > 4.5$ would require corrections of order $>15\%$, which corresponds to modifying the number of peaks by order of $30\%$ (see Section \ref{sec:boost_factors} and Appendix \ref{app:boost_factors}).
Even though the amount of information carried by the high end of the mass function is large, we find that its measurement would be highly dependent on the boost factor and intrinsic allignments modelling.
{\review To avoid this, we choose to use only those \snr\ bins which do not require significant value of boost factor corrections, as compared to the statistical error on the number of peaks in that bin.
We found that when we use $\snr < 4.5$, the measurement of cosmological parameters is not heavily dependent on the application of our estimated corrections.
}

Secondly, as mentioned before, we model the number of peaks as a Gaussian likelihood.
This is only an approximation, as in general the peak count will follow the Poisson distribution,
modified by the impact of sample variance \citep{Hu2003}.
A Gaussian distribution becomes a good approximation to Poisson for mean count of greater than 30.
That is why we require the upper threshold of the highest \snr\ bin to be such that this bin has more than 25 peaks for every cosmological model, including noise peaks.
The highest \snr\ limit was chosen separately for each of the filter scales.
The final upper limits on the $\snr$ was chosen to be $4.4,\ (4.1,\ 4.1)$ for filter scale $\apertsize=12$ (20, 28) arcmin.
The selection of these values was also affected by our choice to keep the \snr\ bin widths constant.

We analysed the maps with a range of filter scales, and found that larger scales ($>10$ arcmin) tend to carry more statistical power.
We decided to use an aperture mass filter with radius of $\apertsize=20$ arcmin as our fiducial model (see Section \ref{sec:aperture_mass_filter}).
Results from other filter scales with $\apertsize=\{ 12, 28\}$ arcmin are also reported, and achieve comparable quality of constraints and similar central value.
We did not attempt to combine different filter scales, as was done in some previous peak statistics works \citep{Marian2012,Liuetal2015Z}; we leave this for the future work.

\subsection{Peak function measurements for DES SV}
\label{sec:peak_functions}

\begin{figure}
\begin{center}
\includegraphics[width=\columnwidth]{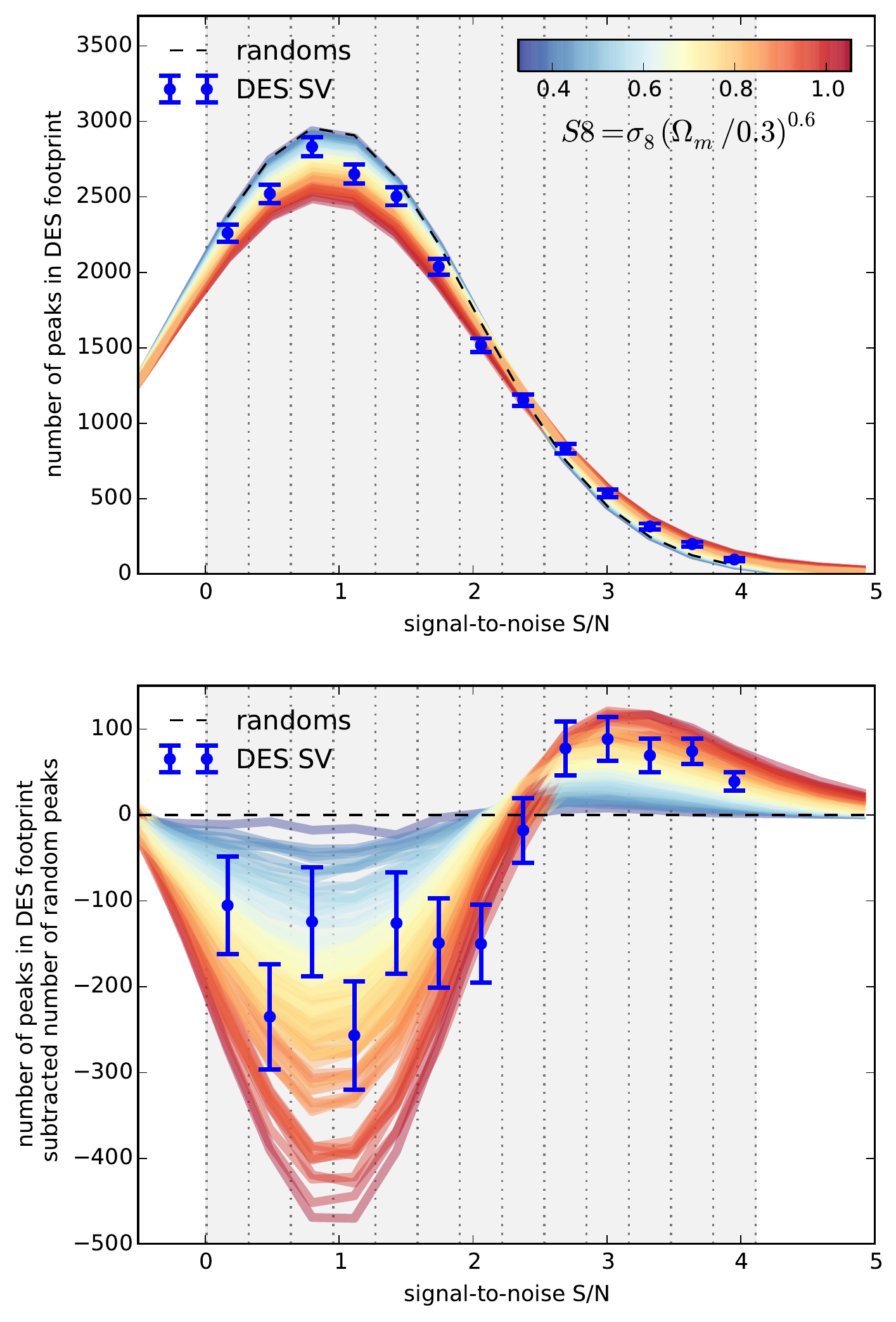}
\end{center}
\caption[]{DES peak functions (blue points) over-plotted on peak functions from simulations with different cosmological models.
The peak count was done using an aperture size of $\apertsize=20$ arcmin.
The blue points with error bars in the top panel show the number of peaks for the full DES SV area.
The black dashed line corresponds to the mean number of random peaks identified in the DES area from 1000 noise realisations of the footprint.
The bottom panel shows the same data, but after subtracting the number of random peaks.
The colour scale shows \SE\ for each cosmological model.
Error bars for the DES measurement are taken from the central model in simulations.
Vertical dotted lines correspond to \snr\ bins boundaries.
The grey shaded area shows the range of \snr\ used in our analysis.
The points on this plot are correlated (see Section \ref{sec:covariances}).
\label{fig:peak_functions}}
\end{figure}

Figure \ref{fig:peak_functions} shows the peak function from the DES data and simulations for the fiducial filter scale of $\apertsize=20$ arcmin.
The top panel shows the number of peaks calculated in the DES footprint for thirteen \snr\ bins.
We also calculated the peak functions from randomised maps.
The bottom panel presents the same peak functions after subtracting the mean number of peaks from the randomised maps, which enables better visual comparison of the DES and simulation results.
The number of noise realisations for randomised maps was 300 times the DES footprint.
The blue points show the DES measurement and the multi-colour lines correspond to peak functions from the simulations, for various combinations of $\Omega_{\rm m}$ and $\sigma_8$ parameters.
The colour corresponds to the value of the \SE\ parameter corresponding to each model.
The error bars on the DES measurement come from the simulations of the central cosmology.

{\review For low \snr, the number of peaks in the randomised shear fields is higher than in fields with cosmological signal.
This can be understood by considering that the observed shear data is a sum of two fields \citep{Liu2014a}:
cosmological shear, with small amplitude and long wavelength; and random shape noise, with large amplitude and small wavelength.
The filter is matched to have a smoothing scale matching the scale of variation of the cosmological shear field.
Consider two extreme cases: a noise-only field and signal-only field, both smoothed with the same filter.
The noise-only field will have more low \snr\ peaks and less high \snr\ peaks than the signal-only field.
The sum of both fields will be a case in between the two extremes: it will have less low \snr\ peaks and more high \snr\ peaks than the noise-only field.
Conversely, in the sum of both fields, we will find more low \snr\ peaks and less high \snr\ peaks than the signal-only field.
\citet{Liu2014a} presents analytical results for number of peaks in the presence of shape noise for the case of convergence fields.
}

\section{Systematics}
\label{sec:systematics}
We consider several systematic effects that may influence our measurement: shear bias, photo-$z$ error, boost factors, intrinsic alignments and the impact of baryons.
We found that shear multiplicative bias and photo-$z$ error have a significant impact on the peak function.
To account for these we marginalise over these systematics in a similar way to \citetalias{TheDarkEnergySurveyCollaboration2015}.
This process is described in sections \ref{sec:shear_bias} and \ref{sec:photoz_error}.
We estimate the impact of the boost factors and intrinsic alignments, describing our results in sections \ref{sec:boost_factors} and \ref{sec:intrinsic_alignments} and Appendices \ref{app:boost_factors} and \ref{app:intrinsic_alignments}.
We found that the potential impact of these effects on high \snr\ peaks can be large. To avoid applying large corrections, we did not include high \snr\ peaks in our analysis.

Baryons will affect the shear peaks in two ways.
Firstly, radiative cooling will cause the enhancement of the halo concentration \citep{Yang2012}, and thus increase the \snr\ of high mass peaks, while leaving the mass of medium peaks mostly unaffected.
The second effect is stellar feedback, which can reduce the mass of small halos and can cause a decrease the count of medium peaks.
\citet{Osato2015a} argue that those effects can partially compensate each other.
Regarding the impact on cosmological parameters, \citet{Osato2015a} showed that omitting the baryonic processes in the emulation of the peak function can cause 1 - 2 \% biases in the $\Omega_{\rm m}$ and $\sigma_8$ parameters.
The level of uncertainty on the measurement from the DES SV data is much larger than the values reported, therefore we do not include the treatment of baryons in our analysis.

We also do not consider corrections for shear additive bias, as we expect the influence of this effect to be small for shear peak statistics.
{\review
Shear additive error is mostly created by leakage of PSF ellipticity into the estimated shear, and it is proportional to PSF ellipticity.
If the PSF ellipticity is constant across the aperture area, then the additive term for all galaxies within that aperture will be the same.
If galaxies are distributed uniformly within the aperture, the aperture mass measurement will not be affected by the additive bias;
the tangential component of the additive systematic with respect to the centre of the aperture will average out.
However, if the galaxies lie on the survey edge mask, a additive systematics may not vanish.
Clampitt et al. (in prep.) presented an analysis of tangential shear around random points in the DES SV data, which does not indicate the presence of additive systematics.
Given that result, we do not apply any further calibrations for additive systematics.
}

\subsection{Shear multiplicative bias}
\label{sec:shear_bias}
Multiplicative shear bias, $m$, is a major systematic error expected in shear catalogues.
It can be caused by various effects arising from the shear measurement process, including noise bias \citep{Kacprzak2012, Refregier2012, Melchior2012}, model bias \citep{Voigt2010, Bernstein2010, Kacprzak2014}, imperfect PSF correction \citep[eg.][]{Paulin-Henriksson2009}.
For an overview of the shear measurement challenges and current state of art methods, see the results from the GREAT3 challenge \citep{Mandelbaum2014}.

Peak statistics will be affected by the shear multiplicative bias, as it directly scales the lensing signal and changes the detection probability of a peak.
\citet{Petri2014} found that it can significantly affect the peak statistics.
In this work, we measure the impact of shear systematic bias, build a model describing the number of peaks as a function of $m$, and then use that model to marginalise the shear bias during the measurement of cosmological parameters.

\citet{Jarvis2015} recommended a Gaussian prior on $m$, centred on 0 with a width of 0.05, be used for the \ngmix\ catalogue.
We include this uncertainty in our cosmological parameter measurement by marginalising out the multiplicative bias with this prior.
First, we have to learn how the peak function reacts to changes in multiplicative bias.
To do this, we create a new suite of simulations for all cosmological models, with a multiplicative shear bias added to the shear, while leaving the sensitivity correction and shape noise the same.
We run these simulations in two configurations, with $m=0.05$ and $m=-0.05$.
We assume a simple first order model, where the change in the peak function following a change in $m$ is linear:
$(N_{\rm peaks}(m) - N_{\rm peaks}(m=0))/N_{\rm peaks}(m=0) = \alpha_{m} m $,
where $\alpha_{m}$ is a factor that we measure from simulations, for each \snr\ bin.
In Appendix \ref{app:systematics_model} we describe this model in more detail.
{\review
In our analysis we define $\alpha_{m}$ to always correspond to the fractional change in number of peaks after subtracting the number of peaks generated from randomised maps.
The number of peaks from randomised maps does not change with varying shear systematics, as any cosmological signal is removed by randomising the shear.
From our simulations we find $\alpha_{m}(\nu) \approx 2$, which is stable across the \snr\ bins.
}
This relation seems to be stable for all cosmological models, as shown in Appendix \ref{app:systematics_model}.
We use this simple model later for creating our likelihood and cosmological parameter inference.

\subsection{Photometric redshift error}
\label{sec:photoz_error}

\citet{Bonnett2015} found that, for the non-tomographic case ($0.3<z<1.3$), the mean redshifts of the lensing sample determined by the four photo-$z$ methods and external data (spectroscopic and COSMOS photometric redshifts) agreed to within a scatter of $\Delta z = 0.02$. When broken-up into tomographic bins the scatter increased to $\Delta z = 0.05$ per bin. These results directly informed the photometric redshift uncertainty prior in the analysis of \citetalias{TheDarkEnergySurveyCollaboration2015}, and must also be taken into account in the present work.
We note that the non-tomographic cosmology from WL 2-pt functions in \citetalias{TheDarkEnergySurveyCollaboration2015} also assumed $\Delta z = 0.05$, and not $\Delta z = 0.02$.
In this work we follow \citetalias{TheDarkEnergySurveyCollaboration2015} and use $\Delta z = 0.05$.

To include the uncertainty on the photo-$z$ estimation we use a similar approach to that applied for shear multiplicative bias.
We aim to create a scaling which relates the change in the mean redshift of the sources to the change in the number of measured peaks.
We run another two configurations of the simulations: with photo-$z$ shifted by $\Delta z= -0.05$ and $\Delta z= +0.05$.
This way we are able to measure the parameter $\alpha_{\Delta z}$, which quantifies this change for each \snr\ bin:
$(N_{\rm peaks}(\Delta z) - N_{\rm peaks}(\Delta z=0))/N_{\rm peaks}(\Delta z=0) = \alpha_{\Delta z} \Delta z$.
Similar to the case of multiplicative bias, we find that a simple first order model is sufficient to describe the impact of the shift of mean $n(z)$.
We find $\alpha_{\Delta z}(\nu) \approx 3$ for four \snr\ bins, and this result is stable across redshift bins and cosmological models.
See Appendix \ref{app:systematics_model} for details.
We use this model in the cosmological inference process.

\subsection{Boost factors}
\label{sec:boost_factors}
{\review
The strength of the shear signal will vary with the distance of source galaxies to the lens, and will be the highest when the distance between the lens and the source galaxy is roughly the same as the distance between the lens and the observer.
This will cause the detection probability of a peak to depend on the distribution of the redshifts of source galaxies at the position of a peak.
}
It is therefore important to make sure that the redshift distribution at the position of the peaks is the same in the DES data and in the simulations.
The procedure described in section \ref{sec:simulations} makes sure that the $n(z)$ for the simulations is the same as in the DES survey, up to photo-$z$ error, for the wide field.
However, the $n(z)$ is varying across the survey, both according to depth and to galaxy clustering.
In this work we address only the latter.
At positions close to clusters, we expect to observe more galaxies than in the wide field.
These extra galaxies are cluster members and do not carry any lensing signal from that cluster, as they reside at the same redshift.
In the simulations the galaxy density is decorrelated from the dark matter density and this can cause a difference between detection probability between the DES survey and the simulations.

Additionally, the large number of cluster galaxies can cause a reduction of the number of source galaxies, due to blending.
When measuring cluster masses in early SV data,
\citet{Melchior2014} used the \balrog{} framework \citep{Suchyta2015} to derive boost factors to correct for this effect.
In this work, we use a similar approach to \citet{Melchior2014} to investigate boost factors for every \snr\ bin,
analysing the number of galaxies surrounding the peaks when using each of three samples: the DES data, the simulations, and the \balrog\ catalogues;
the full procedure is described in Appendix~\ref{app:boost_factors}.

We find that the boost factor corrections would be low, generally below $5\%$.
The dilution of the signal by extra cluster galaxies is minimal ($<2\%$).
The effect of background galaxies lost due to blending is more prominent and can cause a $5\%$ change in the \snr\ of the most massive peaks included in our analysis ($\snr>3.8$).
In our analysis we created a combined correction for boost factors and intrinsic alignments (see Section \ref{sec:intrinsic_alignments}), and limited our analysis to those \snr\ bins for which these corrections do not significantly change the cosmology constraints.

Note that these corrections may be different from those usually reported in cluster lensing science, as they were calculated using sets of peaks which also include spurious peaks from random noise.
{\review Additionally, cluster masses and boost factors used to correct them are derived by looking at the source galaxies, which are selected such that they have a higher redshift than the cluster.}
This removes most of the cluster member galaxies, with only a fraction of the cluster members leaking into the source sample.
Here we do not identify the redshifts of peaks and do not modify the sample of sources according to the position on the sky.
This means that all the cluster galaxies will be included in the estimation of \snr\ of the peak, which can cause the boost factors to be quite different in our work than those calculated in cluster lensing.
Furthermore, many of the peaks will not be placed at the position of a large cluster, and arise due to chance projections of few smaller halos along the line of sight \citep{Yang2011a,Marian2009}.

\subsection{Intrinsic alignments} 
\label{sec:intrinsic_alignments}

The distortion of a galaxy's shape due to WL is very small. The resulting shear is $\sim 1\%$ of the amplitude of a typical galaxy's intrinsic ellipticity. Measurements of WL, whether peak counts or WL 2-pt statistics, rely on the averaging of the shapes of many galaxies. If the galaxies' intrinsic ellipticities are randomly distributed then they will average to zero and the resulting statistic will be sensitive to the WL, as desired.

In reality, it is very possible for processes during the epoch of galaxy formation to produce populations of galaxies whose intrinsic ellipticity is correlated. We call this effect intrinsic alignment (IA) \citep{HRH2000,CKB01,HS04}. These IA correlations will contaminate measured cosmic shear signals. This has been extensively treated in the WL 2-pt case \citep{BK07, JB10, heymans13, KRH+12}, where IAs source two additional terms which sum with the pure WL signal to produce the observed correlation function. One of the terms (the II, or Intrinsic-Intrinsic correlation) is positive and sourced by the correlation of physically close galaxies, while the other (the GI, or Gravitational-Intrinsic cross-correlation) is negative and sourced by the correlation of galaxies on the same patch of sky but separated along the line of sight. IAs have also been considered in WL 3pt measurements and galaxy-galaxy lensing \citep{SJS10,TI12a,TI12b,BMS+12}.

WL peak counts will also experience the effect of IAs. In this case, because peaks are identified through a filtered sum of the lensing signal along a given line of sight, any IA where galaxies align radially with structures along that line of sight will produce a negative contribution to the resulting mass intensity. Imagine a particular line of sight containing a single massive cluster that produces a large integrated mass intensity. If the cluster member galaxies are aligned such that the satellite galaxies point towards the cluster centre they will enter negatively into our filtered sum of ellipticities, reducing the total observed mass intensity. We consider only radial IA within clusters in this paper as this is in keeping with previous attempts to model IAs using a halo model \citep{Schneider2010}, which we build upon here. The assumption of radial alignment is also consistent with the latest observations, which support radial alignment among more luminous (red) galaxies \citep{LWY+13,SMM14} and see no evidence for any alignment of fainter (red and blue) satellite galaxies \citep{CMS+14,SHC+15}. There is no evidence for tangential alignment of satellite galaxies with respect to the central galaxy position among any population.

Of course, for a galaxy's alignment to influence the peaks count, it must be included in the selection of sources. This means it is possible to reduce the impact of IAs by restricting the source selection. For example, a selection of sources which are in a narrow redshift range, centred on a relatively high redshift, would be expected to experience little IA effect because the galaxies associated with the mass fluctuations producing peaks are at significantly lower redshift and not included in the source selection (assuming redshift estimates are accurate for all galaxies). Peaks produced by random noise will not suffer from IAs in any systematic way. In this paper we use a broad redshift range for our source population, therefore it is important to consider the effect of IAs.

We can model the effect of IAs on peak counts by assuming that IAs only affect peak counts through the alignment of satellite galaxies inside individual halos along those lines of sight identified as peaks. In these cases we can follow the IA halo model of \citet{SB10}, where satellite galaxies are assumed to be aligned towards the halo centre with some misalignment angle, $\beta$, between the satellite galaxy major axis and the radial vector of the halo. The distribution of this misalignment angle was derived from simulations and found to reduce the IA of satellite galaxies by a factor of $\bar{\gamma}_{\rm scale}=0.21$, where unity would represent perfectly aligned galaxies. Other measurements from simulations find a similar value \citep[see][for example]{TMM15}. The strength of the IA ellipticities can be assumed to be equal to the intrinsic ellipticity distribution of the satellite population. We describe how this model could be integrated into our peak-counting formalism in Appendix \ref{app:intrinsic_alignments}.
This very simple model calculates an expected change in the \snr\ of a peak, given a number of cluster member galaxies, background galaxies, and a fixed value of the $\bar{\gamma}_{\rm scale}$ parameter.

We use this model to estimate the expected corrections that would have to be applied to modify the value of the peak function in the simulations in order to account for intrinsic alignments.
We find that for the majority of our chosen \snr\ range the corrections would be very low, and for the few highest bins the \snr\ needs to be scaled down by a factor of $\sim 15\%$.
We combine this correction with the boost factor corrections, as shown in Appendices \ref{app:boost_factors} and \ref{app:intrinsic_alignments}.
Figure \ref{fig:boost_corr} shows these corrections, which are most important in relatively high \snr\ bins.
As we combine information from many \snr\ bins, and only the highest bins need boost factor/ intrinsic alignments corrections, we do not expect the final result to heavily depend on these.
With that in mind, we do not apply these corrections for our main cosmological results and other variants.
We explore the analysis variant where the combined correction is applied in Section \ref{sec:constraints_variants}.

\section{Inference}
\label{sec:inference}
In this section we describe the process of inferring cosmological constraints from the measured peak functions.
The steps involved in the inference process are: the construction of the likelihood function, evaluating this function within a specific prior and marginalising the systematic errors.
We consider a Gaussian likelihood with a covariance matrix derived from simulations.
The likelihood is evaluated on a four-dimensional, densely sampled grid.

\subsection{Likelihood analysis and covariance matrix estimation}
\label{sec:covariances}

Let $\hat{\boldsymbol{d}}$ be our vector of measured data points, i.e. the number of shear peaks in different bins of signal-to-noise. In order to derive cosmological parameter constraints from our data we assume that $\hat{\boldsymbol{d}}$ has a multivariate Gaussian distribution (see Section \ref{sec:peak_functions} for the description of the noise model).  Due to shape noise and cosmic variance it fluctuates around a mean value $\boldsymbol{d}$ as
\be
\hat{\boldsymbol{d}} \sim \mathcal{N}(\boldsymbol{d}, \Sigma)\ ,
\ee
where $\Sigma$ is the covariance matrix of our data points. Both $\boldsymbol{d}$ and $\Sigma$ depend on the choice of cosmological parameters and nuisance parameters which we both denote with $\boldsymbol{\pi}$. In our analysis we consider (cf. section \ref{sec:cosmological_constraints})
\begin{equation}
\boldsymbol{\pi} = \{ \Omega_{m}, \sigma_{8}, m, \Delta z \}.
\end{equation}
In a Bayesian approach we assign a posterior probability density to our parameters from our measurement of $\hat{\boldsymbol{d}}$ as
\be
p(\boldsymbol{\pi} | \hat{\boldsymbol{d}}) = \frac{p(\hat{\boldsymbol{d}} | \boldsymbol{\pi}) \ p(\boldsymbol{\pi})}{p(\hat{\boldsymbol{d}})}\ ,
\ee
where $p(\hat{\boldsymbol{d}} | \boldsymbol{\pi})$ is the probability of measuring $\hat{\boldsymbol{d}}$ if the true parameters are $\boldsymbol{\pi}$, $p(\boldsymbol{\pi})$ is a suitably chosen prior density in parameter space (cf. sections \ref{sec:data} and \ref{sec:systematics} for the priors on our nuisance parameters) and $p(\hat{\boldsymbol{d}})$ is just a normalization constant. Under our Gaussian assumption the density $p(\hat{\boldsymbol{d}} | \boldsymbol{\pi})$ is given by
\be
p(\hat{\boldsymbol{d}} | \boldsymbol{\pi}) \sim \exp\left( - \frac{1}{2} \chi^2(\hat{\boldsymbol{d}}, \boldsymbol{\pi}) \right),
\ee
with
\be
\chi^2(\hat{\boldsymbol{d}}, \boldsymbol{\pi}) = \left(\hat{\boldsymbol{d}} - \boldsymbol{d}(\boldsymbol{\pi})\right)^T \Sigma^{-1}(\boldsymbol{\pi}) \left(\hat{\boldsymbol{d}} - \boldsymbol{d}(\boldsymbol{\pi})\right),\
\ee
where $\boldsymbol{d}(\boldsymbol{\pi})$ is the vector of values of our data points for a set of cosmological models $\boldsymbol{\pi}$.
Our modelling of $\boldsymbol{d}(\boldsymbol{\pi})$ is described in Section \ref{sec:interpolation}. In order to estimate the covariance $\Sigma$ we use the N-body simulations that were described in Section \ref{sec:simulations}. Each simulated realisation provides an independent realisation of our data vector $\hat{\boldsymbol{d}_i}$, $i = 1, \dots\ , N_s$, where $N_s = 35 \cdot 300 = 10,500$ since our central cosmology was simulated $35$ times and for each simulation an additional $300$ noise realisations were generated.
In the case where we apply boost factors to our measurement, we also apply randomly drawn boost factors/IA corrections according the error bars shown in Figure \ref{fig:boost_corr} to the signal in our mock catalogues in order to account for our uncertainties in this correction (cf. Appendix \ref{app:boost_factors}). The sample covariance estimate from these realisations is given by
\be
\hat\Sigma = \frac{1}{N_s-1} \sum_{i = 1}^{N_s} \left(\hat{\boldsymbol{d}_i} - \bar{\boldsymbol{d}}\right)\left(\hat{\boldsymbol{d}_i} - \bar{\boldsymbol{d}}\right)^T,
\ee
where $N_s = 10,500$ and $\bar{\boldsymbol{d}}$ is the mean value of all measured data vectors.
Note that in this way we are ignoring the cosmology dependence of the covariance matrix, which, according to \citet{Eifler2009}, can have a significant impact on likelihood contours. However, in the absence of a precise modelling of the covariance matrix or a large set of simulations with different cosmological parameters there is no alternative to our procedure.

Under our Gaussian assumption $\hat\Sigma$ follows a Wishart distribution and hence its inverse $\hat\Sigma^{-1}$ is not an unbiased estimate of the true inverse covariance matrix $\Sigma^{-1}$. This can however be corrected with a multiplicative factor \citep{Kaufman, Hartlap2007, Taylor2013} and an unbiased estimate of $\chi^2(\hat{\boldsymbol{d}}, \boldsymbol{\pi})$ is given by
\be
\label{eq:hartlap_chi_sq}
\hat\chi^2(\hat{\boldsymbol{d}}, \boldsymbol{\pi}) = \frac{N_s-N_d-2}{N_s-1} \left(\hat{\boldsymbol{d}} - \boldsymbol{d}(\boldsymbol{\pi})\right)^T \hat\Sigma^{-1} \left(\hat{\boldsymbol{d}} - \boldsymbol{d}(\boldsymbol{\pi})\right) ,
\ee
where $N_d$ is the dimension of our data vector. Note that we have set the number of realisations in Equation \ref{eq:hartlap_chi_sq} as $N_s=10,500$, which is the total number of shape noise realisations and shape noise is the dominant noise contribution.

We have checked this assumption by making a simple estimate of the fractional uncertainty on the parameter errors derived from our covariance matrix. Using a noisy covariance estimate from a finite number of realisations introduces uncertainties to the constraints derived for cosmological parameters, i.e. uncertainties on the uncertainties we assign to those parameters. Let $p$ be a parameter of our model and $\delta p$ the uncertainty in that parameter derived from our likelihood contours in parameter space. Then, assuming a Gaussian data vector and also a Gaussian likelihood in parameter space, \citet{Taylor2014} derived the fractional uncertainty on $\delta p$ to be
\begin{equation}
\label{eq:uncertainty_on_uncertainty}
\frac{\Delta \delta p}{\delta p} \approx \frac{\sqrt{2(N_s-N_d + N_p -1)}}{(N_s-N_d - 2)},
\end{equation}
where $N_p$ is the overall number of constrained parameters.

Inserting $N_s = 10,500$ into Equation \ref{eq:uncertainty_on_uncertainty} yields a fractional uncertainty of $\sim 1.5\%$. We made an additional estimate of this quantity by using jackknife resampling to estimate the standard deviation of $\delta S_8$, the error on the $S_8$ parameter. In this jackknife we removed 300 of our 10,500 peak function realisations at a time, re-calculating the covariance matrix each time. For each jackknife re-sampled covariance matrix we calculated the central value of $S_8$, as well as the upper and lower $1$-$\sigma$ deviation from that value. With these jackknife estimates we then calculated the standard deviation of $\Delta S_8$, finding a value of $\sim 2\%$. The good agreement between the jackknife estimate and the result of Equation \ref{eq:uncertainty_on_uncertainty} indicates that our modelling of the covariance from the shape noise realisations is accurate.

The clustering of sources with the mass peaks is not realistic in our simulations.
In Section \ref{sec:boost_factors} and Appendix \ref{app:boost_factors} we discussed the impact of this problem on the peak counts.
{\review This mismatch between the simulations and the DES data may also have an impact on the covariance matrix of the peak function.}
We expect it to increase the covariance slightly, but we do not account for this in our analysis.
In principle this is possible to use simulations with realistic clustering, but then a process of applying the DES mask would require more investigation; it will no longer be possible to use exactly the same positions of galaxies in simulations as we observed in the survey, which is what we did in this work.
In the future it will be important to be able to quantify the joint impact of galaxy clustering and mask effects.

\subsection{Interpolation schemes}
\label{sec:interpolation}

The empirical prediction of the number of peaks as a function of cosmological parameters is done on a finitely sampled grid of simulations.
To obtain the likelihood for points in the $\Omega_{\rm m}$ and $\sigma_8$ plane that do not lie on the grid, we have to interpolate and extrapolate from the measured grid points.
This method has been used in previous studies, for example \citet{Liuetal2015Z} used two interpolation methods based on Gaussian process and radial basis functions.
In this work, we obtain the likelihood in two fundamentally different ways: (a) by interpolating the number of peaks for every \snr\ bin through a basis expansion in $\Omega_{\rm m}$ and $\sigma_8$, and (b) by interpolating the $\chi^2$ for each simulation using radial basis functions.
The first approach is the fiducial method, and the second is used to test the robustness of the fiducial result.

Details of these methods are given in appendix \ref{app:peakfun_modelling}.
We find that switching interpolation method makes little difference to our derived cosmology.
The difference in central value of $S_8$ for both schemes is close to 1\%, see Section \ref{sec:constraints_variants} for comparison.

\section{Cosmological constraints}
\label{sec:cosmological_constraints}

With only four parameters to consider, we can calculate the likelihood of our data given the cosmological and systematic parameters on a four dimensional grid.
We model this likelihood as a multivariate Gaussian, with covariance matrix calculated from the simulations; see Section \ref{sec:covariances} for more details.
As detailed in Section \ref{sec:peak-function-measurements}, we chose our highest \snr\ bin such that the Gaussian likelihood will remain a good approximation, {\review which is the case for number of observations greater than 25.}
The size of the parameter grid is chosen to be sufficiently large so that any further increases in its size do not bring any changes to the result.
To marginalise the systematic errors, we sum the probability along the corresponding directions in the grid, having normalised the likelihood cube to unity.

In this section we present the fiducial constraints from shear peaks in \ref{sec:constraints_fiducial} before examining the effect of different analysis assumptions in \ref{sec:constraints_variants} and comparing results from the peaks analysis to those from DES WL 2-pt statistics in \ref{sec:constraints_variants_2-pt}.

\subsection{Fiducial result}
\label{sec:constraints_fiducial}

\begin{figure}
\begin{center}
\includegraphics[width=\columnwidth]{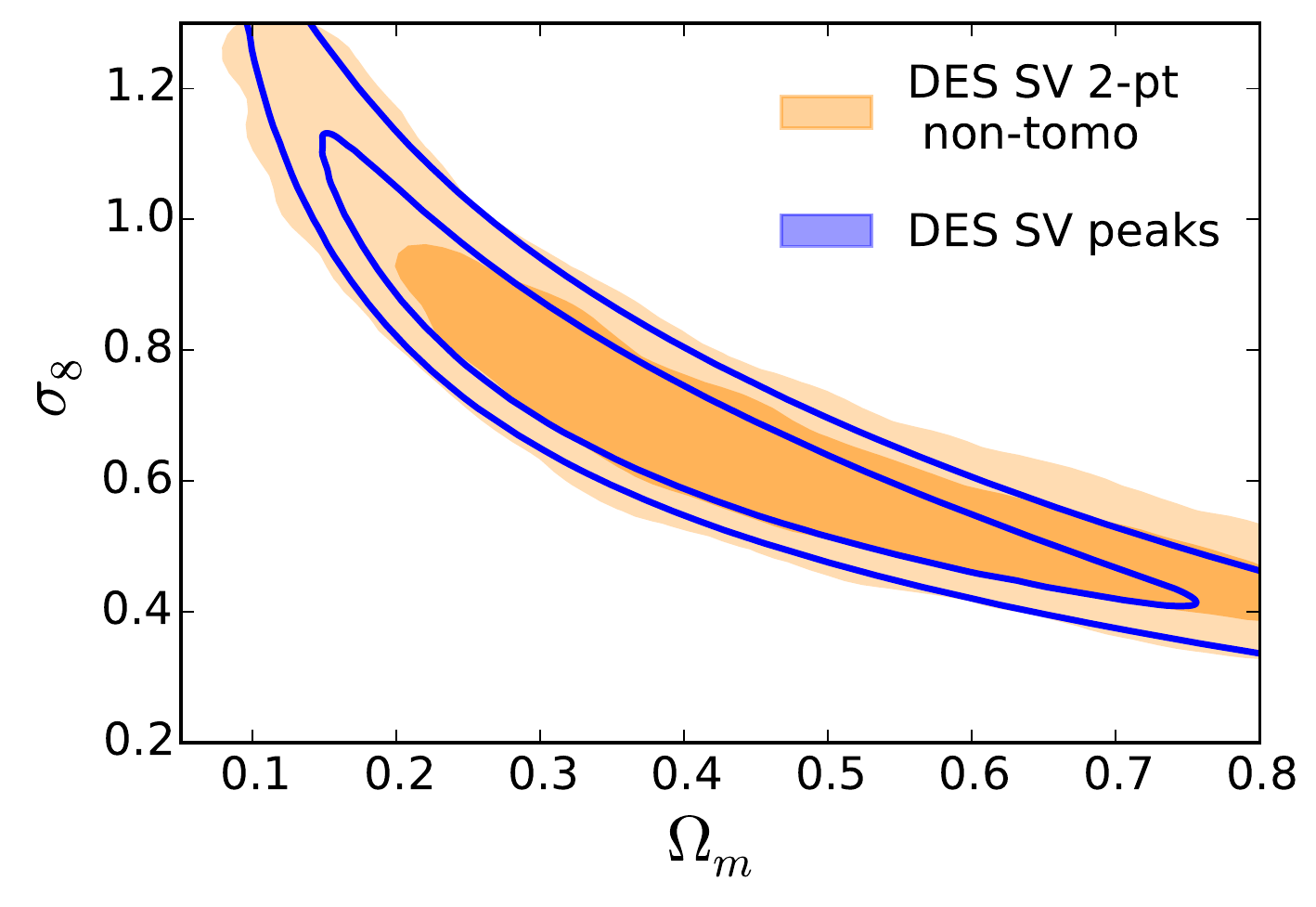}
\end{center}
\caption[]{Constraints on $\Omega_{\rm m}$ and $\sigma_8$ from peak statistics in DES SV (blue contour), compared to equivalent constraint from DES {\review cosmic shear} 2-pt functions (orange).
The contours represent the 68\% and 95\% confidence limits.
Across the $\Omega_{\rm m} / \sigma_8$ degeneracy, the uncertainty on the measurement with peak statistics is $S_8 = \sigma_{8}(\frac{\Omega_{\rm m}}{0.3})^{\alpha}=0.77 \pm 0.07$ with best fitting $\alpha=0.6$.
These constraints include marginalisation over systematic errors: shear bias and error in the mean of the redshift distribution.
The orange contours show the constraints from the non-tomographic DES SV WL 2-pt measurement, with other cosmological parameters set to the same values as used in the simulations for peak statistics:
$h=0.7$, $\Omega_{\rm b}=0.04$ and $n_s=1$.
They also include marginalisation of the systematic errors with the same priors.
\label{fig:constraints_fiducial}}
\end{figure}

Figure \ref{fig:constraints_fiducial} shows the fiducial constraints on $\Omega_{\rm m}$ and $\sigma_8$ from shear peak statistics using our main analysis pipeline, marginalised over both photo-$z$ and shear measurement nuisance parameters. The corresponding measurement of \SE\ is shown in Figure \ref{fig:constraints_S8}. The maximal constraint on the $\sigma_8$-$\Omega_{\rm m}$ degeneracy in the case of our peaks analysis is given by $S_8 = \sigma_{8}(\frac{\Omega_{\rm m}}{0.3})^{\alpha}=0.77 \pm 0.07$, with $\alpha=0.6$.
We found the best fitting $\alpha=0.58$, and we set it to $\alpha=0.6$ for the rest of the analysis.
Changing the slope of \SE\ to $\alpha=0.5$, the direction of maximal sensitivity in the WL 2-pt analysis, has very little effect on the main peaks analysis, changing the best-fit value from $S_8 = 0.77$ to $S_8 = 0.76$ and increasing the error bars by $5.7\%$.

\subsection{Comparison of results for different variants}
\label{sec:constraints_variants}

\begin{figure*}
\begin{center}
\includegraphics[width=\textwidth]{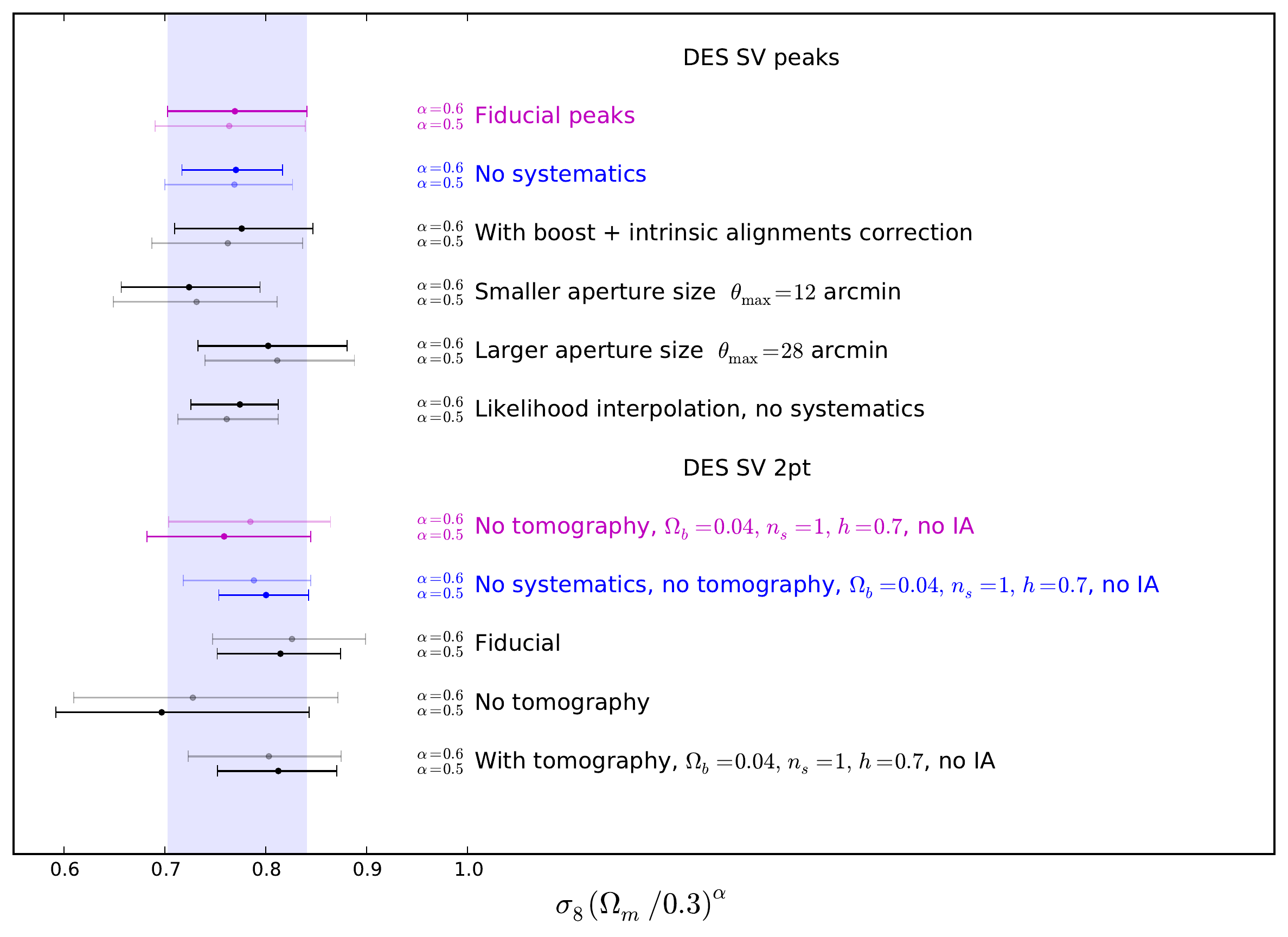}
\end{center}
\caption[]{Constraints on $S_8 = \sigma_{8} (\Omega_{m}/0.3)^{\alpha}$ from the DES SV shear peaks and cosmic shear 2-pt analyses for the fiducial analysis method and several variants. Each analysis variation has 68\% confidence limits shown by horizontal lines, with the best-fit values identified by dots. The vertical blue region is used to highlight the parameter range in agreement with our fiducial peaks analysis. Each analysis variant is described by text on the right hand side of the plot. For each analysis variant results are shown for both $\alpha=0.6$ (the direction of optimal constraint for the peaks analysis) and $\alpha=0.5$ (the same for the 2-pt analysis). The more (less) constraining choice of $\alpha$ for each observable is presented in bold (faint). The colour-coding of the results is a guide to the most comparable constraints between the two observables.
\label{fig:constraints_S8}}
\end{figure*}

We have tested a number of variants to our main peaks analysis. The main alternate analysis methods are displayed in Figure \ref{fig:constraints_S8}. Unsurprisingly, the greatest change in constraining power comes when we choose to fix the nuisance parameters for both photo-$z$ errors and shear measurement bias at zero, rather than marginalising over them. This decreases the errors by $29\%$ ($15\%$) for $\alpha=0.6$ (0.5).
The best-fit values of \SE\ are almost unchanged when we fix our nuisance parameters, as expected when Gaussian priors with zero mean are applied.

The implementation of the boost/IA corrections, described in Section \ref{sec:boost_factors}, has a very limited impact on the peaks constraints, as designed in the analysis process;
we purposefully limited the range of \snr, such that the clusters with possible high and uncertain boost factors are excluded from our analysis.
We compare the systematics-free constraints to equivalent calculated with the alternate interpolation, described in Section \ref{sec:interpolation}.
The results are consistent, with the $\Delta \SE < 0.01$ change in the best-fit value.
The error contours are slightly smaller for the alternate interpolation.

Changes in filter size have a more noticeable effect on the constraints on \SE. A reduction from $\apertsize=20$ arcmin to $\apertsize=12$ arcmin changes the best-fit value to $\SE = 0.72 \pm 0.07$, while an increase to $\apertsize=28$ produces a constraint of $\SE = 0.80^{+0.08}_{-0.07}$.
To assess the significance of these differences, we investigated the expected level of correlation between constraints from these filter sizes, by looking at the results of simulation from the central model.
We found that the correlation coefficient between measurements of the $S_8$ parameter from filter sizes of $\apertsize=20$ arcmin to $\apertsize=12$ and $\apertsize=28$ to be both $\approx 0.5$.
Accounting for that correlation we estimate this difference to be on the $1\sigma$ significance level.
This indicates that the results from both smaller and larger filter sizes are entirely consistent with the result from our main analysis.

\subsection{Comparison of peaks and 2-pt measurements}
\label{sec:constraints_variants_2-pt}

We also use Figure \ref{fig:constraints_fiducial} to compare the constraints from our peaks analysis to similar measurements using DES SV WL 2-pt measurements. These measurements were run especially for this current analysis but, to make a fair comparison, we make sure we use exactly the same configuration of input parameters for the 2-pt chains as those used in the shear peaks analysis, setting $h=0.7$, $\Omega_{\rm b}=0.04$ and $n_s=1$, and ignoring intrinsic alignments.
For that configuration of the 2-pt analysis, we obtain $\SE=0.78 \ (0.76) \pm 0.08 \ (0.08)$ for $\alpha=0.6$ (0.5).
The constraints from peak and 2-pt statistics are very close: the best-fit values from the two different observables differ by $\Delta S_8 \approx 0.015$.

\begin{figure}
\begin{center}
\includegraphics[width=\columnwidth]{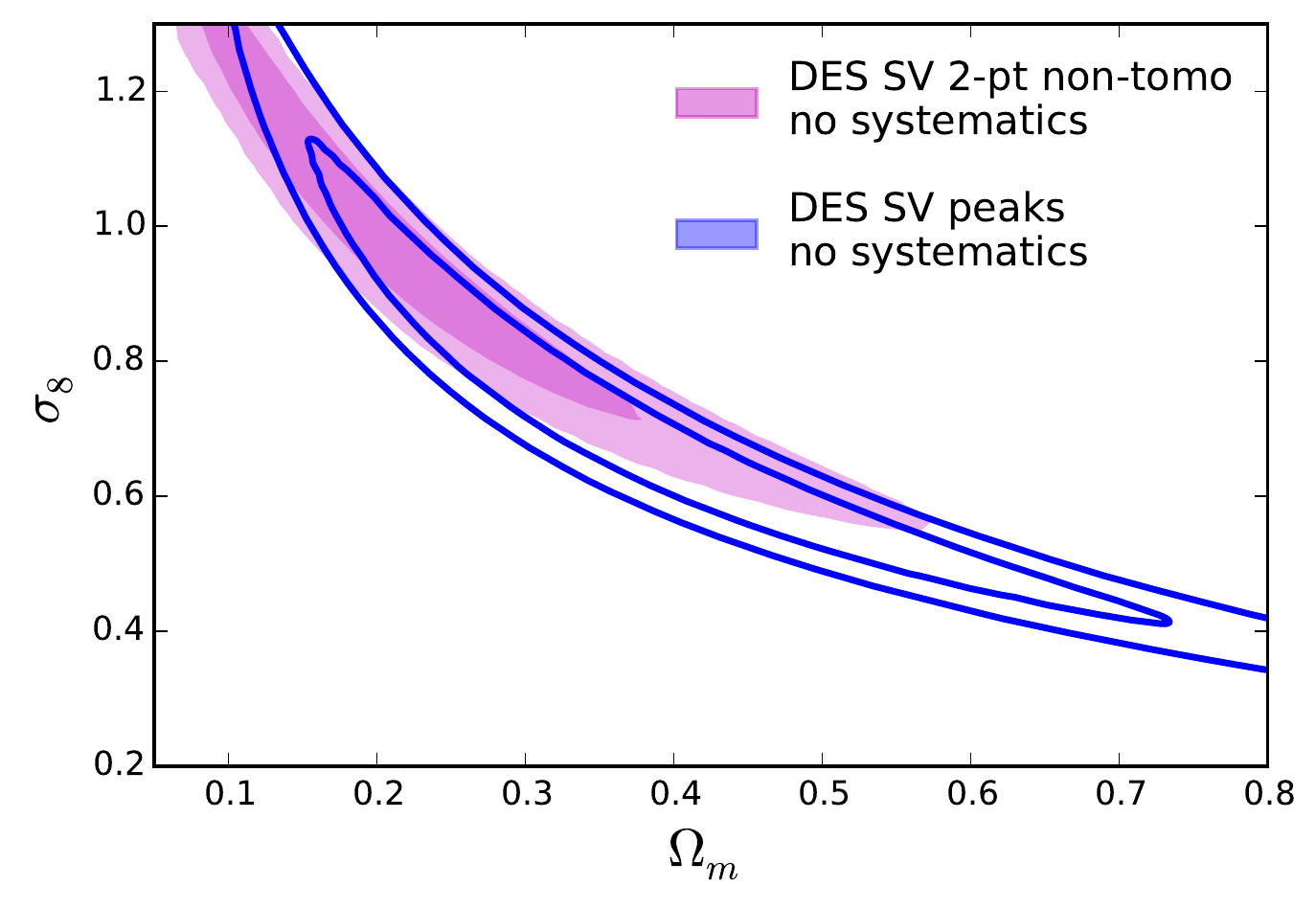}
\end{center}
\caption[]{Comparison between constraints from peak statistics (blue solid lines) and {\review shear} 2-pt functions (pink contours), with systematics excluded.
Both constraints show the 68\% and 95\% confidence limits.
To make a fair comparison, we did not marginalise other parameters for the 2-pt constraints and we set $h=0.7$, $\Omega_{\rm b}=0.04$ and $n_s=1$, to match the configuration for simulations.
\label{fig:constaints_variants_nosyst}}
\end{figure}

We compare the results with systematics ignored in Section \ref{sec:constraints_variants}, Figure \ref{fig:constaints_variants_nosyst}.
For the 2-pt constraint, other cosmological parameters were fixed as before and intrinsic alignments were ignored, which makes it a fair comparison.
{\review
Fixing the systematic nuisance parameters to zero, rather than marginalising over them, has a somewhat different effect for WL 2-pt and peak statistics.
}
For the 2-pt analysis, the constraints shrink by $\sim 20\%$ ($\sim 46\%$) when $\alpha=0.6 \ (0.5)$ which produces a shift in the best-fit \SE\ of $<12\%$ ($<40\%$).
For peak statistics, the central value remains almost unchanged, and the error shrink by 25\% (15\%) for $\alpha=0.6$ (0.5).

{\review Let's now compare the peak statistics results with the fiducial tomographic constraints in \citetalias{TheDarkEnergySurveyCollaboration2015}}.
The fiducial result from the 2-pt analysis uses three-bin tomography ($0.3<z<0.55,\ 0.3<z<0.55,\ 0.3<z<1.3$) and varies five cosmological parameters ($\Omega_{m},\ \sigma_{8},\ \Omega_{b},\ h,\ n_{s}$) and seven nuisance parameters ($m_{1},\ m_{2},\ m_{3},\ \delta{z_{1}},\ \delta{z_{2}},\ \delta{z_{3}},\ A_{\rm IA} $).
This produced a fiducial constraint of $S_8 = \sigma_{8}(\frac{\Omega_{\rm m}}{0.3})^{0.5}=0.81^{+0.062}_{-0.060}$.
When we re-analyse this chain with $\alpha=0.6$ we see slightly larger errors and a slightly higher best-fit value, $S_8 =0.83^{+0.08}_{-0.07}$. Both of these constraints are compatible with our fiducial peaks analysis, showing best-fit \SE\ values at the upper end of the peaks 68\% confidence region.
However, we should not expect the fiducial results from the WL 2-pt analysis to be entirely consistent with that of the peaks analysis. For one thing, the 2-pt analysis is tomographic, while the peaks analysis is an integral along the entire line of sight. Secondly, the marginalisation over intrinsic alignments uncertainty is included in the fiducial WL 2-pt analysis.
Finally, the shear peaks analysis is using a cosmological model with fixed $\Omega_{\rm b}$, $h_{70}$ and $n_s$.

Our analysis was done for fixed $\Omega_{\rm b}$, $h_{70}$ and $n_s$.
It is interesting to check how much of an impact it would have if these parameters were marginalised, in a similar way as it is done for the WL 2-pt.
As this is not available for us due to limited simulations space, we can only investigate the impact it has on the 2-pt function.
We ran additional chains for the fiducial setup of the 2-pt analysis, but with $\Omega_{\rm b}=0.04$, $h=0.7$ and $n_s=1$.
The \SE\ measurement for this configuration is shown in Figure \ref{fig:constraints_S8} under the entry: `with tomography, $\Omega_{\rm b}=0.04$, $h=0.7$ and $n_s=1$'.
Both central values for that configuration are very close to those from the fiducial one, and the errors are only slightly improved.
That indicates that marginalising over these parameters, with the priors used in the 2-pt analysis, does not have much impact on the constraints.
If we assume that the shear peak statistics respond similarly to changes in these parameters, we can expect the errors on \SE\ to be increased only by a small amount.

\section{Conclusions}
\label{sec:conclusions}

We performed a shear peak statistics analysis of the Dark Energy Survey Science Verification data set, described in \citet{Jarvis2015}.
We created aperture mass maps from the DES area and from the set of N-body simulations from \citet{Dietrich2009}, edited to replicate the DES mask, shape noise and galaxy redshift distribution.
Mass peaks were counted in bins of low and medium \snr, spanning the range between 0 and $\sim 4.5$.
We did not use the high \snr\ peaks, despite the fact that we found them to carry a large amount of cosmological information.
This is because the boost factor and intrinsic alignment corrections estimated in our analysis (see sections \ref{sec:boost_factors} and \ref{sec:intrinsic_alignments} and appendices \ref{app:boost_factors} and \ref{app:intrinsic_alignments}) are larger and more uncertain for high \snr\ peaks.
These boost factors capture the effects of cluster member galaxies and loss of source galaxies due to enhanced blending at the positions of most massive clusters.
Intrinsic alignment will further decrease the \snr\ of a peak.
These effects were not modelled in the simulations available for use.

We include uncertainties on shear multiplicative bias and the mean of the redshift distribution in our analysis.
We found both these systematics affected the observed peak function significantly: adding 5\% multiplicative bias changes the peak count by roughly 10\%, and changing the mean redshift of sources by $\Delta z= 0.05$ induces a $\sim 15\%$ change in the value of the peak function.
The effect of these systematics is marginalised in the cosmological inference process.
Their overall impact on cosmological constraints seems to be comparable to the one induced by them on the WL 2-pt functions.

The cosmological constraints for the $\Lambda$CDM model with fixed $h=0.7$, $\Omega_{\rm b}=0.04$ and $n_s=1$ from DES SV peak statistics are
$S_8 = \sigma_{8}(\Omega_{\rm m}/0.3)^{0.6}=0.77 \pm 0.07$.
We checked the robustness of this result against the choice of interpolation scheme used to create the likelihood and the impact of our estimated boost factor correction, finding the results to be stable.
Results for varying filter scale away from the fiducial $\apertsize=20$ to $\apertsize=12$ and $\apertsize=28$ (arcmin) showed slight deviation, on the level of $1\sigma$, which is expected given the level of correlation in signal coming from these aperture sizes.

We compare the peak statistics results to the equivalent constraints from the DES SV WL 2-pt analysis.
With this comparison in mind, we used the same data as the DES SV WL 2-pt \citepalias{TheDarkEnergySurveyCollaboration2015}, using the same shear data and galaxy $n(z)$.
To make a fair comparison, we ran additional chains using the 2-pt statistics to enforce $h=0.7$, $\Omega_{\rm b}=0.04$ and $n_s=1$, the use of non-tomographic correlation function and absence of modelling of intrinsic alignments.
The results from our peak statistics are consistent with, and of similar constraining power as, the one from the 2-pt.
The impact of shear and photo-$z$ systematics is comparable and increases the error bars by $\sim 30\%$ for both probes.

DES will deliver $\sim 5000$ deg$^2$ of lensing data with similar depth.
In this work we demonstrated the feasibility of cosmological analysis with shear peak statistics in DES, which gives a promising outlook for this type of analysis for upcoming DES data.
However, more investigation into systematics will be required in order to fully utilise the statistical power of DES shear peak statistics.
In our analysis, we found that the multiplicative shear bias and redshift errors are the limiting systematics.
Just as is the case with WL 2-pt functions, these systematics must be controlled well in future analyses.
Additionally, future studies with peak statistics could potentially gain much more constraining power by including high \snr\ peaks.
These peaks carry non-Gaussian information and are sensitive to the high mass end of the halo mass function.
However, in order to capitalise on this potential information gain, effects appearing for high mass peaks will have to be accounted for: both the loss of galaxies due to blending and intrinsic alignments can cause significant differences in the \snr\ of peaks, if these effects are not modelled in simulations.
For deeper surveys and/or tomographic peak statistics measurements, the impact of intrinsic alignments may actually be smaller, as the number of background sources compared to the number of cluster members may be larger than in this study.

\section*{Acknowledgements}

We are grateful for the extraordinary contributions of our CTIO colleagues and the DECam Construction, Commissioning and Science Verification
teams in achieving the excellent instrument and telescope conditions that have made this work possible.  The success of this project also
relies critically on the expertise and dedication of the DES Data Management group.

Funding for the DES Projects has been provided by the U.S. Department of Energy, the U.S. National Science Foundation, the Ministry of Science and Education of Spain,
the Science and Technology Facilities Council of the United Kingdom, the Higher Education Funding Council for England, the National Center for Supercomputing
Applications at the University of Illinois at Urbana-Champaign, the Kavli Institute of Cosmological Physics at the University of Chicago,
the Center for Cosmology and Astro-Particle Physics at the Ohio State University,
the Mitchell Institute for Fundamental Physics and Astronomy at Texas A\&M University, Financiadora de Estudos e Projetos,
Funda{\c c}{\~a}o Carlos Chagas Filho de Amparo {\`a} Pesquisa do Estado do Rio de Janeiro, Conselho Nacional de Desenvolvimento Cient{\'i}fico e Tecnol{\'o}gico and
the Minist{\'e}rio da Ci{\^e}ncia, Tecnologia e Inova{\c c}{\~a}o, the Deutsche Forschungsgemeinschaft and the Collaborating Institutions in the Dark Energy Survey.

The Collaborating Institutions are Argonne National Laboratory, the University of California at Santa Cruz, the University of Cambridge, Centro de Investigaciones Energ{\'e}ticas,
Medioambientales y Tecnol{\'o}gicas-Madrid, the University of Chicago, University College London, the DES-Brazil Consortium, the University of Edinburgh,
the Eidgen{\"o}ssische Technische Hochschule (ETH) Z{\"u}rich,
Fermi National Accelerator Laboratory, the University of Illinois at Urbana-Champaign, the Institut de Ci{\`e}ncies de l'Espai (IEEC/CSIC),
the Institut de F{\'i}sica d'Altes Energies, Lawrence Berkeley National Laboratory, the Ludwig-Maximilians Universit{\"a}t M{\"u}nchen and the associated Excellence Cluster Universe,
the University of Michigan, the National Optical Astronomy Observatory, the University of Nottingham, The Ohio State University, the University of Pennsylvania, the University of Portsmouth,
SLAC National Accelerator Laboratory, Stanford University, the University of Sussex, and Texas A\&M University.

The DES data management system is supported by the National Science Foundation under Grant Number AST-1138766.
The DES participants from Spanish institutions are partially supported by MINECO under grants AYA2012-39559, ESP2013-48274, FPA2013-47986, and Centro de Excelencia Severo Ochoa SEV-2012-0234.
Research leading to these results has received funding from the European Research Council under the European Union’s Seventh Framework Programme (FP7/2007-2013) including ERC grant agreements
 240672, 291329, and 306478.

This research used resources of the Calcul Quebec computing consortium, part of the Compute Canada network.

DK acknowledges support from a European Research Council Advanced Grant FP7/291329.
TK thanks the support of ETHZ ISG and the Brutus cluster team.
OF was supported by SFB-Transregio 33 `The Dark Universe' by the Deutsche Forschungsgemeinschaft (DFG).

\appendix

\section{Modelling the number of peaks as a function of shear multiplicative bias and redshift error}
\label{app:systematics_model}

In order to to accurately account for the multiplicative shear bias and redshift error in the peak statistics analysis, we have to understand how the peak abundance function reacts to changes in the multiplicative bias and redshift error.
To describe this mapping we assume a simple first order model, where the fractional change in the peak function is related linearly to the change in multiplicative bias, $m$, or error on mean of the redshift distribution, $\Delta z$.
The model for these systematics is:
\begin{align}
\frac{N_{\rm{peaks}}(m) - N_{\rm{peaks}}(m=0)}{N_{\rm{peaks}}(m=0)} &= \alpha_{m}(\nu) \cdot \nu \cdot (1+m)
\label{eqn:frac_systematics1}
\\
\frac{N_{\rm{peaks}}(\Delta z) - N_{\rm{peaks}}(\Delta z=0)}{N_{\rm{peaks}}(\Delta z=0)} &= \alpha_{\Delta z}(\nu) \cdot \nu  \cdot (1+\Delta z),
\label{eqn:frac_systematics2}
\end{align}
where $\nu=\snr$ ratio, $\alpha_{m}(\nu)$ and $\alpha_{\Delta z}(\nu)$ are \snr\ dependent scaling factors that can be measured from simulations.
To do this, we run simulations with added systematic effects.
In total, we analyse five configurations of $\{m, \Delta z \}$: $\{0, 0\}$, $\{-0.05, 0\}$, $\{0.05, 0\}$, $\{0, -0.05\}$, $\{0, 0.05\}$.
{\review The multiplicative bias was added by multiplying the shears in the simulations by a factor of (1-m), and redshift error by shifting the mean of the $n(z)$ distribution during the process of applying the DES mask and $n(z)$, described in Appendix \ref{app:des_mask}.
}

\begin{figure}
\begin{center}
\includegraphics[width=\columnwidth]{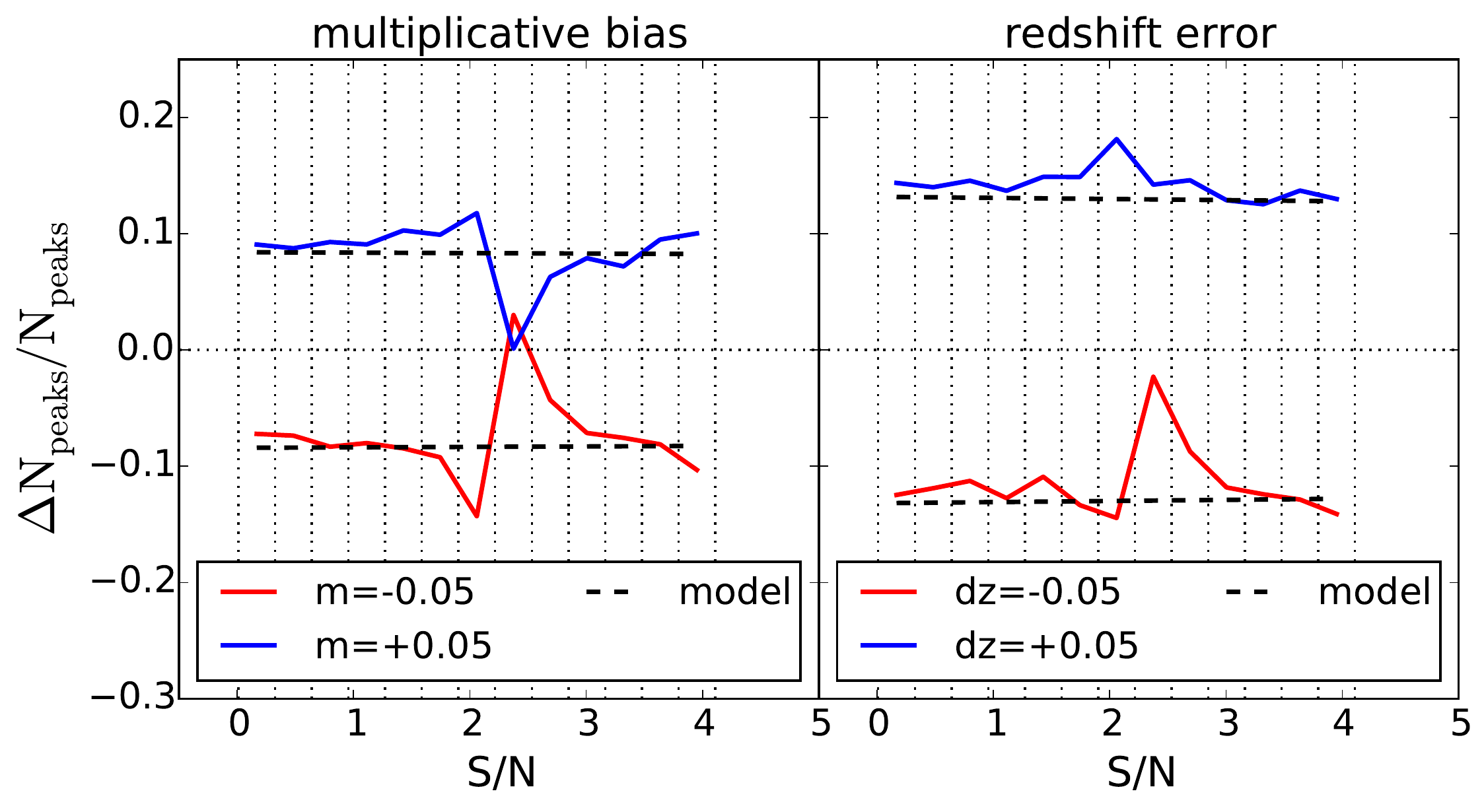}
\end{center}
\caption[]{Effect of the systematic errors in shear and $n(z)$ on the peak function.
Left and right panels depict the case when shear multiplicative bias and redshift errors are added, respectively.
They show the fractional change in number of peaks after subtracting the number of peaks from random maps.
The dependence on \snr\ is modelled with a linear fit, marked with the dashed line.
For bins around $\snr \approx 2$ the measurement is very noisy, as the difference between peaks from random maps and maps containing shear signal is close to zero.

\label{fig:systematics_models}}

\begin{center}
\includegraphics[width=\columnwidth]{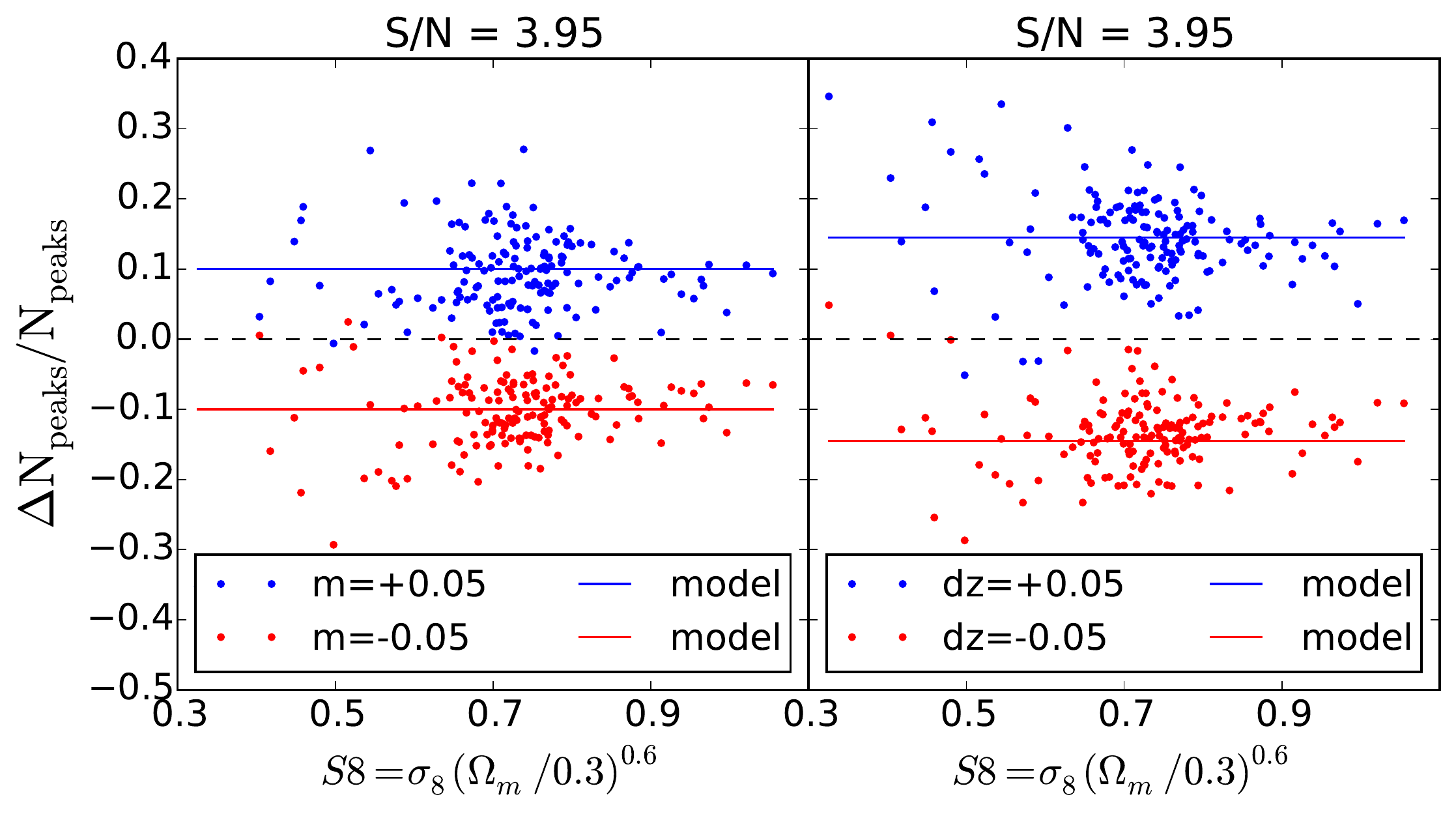}
\end{center}
\caption[]{Fractional change in number of observed peaks for two systematics: shear multiplicative bias (left) and mean redshift error (right), as function of \SE.
The measurements are shown only for the highest \snr\ bin; the behaviour for other bins is very similar.
Blue and red points show the result for positive and negative change in the value of the systematic, respectively.
Each point corresponds to a cosmological model.
For low \SE\ models the ratio becomes noisier.
The solid lines show the value of the systematics model used.
The fractional change in the number of peaks is close to constant across cosmological models, which allows us to use the simple linear systematics model for all cosmologies.
\label{fig:systematics_S8}}
\end{figure}

The comparison of results from these runs are presented in Figure \ref{fig:systematics_models}.
These figures show the fractional change in the peak function after the systematic is applied, for all \snr\ bins, as in Equation \ref{eqn:frac_systematics1} and \ref{eqn:frac_systematics2}.
The left panel shows the impact of shear multiplicative bias and right panel of redshift error.
This fraction is calculated after subtracting the expected number of peaks from maps created from randomised shapes.
The error is dominated by the number of simulations we were able to run; measurement of this small deviation requires many noise realisations.
We used 300 noise realisations for each cosmology and we consider the accuracy on the measurement of $\alpha_{\Delta z}$ and $\alpha_m$ to be sufficient for this data set.
The measurement for \snr\ bins in the middle of the range are very noisy, as the number of peaks from real and randomised maps is almost equal.
These results are created using the default aperture size of 20 arcmin.

The fractional change in number of peaks calculated this way is linear with \snr\ bin, which greatly simplifies our model.
From the multiplicative bias runs for the central cosmological model, we measure $\alpha_{m}(\nu) \approx 2$ and $\alpha_{\Delta z}(\nu) \approx 3$ for all \snr\ bins.
{\review We found a similar relationship for other aperture sizes, with  $\alpha_{m}(\nu) \approx 1.8, \ \alpha_{\Delta z}(\nu) \approx 2.8$ and $\alpha_{m}(\nu) \approx 2,\ \alpha_{\Delta z}(\nu) \approx 3$ for $\apertsize=12$ and $\apertsize=28$ arcmin, respectively.}
More detailed modelling of that function may be necessary for future peak statistics studies, perhaps beyond the first order model.

We use these measurements for cosmological parameter estimation, where the peak function in the presence of systematics is calculated as
\begin{align}
N_{\rm{peaks}}(m, \Delta z, \nu) = &N_{\rm{peaks}}(m=0,  \ \Delta z=0,  \ \nu) \nonumber \\  & \cdot \alpha_m(\nu) \cdot (1+m)   \cdot \alpha_{\Delta z}(\nu) \cdot  (1+\Delta z) \cdot
\end{align}

\noindent
We derived our systematics model from the measurements of peak functions using the central cosmological parameter configuration.
For this configuration, we analysed 35 times more data than for other parameter sets, which makes our measurement of $\alpha_m$ and $\alpha_{\Delta z}$ more accurate than for other models.
It is also important to validate that this model can be used for other cosmological parameter sets.
We check this by plotting the fractional change in number of peaks, given a change in systematic, as a function of the \SE\ parameter corresponding to other cosmological models.
Figure \ref{fig:systematics_S8} shows the results for the highest \snr\ bin.
We found very similar behaviour for other \snr\ bins.
Each point on the plot corresponds to a different configuration of $\Omega_{\rm m}$ and $\sigma_8$, as shown in Figure \ref{fig:sim_grid}.
The left panel shows the result for shear multiplicative bias and the right panel for redshift error.
Red points are measured from a simulation with positive $m$ or $\Delta z$, and blue with negative $m$ or $\Delta z$.
It is noticeable that the scatter on the fractional change in number of peaks increases with decreasing \SE.
This is expected as this is a measurement of a ratio, which becomes more noisy when a low number of peaks are detected above the random peaks.
Cosmological models with low \SE\ have in general a smaller number of peaks.
The fractional change in number of peaks seems to be constant for all values of the \SE\ parameter, which again simplifies the modelling of systematics.
That allows us to use the first order model derived here to calculate peak functions for parameter sets in $\Omega_{\rm m}$, $\sigma_8$, $m$ and $\Delta z$.

\section{Interpolating from the simulation grid}
\label{app:peakfun_modelling}

The peak functions are only calculated on a finite grid of points in the $\Omega_{\rm m}$ - $\sigma_8$ plane.
In order to calculate the likelihood of the data given a cosmological model, for all combinations of $\Omega_{\rm m}, \sigma_8$, we need to use an interpolation scheme.
In this work, we use two interpolation schemes and verify that the results obtained by each are consistent.
The default scheme creates a function which maps the cosmological parameters into number of peaks, one for each \snr\ bin separately.
This scheme uses a basis expansion to 40 basis functions.
Coefficients for basis functions are then fitted to the number of peaks for each \snr\ bin, requiring all of the coefficients to be either positive or negative.

This constraint enforces the expected monotonicity of the number of peaks as a function of cosmological model.
The basis functions are created using a mixture of polynomials and the degeneracy parameter $S8 = \sigma_8 (\Omega_{\rm m}/0.3)^{\alpha}$.
We use 2D polynomials in $\sigma_8$ and $\Omega_{\rm m}$ up to 4th order, together with eight \SE\ functions with $\alpha \in \{0.40, 0.45, 0.50, 0.55, 0.60, 0.65, 0.70, 0.75\}$, their squares and cubes.
{\review
The basis function has the form:
\begin{align}
\phi(\sigma_8, \Omega_{\rm m}) = [1,\ -1,\ \sigma_8,\ \Omega_{\rm m},\ \sigma_8^2,\ \Omega_{\rm m}^2,\ \sigma_8 \Omega_{\rm m}^2, \ \sigma_8^2 \Omega_{\rm m},\ ...\ , \nonumber \\  S_8^{\alpha=0.4},\ ...\ ,\ S_8^{\alpha=0.75}, (S_8^{\alpha=0.4})^2,\ ...\ ,\ (S_8^{\alpha=0.75})^2, \ ... \ ].
\end{align}
Many of these coefficients are found to be close to zero during the fitting process; for example, for the highest \snr\ bin, only 7 coefficients not very close to zero.
}
The fitting process is done using the convex optimization package \textsc{Cvxpy} \citep{cvxpy}.

\begin{figure}
\begin{center}
\includegraphics[width=\columnwidth]{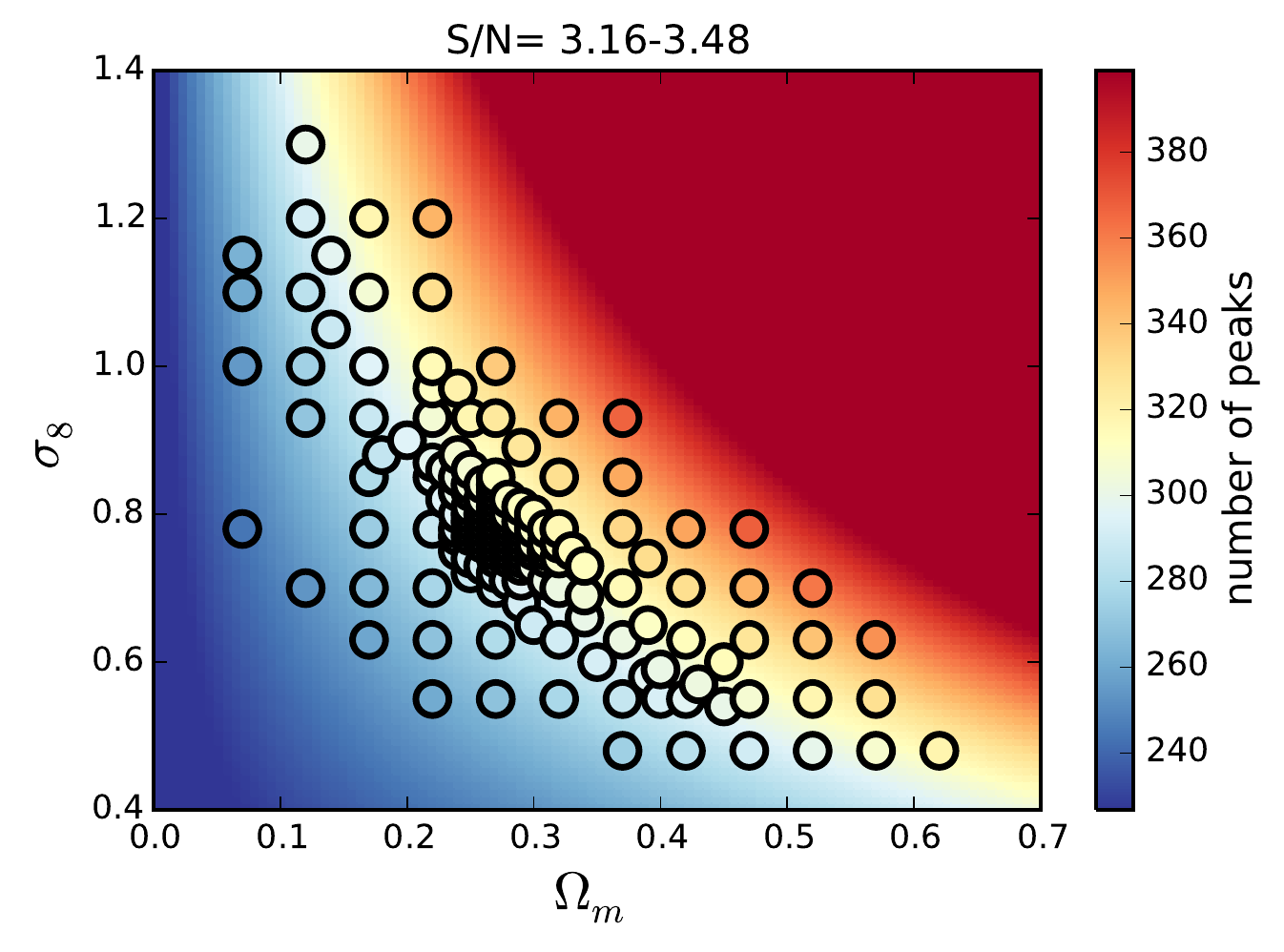}
\includegraphics[width=\columnwidth]{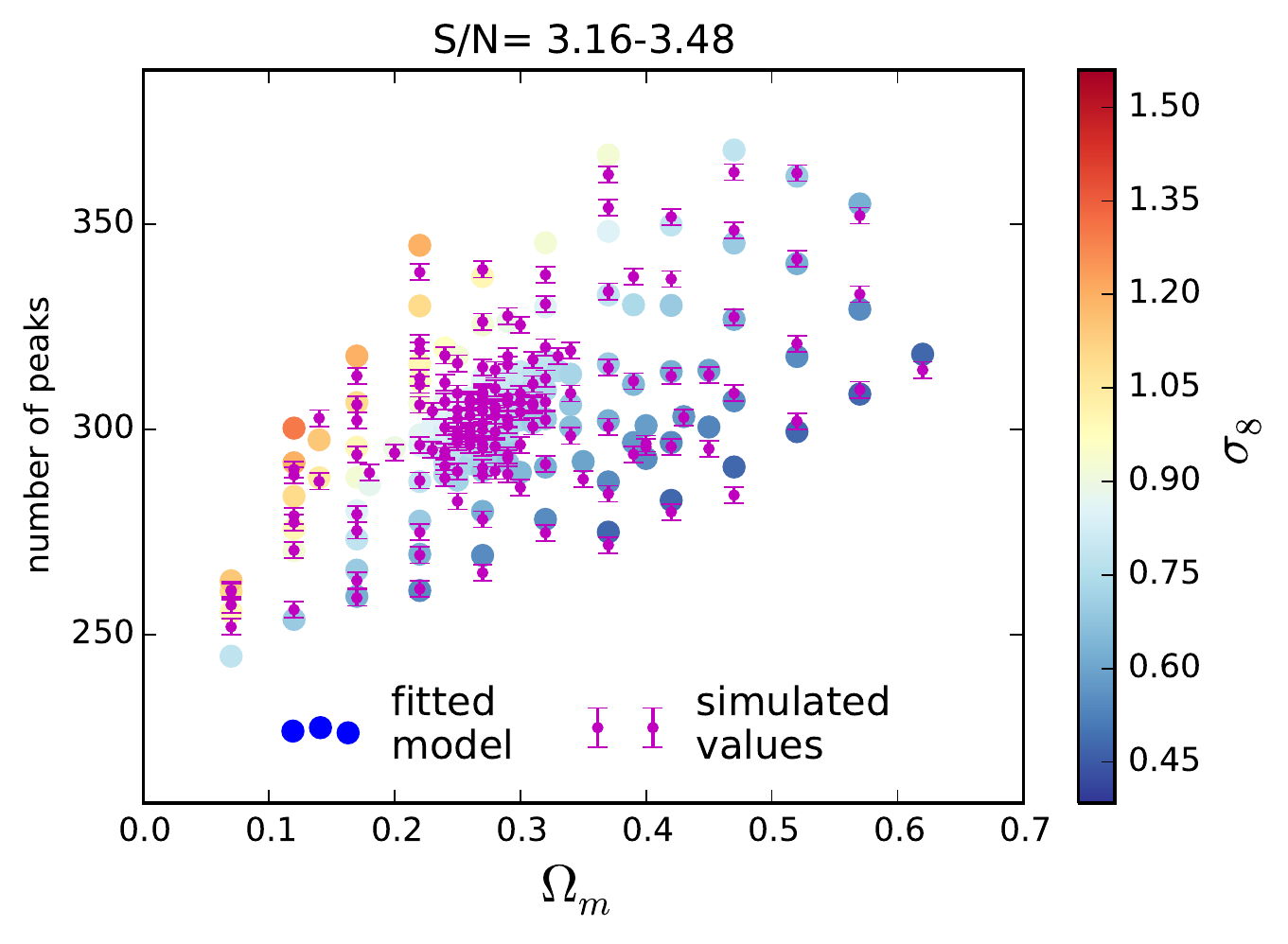}
\includegraphics[width=\columnwidth]{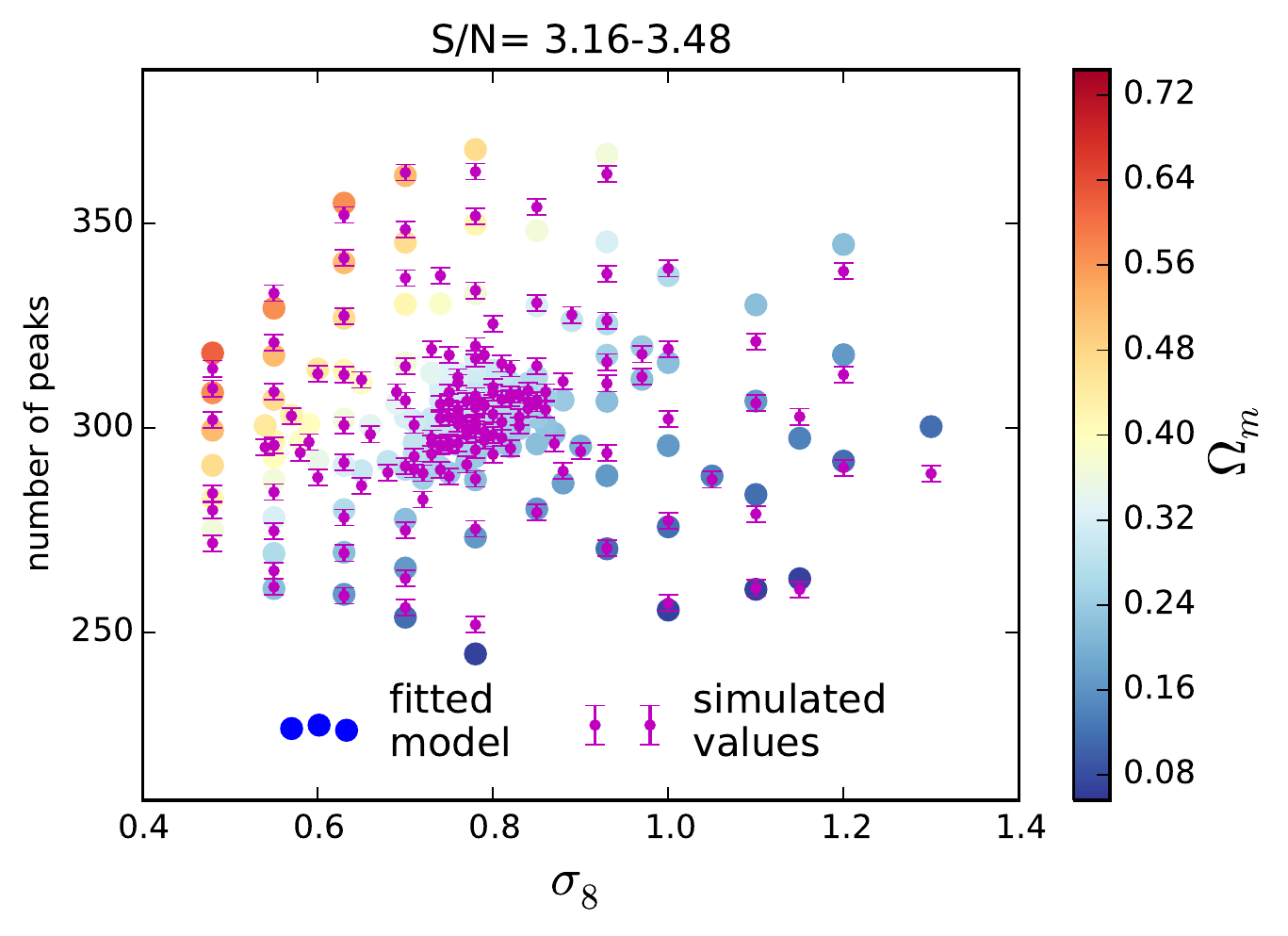}
\end{center}
\caption[]{
Modelling the number of peaks as a function of cosmology.
The top panel shows the peak count for one \snr\ bin (3.16-3.48), as a function of the $\Omega_{\rm m}$ - $\sigma_8$ parameters.
The grid of simulations is shown with open circles.
The colour shows the interpolated peak count.
Middle and bottom panels show the peak count as a function of a single cosmological parameter ($\Omega_{\rm m}$ and $\sigma_8$ for middle and bottom panels respectively), with the other marked in the colourscale.
The magenta points mark the number of peaks calculated from simulations, with error bars representing the uncertainty on the mean from many noise realisations.
\label{fig:interp_npeaks}}
\end{figure}

{\review
We verify that these interpolation schemes work well by inspecting the difference between the model and the simulated peak counts for each \snr\ bin.
}
Figure \ref{fig:interp_npeaks} shows the simulated peak functions and the fitted models for an example \snr\ bin, using the default method.
The upper panel shows the results of interpolation on the $\Omega_{\rm m}$ - $\sigma_8$ plane, based on simulation points marked by the open circles.
{\review
To better visualise the differences between the fitted model and the simulated number of peaks we plot, in the middle and bottom panels, the peak count measurements as a function of $\Omega_{\rm m}$ and $\sigma_8$, respectively.
These measurements are marked with magenta points and error bars corresponding to the error on the mean of the noise realisations.
The coloured points are the peak counts as predicted by our fitted model at the 158 cosmological parameter sets which were used in simulations.
The colour corresponds to the value of the other cosmological parameter: $\Omega_{\rm m}$ and $\sigma_8$ for the middle and bottom panels, respectively.
This allows us to verify that the fitted model neither over-fits nor under-fits the simulated peak counts.
}
We consider the fit to be sufficiently good for the accuracy of our current SV data.
In the future it may be important to improve the fitting scheme to assure that it does not introduce systematics on the cosmological parameters due to inaccurate modelling.

We use the alternate approach to measure the constraint on $\SE$ for the variant without systematics.
This approach uses radial basis functions to interpolate linearly in $\chi^{2}$, rather than in the peak function itself, and a small amount of smoothing is simultaneously applied to the interpolated values of $\chi^2$. The interpolation is carried out in two dimensions on the $\Omega_{\rm m}$-$\sigma_8$ plane.
We then directly compute the errors on $\SE$.
The alternate interpolation approach gives results which are consistent with the default method (Section \ref{sec:constraints_variants}).
This gives us confidence that the constraints on $S_8$ we present are robust to the method of interpolation.

\section{Boost factors}
\label{app:boost_factors}

A massive cluster can have many member galaxies, which reside at the same redshift as the dark matter halo.
Thus, at positions of clusters, the redshift distribution of source galaxies, $n(z)$, will be modified as compared to other areas in the survey.
The presence of extra cluster member galaxies will cause an excess $n(z)$ at the cluster redshift.
Additionally, due to crowding of the field, the fraction of blended objects will be increased.
This will cause some of the source galaxies, as well as cluster members, to be cut out of the catalogues by the shear analysis pipeline.

\begin{figure}
\begin{center}
\includegraphics[width=\columnwidth]{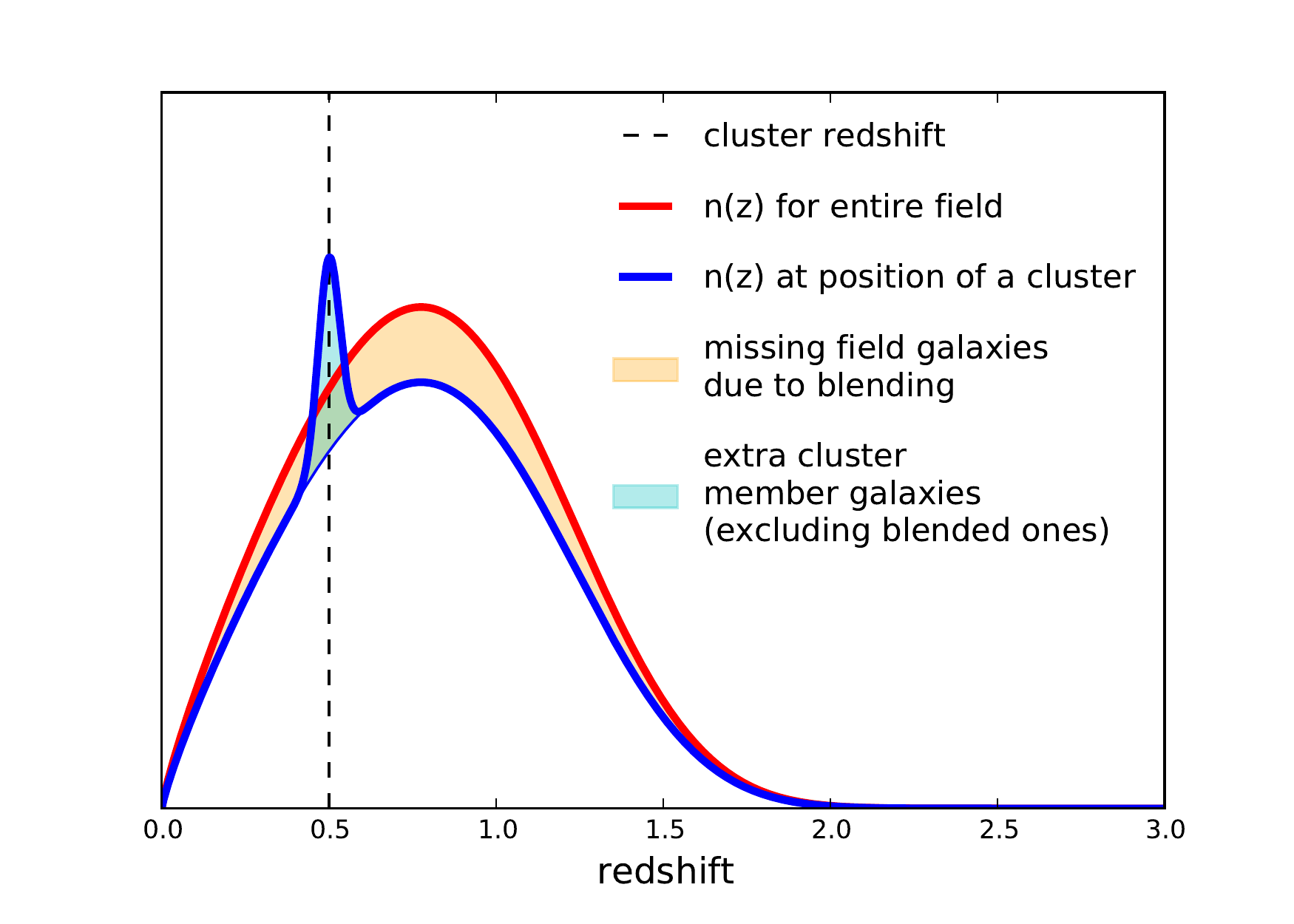}
\end{center}
\caption[]{
Cartoon depicting the difference between the $n(z)$ in survey data (blue solid line) and the simulation data (red solid line) at a position of a hypothetical peak corresponding to a cluster at redshift $z=0.5$ (black dashed line).
The number of galaxies removed by blending is shown by the cyan area and the extra cluster galaxies in peach area.
Both extra galaxy clusters and missing background galaxies will create a difference in peak signal strength between the DES survey data and simulations, where the position of dark matter peaks and galaxies are decorrelated.
\label{fig:nz_cartoon}}
\end{figure}

In simulations we use exactly the same galaxy positions as in the DES data, as explained in Section \ref{sec:simulations} and Appendix \ref{app:des_mask}.
However, the redshift distribution is homogeneous across the field, and thus the relation between the spatial position and redshift is broken.
In simulations, this will cause the over-densities of galaxies to be decorrelated from over-densities in dark matter.
A cartoon in Figure \ref{fig:nz_cartoon} compares the $n(z)$ at the position of a peak between DES data and simulations.
For a cluster at redshift of $z=0.5$ (marked by a dashed line), we observe galaxies associated with the cluster at its redshift (cyan area).
We also observe lost galaxies at all redshifts due to blending (peach area).
The total $n(z)$ in the DES survey data at the position of a peak is marked as a blue solid line, and the $n(z)$ in simulations, at the positions of a peak in simulations, is marked as a red solid line.

Both these effects - presence of cluster members and losing source galaxies due to blending - will impact peak statistics.
The extra cluster galaxies will dilute the shear signal at the position of the cluster in DES data compared to the simulations.
Similarly, background galaxies lost due to blending will cause the statistical power of the lensing signal to be decreased.
It is important to calculate how large an impact these effects have on the peak number counts in each SNR bin.
Then a correction can be applied to the peak function that is analogous to the boost factors in cluster lensing studies \citep{Applegate2012, Melchior2014, Sheldon2009}

One way to create such a correction is to look at the number of galaxies as a function of radius for peaks in DES data and simulations.
To capture the effects due to blending we use the \balrog\ catalogue, which maps the survey selection function as a function of position on the sky.
Details of creation of \balrog\ catalogues are presented in Appendix \ref{app:balrog}, and by \citet{Suchyta2015}.

\begin{figure*}
\begin{center}
\includegraphics[width=\textwidth]{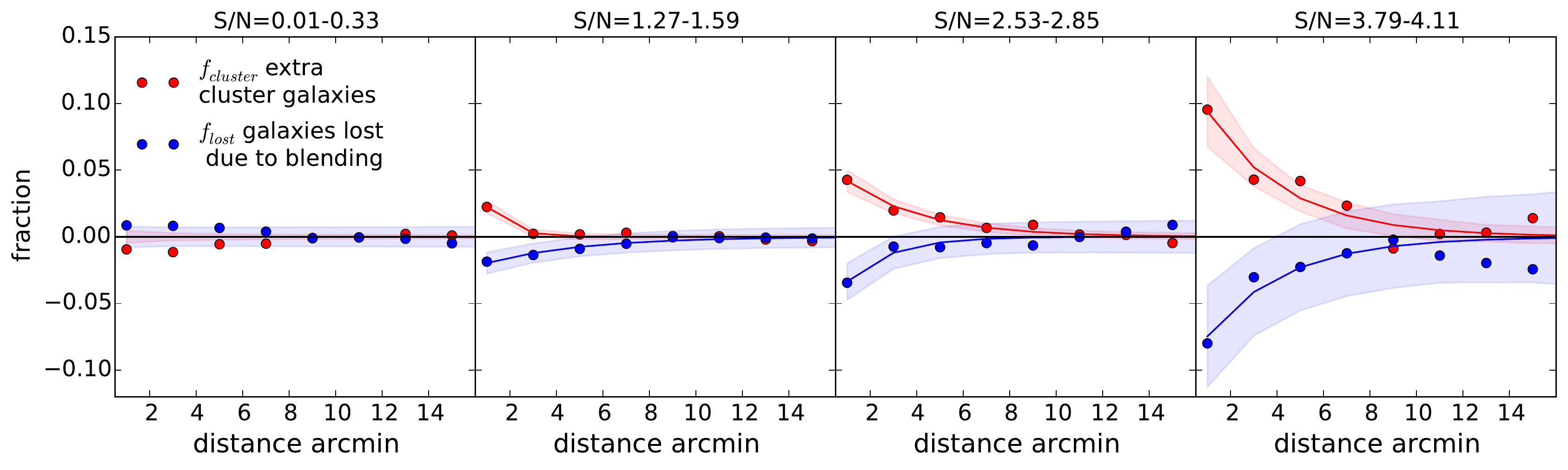}
\end{center}
\caption[]{
Estimate of the mean number of galaxies belonging to a cluster (red) and lost due to blending (blue), as a fraction of the number of galaxies in the simulations.
The distance is the radius away from the peak centre. The peaks were identified using a fiducial filter size of 20 arcmin.
The four panels correspond to \snr\ bins for low, medium and high \snr\ peaks.
Error bars are calculated by propagating the uncertainty on the mean of $N_{\rm{BALROG}}$ and $N_{\rm{DES}}$-$N_{\rm{BALROG}}$.
\label{fig:nz_corrections}}
\end{figure*}

Let's start with a simple description of the problem.
Consider an aperture positioned at the centre of a peak in the DES survey.
The number of galaxies in that aperture will be
\be
    N_{\rm{DES}} = N_{\rm{nz}} - N_{\rm{nz}}^{\rm{blended}} + N_{\rm{cluster}} - N_{\rm{cluster}}^{\rm{blended}},
\ee
where $N_{\rm{nz}}$ is the number of field galaxies at other redshifts distributed as $n(z)$,
$N_{\rm{nz}}^{\rm{blended}}$ is the number of galaxies at other redshifts lost due to blending,
$N_{\rm{cluster}}$ is the number of cluster member galaxies and
$N_{\rm{cluster}}^{\rm{blended}}$ is the number of cluster members galaxies lost due to blending.
In simulations, the number of galaxies at the position of a peak is, by construction
\be
    N_{\rm{SIM}} = N_{\rm{nz}}.
\ee
In \balrog\ catalogues, the number of galaxies around peaks identified in DES data is
\be
    N_{\rm{BALROG}} = N_{\rm{nz}} - N_{\rm{nz}}^{\rm{blended}}.
\ee

\noindent
Firstly, let's calculate how many cluster member galaxies we observe in DES for a particular \snr\ bin, as compared to the wide-field, ignoring the blended cluster members.
In fractional terms we can express it as a ratio $f_{\rm{cluster}}$, and it can be calculated using available catalogues in the following way
\be
    f_{\rm{cluster}} = \frac{N_{\rm{cluster}} - N_{\rm{cluster}}^{\rm{blended}}}{N_{\rm{nz}}} = \frac{N_{\rm{DES}} - N_{\rm{BALROG}}}{N_{\rm{SIM}}}.
\ee

\noindent
Secondly, the fraction of field galaxies lost due to blending $f_{\rm{lost}}$ can be calculated using the following combination
\be
    f_{\rm{lost}} = \frac{N_{\rm{nz}}^{\rm{blended}}}{N_{\rm{nz}}} = \frac{N_{\rm{BALROG}}-N_{\rm{SIM}}}{N_{\rm{SIM}}}.
\ee

\noindent

\begin{figure}
\begin{center}
\includegraphics[width=\columnwidth]{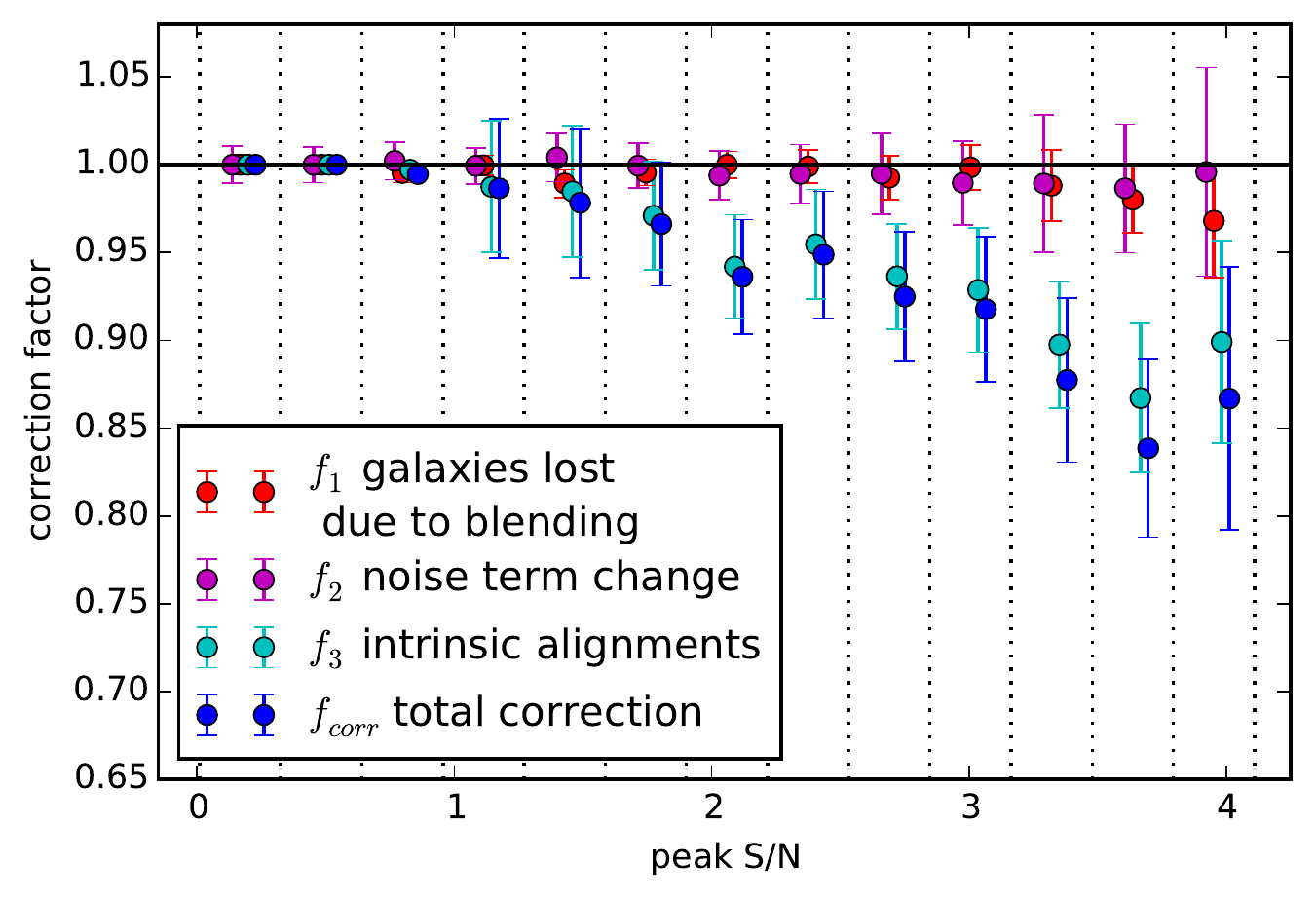}
\end{center}
\caption[]{
Boost factor and intrinsic alignment corrections for all \snr\ bins used in the analysis.
Error bars on these factors are propagated from the errors on $f_{\rm{cluster}}$ and $f_{\rm{lost}}$.
\label{fig:boost_corr}}
\end{figure}

We estimate these factors by looking at the number of galaxies surrounding peaks in the DES data, simulations, and \balrog\ catalogues, as a function of their radius away from the centre of a peak.
Figure \ref{fig:nz_corrections} presents calculated extra cluster member fraction $f_{\rm{cluster}}$ (red line) and lost field galaxies $f_{\rm{lost}}$ (blue line), using a filter with size 20 arcmin.
The error bars are calculated by forward-propagating the error on the mean on $N_{\rm{BALROG}}$ and $N_{\rm{DES}}$-$N_{\rm{BALROG}}$.
As we expect, the number of cluster galaxies increases towards the centre of the peak, reaching a $10\%$ increase for the highest \snr\ bin.
We also find this effect to be stronger for higher \snr\ peaks, which correspond to more massive clusters.
The fraction of missing galaxies due to blending also increases towards the centre of a peak, reaching $-10\%$ very close to the centre for high \snr\ bins.
As expected, this fraction increases with increasing \snr.
Note that these calculations are done using {\em all} identified peaks, including both real ones, corresponding to clusters, and spurious ones, corresponding to noise fluctuations.
Therefore the results are not directly comparable with those reported in the works on cluster lensing because our measurement is diluted by the spurious peaks.

Using these functions, we estimate their impact on the detection probability of a peak.
Let's consider the \snr\ of a peak, replacing the sum over galaxies with sum over the number of galaxies in bins of radius away from the centre (in the limit of infinitely small bins these operations are equivalent),
\be
\label{eqn:snr_corr}
\snr \approx \frac
{
\sum_r [ N_{\rm{nz}}(r) - N_{\rm{nz}}^{\rm{bl}}(r) ] \mathcal{Q}(r) g^{t}(r) + noise
}
{
\sqrt{ \epsilon^2 \sum_r [ N_{\rm{nz}}(r) - N_{\rm{nz}}^{\rm{bl}}(r) + N_{\rm{cl}}(r) - N_{\rm{cl}}^{\rm{bl}}(r)]  \mathcal{Q}^2(r) },
}
\ee
where $g^t$ is the tangential shear and $noise$ is the noise contribution with zero mean, which we ignore here, as we are concerned only by the mean of the \snr\ estimate.
We can write this equation in terms of $f_{\rm{cluster}}$ and $f_{\rm{lost}}$
\be
\hspace*{-1cm}
\label{eqn:snr_corr_f}
\snr \approx \frac
{
\sum_r [ N_{\rm{nz}}(r) - N_{\rm{nz}}(r)f_{\rm{lost}}(r) ] \mathcal{Q}(r) g^{t}(r)
}
{
\sqrt{ \epsilon^2 \sum_r [ N_{\rm{nz}}(r) - N_{\rm{nz}}(r)f_{\rm{lost}}(r) + N_{\rm{nz}}(r)f_{\rm{cluster}}(r) ]  \mathcal{Q}^2(r) },
}
\ee
where $\epsilon$ is the ellipticity standard deviation.
We can treat the numerator and denominator separately.
Let's start with the numerator.
We can replace the tangential shear profile, $g^t(r)$, by the filter profile scaled by a factor, $c$, so $g^t = c \mathcal{Q}(r)$.
The filter profile is designed to match the shear signal, so using it instead as $g^t$ actually represents the worst case scenario.
We can create the correction factor, $f_1$, which represents how much the numerator part of the equation changes when the blended galaxies are included.
If we write
\be
f_1 \equiv  \frac
{
\sum_r [N_{\rm{nz}}(r) - N_{\rm{nz}}(r)f_{\rm{lost}}(r)] \mathcal{Q}^2(r) c
}
{
\sum_r [N_{\rm{nz}}(r)] \mathcal{Q}^2(r) c
},
\label{eqn:snr_f1}
\ee
then we notice that the scaling factor, $c$, cancels out.
Now let's consider a change, $f_2$, in the noise term in the denominator
\be
f_2 \equiv  \frac
{
\sqrt{ \epsilon^2 \sum_r [ N_{\rm{nz}}(r) - N_{\rm{nz}}(r)f_{\rm{lost}}(r) + N_{\rm{nz}}(r)f_{\rm{cluster}}(r) ]  \mathcal{Q}^2(r) }
}
{
\sqrt{ \epsilon^2 \sum_r [ N_{\rm{nz}}(r) ]  \mathcal{Q}^2(r) }
}
\label{eqn:snr_f2}
\ee
and here the $\mathcal{Q}^2$ is taken from the definition of \snr, not a replacement of $g^t$.
The combined correction factor is then $f_{\rm{corr}} = f_1 / f_2$.

Error bars on this correction are calculated by propagating the uncertainty on $f_{\rm{cluster}}$ and $f_{\rm{lost}}$.
Figure \ref{fig:boost_corr} shows the resulting corrections for four \snr\ bins, with a filter size of 20 arcmin.
The magnitude of the correction increases with increasing \snr, as expected.

These corrections are not applied for the fiducial analysis. By limiting the \snr\ range to low and medium peaks we made sure that the boost factors do not play a dominant role in our analysis.
We report the results with this calibration as one of the analysis variants, presented in Section \ref{sec:constraints_variants}.
In this variant, we apply these corrections to the results from simulations.
To do this, we use the derivative of number of peaks with respect to shear multiplicative bias described in Section \ref{sec:shear_bias}, which can be also understood as the derivative of number of peaks with respect to the \snr.
This will not be a completely accurate way to apply this correction, but it can be a good approximation.
This is because, for the multiplicative bias case, we assume the same shear bias for all \snr\ bins.
Here the bias is varying across bins.
However, if we assume that most of the change in the number of peaks is due to peaks moving out from given \snr\ bin to its lower neighbour, and the number of peaks flowing from a higher neighbour bin is comparatively low, then the model assuming constant \snr\ change in all bins should be a decent approximation.
Therefore the corrected number of peaks in \snr\ bin is $N_{\rm{corrected}} = \frac{dN}{d \snr} \cdot (1-f_{corr}) \cdot N_{\rm{peaks}}$.
The derivative $\frac{dN}{d \snr}$ can be approximated by $\frac{dN}{d \snr} \approx \frac{dN}{d m}$, which in our case was $\frac{dN}{d m} \approx 2$ for most of the bins, which is what we use for this correction too.
The errors on corrections propagated further to the covariance matrix.

More detailed studies of the impact of cluster members and blending on shear peak statistics can be conducted in the future, as well as investigations of schemes to calibrate this statistic.
For this work we found that these corrections do not change our cosmological results significantly.
In future experiments with improved constraining power, it may be important to take these effects into account on a very precise level.

\section{Intrinsic alignments}
\label{app:intrinsic_alignments}

In Section \ref{sec:intrinsic_alignments} we described the possible physical origin of an IA signal, which would affect our peak count measurement. This would arise if satellite galaxies in the halos of lensing clusters are radially aligned with the halo centre and those galaxies are included in the source selection used in peak-finding.

We can describe a new conversion ratio, $f_{\rm IA}$, as the fractional difference in $\snr$ due to IAs,

\begin{equation}
f_{\rm{IA}} =  \frac{\snr\ -
\frac{
\sqrt{2} \sum_r \mathcal{Q}(r) N_{\rm{nz}}(r) f_{\rm{cluster}}(r) \bar\gamma_{\rm{scale}}\epsilon
}{
\sqrt{ \sum_r \mathcal{Q}(r)^2 N_{\rm{nz}}(r) \epsilon^2} }
}{\snr},
\end{equation}
where $\epsilon=0.36$ is the typical intrinsic ellipticity modulus of a galaxy and $\bar\gamma_{\rm{scale}}$ is the parameter controlling the strength of the alignment. We have assumed that only galaxies that are members of foreground halos suffer from IA and that IAs act to reduce the signal observed along the line of sight without affecting the noise level. This assumes that satellite galaxies are radially aligned towards their halo centre, a conservative approach consistent with existing measurements and previous work modelling IAs at the halo level \citep{LWY+13,SMM14,CDM+15,SHC+15,Schneider2010}.

As in our boost factor predictions, the fraction of galaxies along the line of sight which are cluster members is given by $f_{\rm{cluster}}(r)$. We assume that the ellipticity of the cluster member galaxies is given by $\epsilon$, the dispersion of the intrinsic shape distribution of our sources. $\bar\gamma_{\rm{scale}}$ is a scaling factor, corresponding to the level of alignment of galaxies within clusters. We use a value of $\bar\gamma_{\rm{scale}}=0.21$, following the value derived in \citet{SB10}, from where we have taken much of the inspiration for our simple halo model of IAs. This value is consistent with recent estimates of the misalignment angle from state of the art hydrodynamical simulations \citep{TMM15}.

When we include this model of IAs, as well as the boost factors described in appendix \ref{app:boost_factors}, we see a relatively minor shift in derived cosmological parameters. When the boost factor and IAs were ignored we measured $S_8 = 0.77 \pm 0.07$ and a value of $S_8 = 0.78 \pm 0.07$ when both effects were included. As the overall shift is relatively minor, we feel justified in ignoring both the boost factors and IAs when quoting our headline cosmology constraints. The uncertainty in the factors contributing to the correction factors are more significant than the subsequent shift in cosmology.

\section{Balrog catalogues}
\label{app:balrog}

We use simulations generated by the \balrog{} pipeline, described in \citet{Suchyta2015}.
The software inserts simulated objects into the real DES images,
convolving each object with the measured PSF and scaling the object flux values to the measured photometric calibration.
Following this, \balrog{} runs the DES detection and measurement pipeline (described in \citealt{mohr2012, desai2012}) on the images.

\citet{Suchyta2015} show that the output from these simulations is representative of the DES data and
can be used to model systematic biases present in that data.
We construct a \balrog{} sample for use in determining the boost factor corrections in Appendix~\ref{app:boost_factors},
where we are using the \balrog{} galaxies to analyse the systematic effects of how
shear peaks are diluted by systematic effects such as foreground (cluster) contamination and blending.
The methodology is very similar to that outlined in \citet{Melchior2014}.

Our simulation strategy uses the same basic approach as presented in \citet{Suchyta2015}.
Galaxies are simulated as single-component S\'ersic profiles,
where the physical properties (brightness, size, axis ratio, S\'ersic index)
are sampled from a catalogue based on COSMOS data \citep[][see Section~3.1 of \citealt{Suchyta2015}]{jouvel2009, great3}.
This sampling catalogue is identical to the one from \citet{Suchyta2015},
except that photometric measurements in the DES filters have been added to the catalogue of \citet{jouvel2009},
and we have substituted these magnitudes for the Subaru ones used in \citet{Suchyta2015}.

We add objects to the coadds, self-consistently building a new $riz$ detection image
for each simulation realisation, configuring the \swarp{} \citep{swarp} and \sex{} \citep{sextractor}
calls in the same manner as was done for the DES SV processing.
For this analysis, we have extended the \balrog{} coverage further south to include
the full area of the DES shear catalogue constructed in \citet{Jarvis2015}.

Where possible, we apply the selection cuts described in \citet{Jarvis2015} to the \balrog{} sample.
This includes the masking scheme and several selections based on \sex{} quantities, such as star-galaxy separation.
However, the DES shear measurements have been made using the single-epoch images,
whereas \balrog{} has only been run over the coadds, as a result we do not have shear measurements for \balrog{} objects.
Hence, we cannot directly apply cuts based on outputs of the shear measurement code to the \balrog{} sample.

To approximate the effects of the shape measurement selections, we use nearest neighbour reweighting
(as described in \citealt{lima2008} and applied to DES photometric estimation in \citet{sanchez2014} and \citet{Bonnett2015})
to match the \balrog{} catalogue to the final \ngmix{} catalogue in \citet{Jarvis2015},
applying the weights to the two-dimensional space of $i$-band \texttt{MAG\_AUTO} and \texttt{FLUX\_RADIUS} \sex{} measurements.
These quantities were chosen based on the motivation that size and magnitude
primarily govern whether one can make a successful shape measurement for a galaxy,
and that $i$-band is the central band for the $riz$ shape measurements.

The \balrog{} galaxies also do not have photo-$z$ measurements.
Consequently, we chose to add three photo-$z$ bins as an extra parameter in the reweighting.
We bin the \ngmix{} catalogue into the three tomographic bins used in DES two-point
shear tomography analysis \citep{Becker2015} and match a \balrog{} sample to each,
sampling such that the output \balrog{} catalogues number density matches that of the \ngmix{} bin.
Hence, by construction, the output \balrog{} catalogue is matched to the same total $n(z)$ as the shape catalogue,
with the same $i$-band size and magnitude distributions.
We employ this catalogue for our boost factor tests in Appendix~\ref{app:boost_factors}.

\section{Interpolation scheme for applying the DES $n(z)$ and mask}
\label{app:des_mask}
Our goal is to create simulated shear catalogues that have exactly the same galaxy positions, shape noise, weights and multiplicative shear corrections as the DES data.
We would like the simulations to have the same source redshift distribution as DES, calculated by the \skynet\ photo-$z$ code and described in \citet{Bonnett2015}.

\begin{figure}
\begin{center}
\includegraphics[width=\columnwidth]{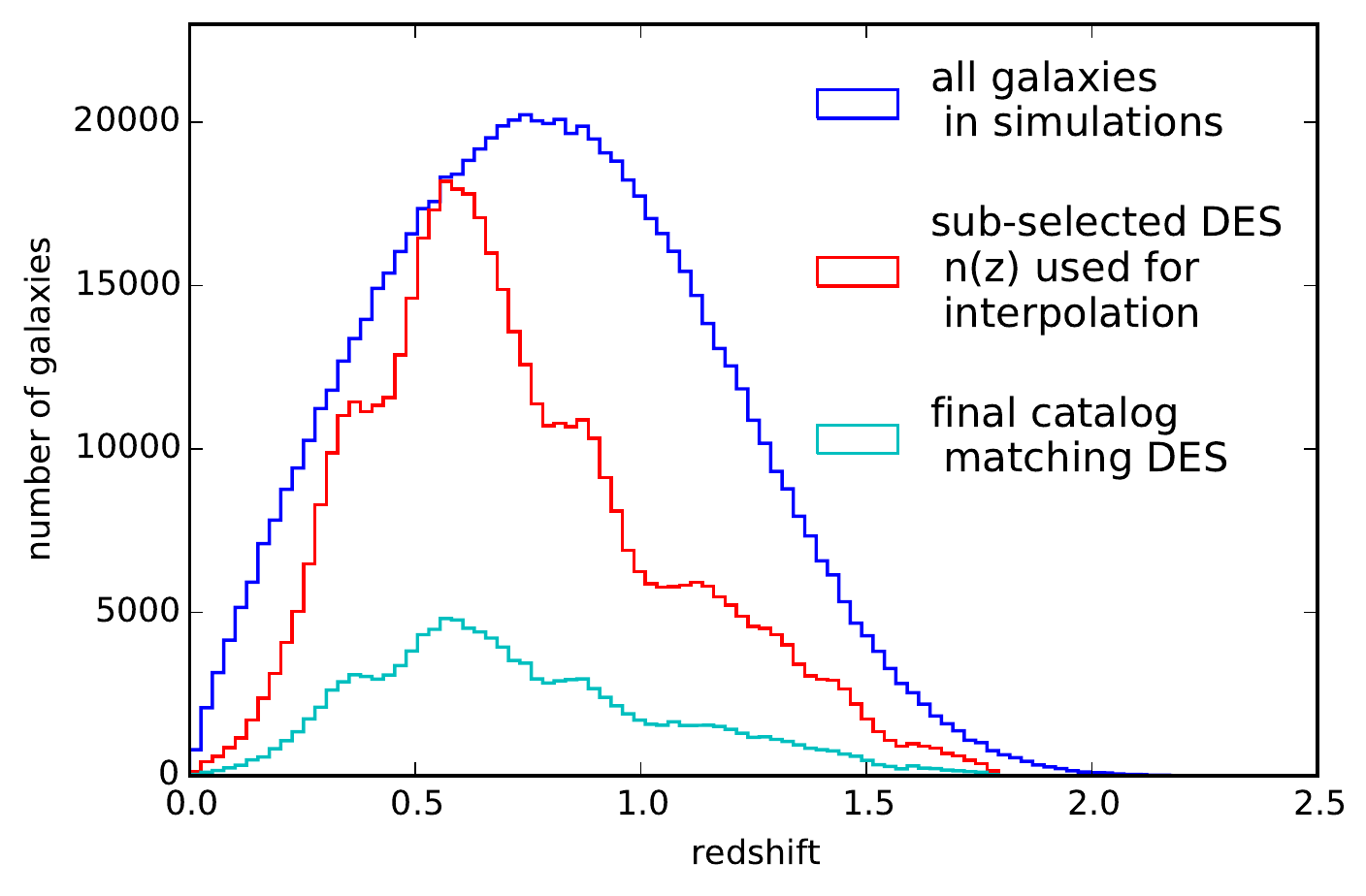}
\end{center}
\caption[]{Selection of $n(z)$ during the process of creating simulation catalogues with DES properties.
The blue histogram shows the full set of galaxies in the simulations, the red histogram shows the training set with applied $n(z)$ from DES, and cyan distribution shows the final catalogue.
\label{fig:nz_mask_interp}}

\begin{center}
\includegraphics[width=\columnwidth]{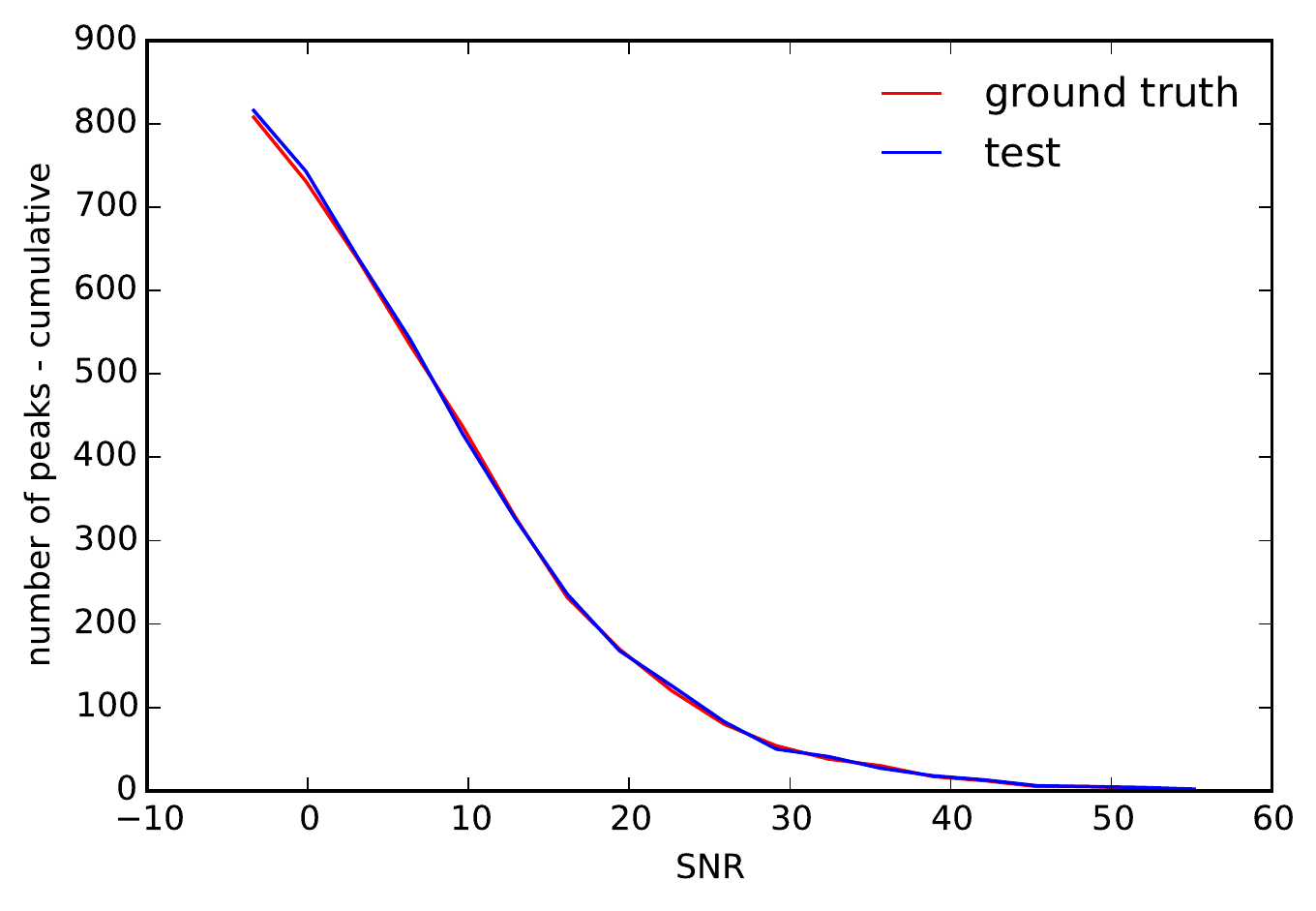}
\end{center}
\caption[]{Results of the test verifying the performance of the interpolation.
The red line is the cumulative peak function from the truth set, which is created using original galaxies from the training set, selected to have the density of 7 arcmin$^2$.
The blue line shows the cumulative peak function from the maps created using shears interpolated using the training set, located at new test positions.
The truth set and test set agree to good precision, confirming the good performance of the scheme.
No shape noise was used in this test because it causes the peaks to have very high \snr.
\label{fig:nz_mask_interp_test}}
\end{figure}

The simulation catalogues by \citet{Dietrich2009} are stored in the form of galaxy catalogues, with positions, redshift and shear, sampled uniformly across the $6 \times 6$ deg patches with a specific $n(z)$, which is described as $p(z) = (z/1.171)^{0.836}\exp\left[ - (z/1.171)^{3.425} \right]$.
To create a catalogue with $n(z)$ from DES, we sub-select from the simulations catalogue in a way that achieves the maximum number of objects.
Figure \ref{fig:nz_mask_interp} shows the $n(z)$ through the selection process, for one of the $3 \times 3$ deg patches.
Starting with the full simulation catalogue (blue), we sub-select galaxies randomly creating a new set which has the \skynet\ $n(z)$.
This set will be used for interpolating the shear to the position of DES galaxies, which have the $n(z)$ showed in cyan line.
To assign a shear according to the simulations to the position of DES galaxy, we interpolated from the sub-selected simulation catalogue to the positions of DES galaxies, using a nearest neighbour interpolation.
In the end we obtain a simulation catalogue matching DES, with shear taken from simulations according to DES $n(z)$.
The key to the performance of this procedure is the fact that the simulations have much larger galaxy density than the DES data, 25 galaxies/arcmin$^2$ versus 7 galaxies/arcmin$^2$.

To verify that this configuration can be used for our analysis, we ran a simple test.
We use the full sub-selected simulation with DES $n(z)$ as the training set and a random fraction of that catalogue chosen to have 7 galaxies per arcmin$^2$ as a truth set.
To create a test set we sample the area uniformly.
Then, using the training set, we interpolate to the newly sampled positions in the test set.
We create maps and count peaks from the truth set and test set.
This test is performed without adding shape noise.
The comparison of the maps and peak functions measured from the test and truth sets will inform us about the performance of this interpolation scheme.

Figure \ref{fig:nz_mask_interp_test} shows the resulting cumulative peak functions.
The red line shows the peak function measured from the truth set.
The blue line is the peak function counted from maps which were created by interpolating from the training set galaxies to newly drawn positions in the test set.
Both functions are very similar, which indicates that the interpolation scheme is working as expected.

\bibliographystyle{mn2e}
\bibliography{shear-peaks-paper,laura,donnacha,oliver,suchyta,will,tomek}

\begin{thebibliography}{134}
\expandafter\ifx\csname natexlab\endcsname\relax\def\natexlab#1{#1}\fi

\bibitem[{{Abate} {et~al}\mbox{.}(2009){Abate}, {Wittman}, {Margoniner},
  {Bridle}, {Gee}, {Tyson}, \& {Dell'Antonio}}]{Abateetal2009}
{Abate} A., {Wittman} D., {Margoniner} V.~E., {Bridle} S.~L., {Gee} P., {Tyson}
  J.~A., {Dell'Antonio} I.~P., 2009, \apj, 702, 603

\bibitem[{Allen, Evrard \& Mantz(2011)Allen, Evrard, \& Mantz}]{Allen2011}
Allen S.~W., Evrard A.~E., Mantz A.~B., 2011, Annual Review of Astronomy and
  Astrophysics, 49, 409

\bibitem[{Applegate {et~al}\mbox{.}(2012)Applegate, von~der Linden, Kelly,
  Allen, Allen, Burchat, Burke, Ebeling, Mantz, \& Morris}]{Applegate2012}
Applegate D.~E. {et~al.}, 2012, Monthly Notices of the Royal Astronomical
  Society, 439, 48

\bibitem[{{Bard} {et~al}\mbox{.}(2013){Bard}, {Kratochvil}, {Chang}, {May}, \&
  {et.al.}}]{Bardetal2013}
{Bard} D., {Kratochvil} J.~M., {Chang} C., {May} M., {et.al.}, 2013, \apj, 774,
  49

\bibitem[{Bard, Kratochvil \& Dawson(2014)Bard, Kratochvil, \&
  Dawson}]{Bard2014}
Bard D., Kratochvil J.~M., Dawson W., 2014, ArXiv e-prints

\bibitem[{Becker {et~al}\mbox{.}(2015)Becker, Troxel, MacCrann, Krause, Eifler,
  Friedrich, Nicola, Refregier, Amara, Bacon, Bernstein, Bonnett, Bridle,
  Busha, Chang, Dodelson, Erickson, Evrard, Frieman, Gaztanaga, Gruen, Hartley,
  Jain, Jarvis, Kacprzak, Kirk, Kravtsov, Leistedt, Rykoff, Sabiu, Sanchez,
  Seo, Sheldon, Wechsler, Zuntz, Abbott, Abdalla, Allam, Armstrong, Banerji,
  Bauer, Benoit-Levy, Bertin, Brooks, Buckley-Geer, Burke, Capozzi, Rosell,
  Kind, Carretero, Castander, Crocce, Cunha, D'Andrea, da~Costa, DePoy, Desai,
  Diehl, Dietrich, Doel, Neto, Fernandez, Finley, Flaugher, Fosalba, Gerdes,
  Gruendl, Gutierrez, Honscheid, James, Kuehn, Kuropatkin, Lahav, Li, Lima,
  Maia, March, Martini, Melchior, Miller, Miquel, Mohr, Nichol, Nord, Ogando,
  Plazas, Reil, Romer, Roodman, Sako, Sanchez, Scarpine, Schubnell,
  Sevilla-Noarbe, Smith, Soares-Santos, Sobreira, Suchyta, Swanson, Tarle,
  Thaler, Thomas, Vikram, Walker, \& Collaboration}]{Becker2015}
Becker M.~R. {et~al.}, 2015, ArXiv e-prints

\bibitem[{{Berg{\'e}}, {Amara} \& {R{\'e}fr{\'e}gier}(2010){Berg{\'e}},
  {Amara}, \& {R{\'e}fr{\'e}gier}}]{Berge2010}
{Berg{\'e}} J., {Amara} A., {R{\'e}fr{\'e}gier} A., 2010, \apj, 712, 992

\bibitem[{{Berg{\'e}} {et~al}\mbox{.}(2008){Berg{\'e}}, {Pacaud},
  {R{\'e}fr{\'e}gier}, {Massey}, {Pierre}, {Amara}, {Birkinshaw},
  {Paulin-Henriksson}, {Smith}, \& {Willis}}]{Berge2008}
{Berg{\'e}} J. {et~al.}, 2008, \mnras, 385, 695

\bibitem[{Bernstein(2010)}]{Bernstein2010}
Bernstein G.~M., 2010, Monthly Notices of the Royal Astronomical Society, 406,
  2793

\bibitem[{{Bertin} \& {Arnouts}(1996)}]{sextractor}
{Bertin} E., {Arnouts} S., 1996, \aaps, 117, 393

\bibitem[{{Bertin} {et~al}\mbox{.}(2002){Bertin}, {Mellier}, {Radovich},
  {Missonnier}, {Didelon}, \& {Morin}}]{swarp}
{Bertin} E., {Mellier} Y., {Radovich} M., {Missonnier} G., {Didelon} P.,
  {Morin} B., 2002, in Astronomical Society of the Pacific Conference Series,
  Vol. 281, Astronomical Data Analysis Software and Systems XI, {Bohlender}
  D.~A., {Durand} D., {Handley} T.~H., eds., p. 228

\bibitem[{Bertin(2011)}]{Bertin2011}
Bertin Â., 2011, Astronomical Data Analysis Software and Systems XX. ASP
  Conference Proceedings, 442

\bibitem[{Blandford \& Narayan(1986)}]{Blandford1986}
Blandford R., Narayan R., 1986, The Astrophysical Journal, 310, 568

\bibitem[{{Blazek} {et~al}\mbox{.}(2012){Blazek}, {Mandelbaum}, {Seljak}, \&
  {Nakajima}}]{BMS+12}
{Blazek} J., {Mandelbaum} R., {Seljak} U., {Nakajima} R., 2012, \jcap, 5, 41

\bibitem[{Bonnett {et~al}\mbox{.}(2015)Bonnett, Troxel, Hartley, Amara,
  Leistedt, Becker, Bernstein, Bridle, Bruderer, Busha, Kind, Childress,
  Castander, Chang, Crocce, Davis, Eifler, Frieman, Gangkofner, Gaztanaga,
  Glazebrook, Gruen, Kacprzak, King, Kwan, Lahav, Lewis, Lidman, Lin, MacCrann,
  Miquel, O'Neill, Palmese, Peiris, Refregier, Rozo, Rykoff, Sadeh,
  S{\'{a}}nchez, Sheldon, Uddin, Wechsler, Zuntz, Abbott, Abdalla, Allam,
  Armstrong, Banerji, Bauer, Benoit-L{\'{e}}vy, Bertin, Brooks, Buckley-Geer,
  Burke, Capozzi, Rosell, Carretero, Cunha, D'Andrea, da~Costa, DePoy, Desai,
  Diehl, Dietrich, Doel, Neto, Fernandez, Flaugher, Fosalba, Gerdes, Gruendl,
  Honscheid, Jain, James, Jarvis, Kim, Kuehn, Kuropatkin, Li, Lima, Maia,
  March, Marshall, Martini, Melchior, Miller, Neilsen, Nichol, Nord, Ogando,
  Plazas, Reil, Romer, Roodman, Sako, Sanchez, Santiago, Smith, Soares-Santos,
  Sobreira, Suchyta, Swanson, Tarle, Thaler, Thomas, Vikram, \&
  Walker}]{Bonnett2015}
Bonnett C. {et~al.}, 2015, ArXiv e-prints

\bibitem[{{Bridle} \& {King}(2007)}]{BK07}
{Bridle} S., {King} L., 2007, New Journal of Physics, 9, 444

\bibitem[{{Catelan}, {Kamionkowski} \& {Blandford}(2001){Catelan},
  {Kamionkowski}, \& {Blandford}}]{CKB01}
{Catelan} P., {Kamionkowski} M., {Blandford} R.~D., 2001, \mnras, 320, L7

\bibitem[{{Chang} {et~al}\mbox{.}(2015){Chang}, {Vikram}, {Jain}, {Bacon},
  {Amara}, {Becker}, {Bernstein}, {Bonnett}, {Bridle}, {Brout}, {Busha},
  {Frieman}, {Gaztanaga}, {Hartley}, {Jarvis}, {Kacprzak}, {Kov{\'a}cs},
  {Lahav}, {Lin}, {Melchior}, {Peiris}, {Rozo}, {Rykoff}, {S{\'a}nchez},
  {Sheldon}, {Troxel}, {Wechsler}, {Zuntz}, {Abbott}, {Abdalla}, {Allam},
  {Annis}, {Bauer}, {Benoit-L{\'e}vy}, {Brooks}, {Buckley-Geer}, {Burke},
  {Capozzi}, {Carnero Rosell}, {Carrasco Kind}, {Castander}, {Crocce},
  {D'Andrea}, {Desai}, {Diehl}, {Dietrich}, {Doel}, {Eifler}, {Evrard}, {Fausti
  Neto}, {Flaugher}, {Fosalba}, {Gruen}, {Gruendl}, {Gutierrez}, {Honscheid},
  {James}, {Kent}, {Kuehn}, {Kuropatkin}, {Maia}, {March}, {Martini},
  {Merritt}, {Miller}, {Miquel}, {Neilsen}, {Nichol}, {Ogando}, {Plazas},
  {Romer}, {Roodman}, {Sako}, {Sanchez}, {Sevilla}, {Smith}, {Soares-Santos},
  {Sobreira}, {Suchyta}, {Tarle}, {Thaler}, {Thomas}, {Tucker}, \&
  {Walker}}]{Changetal2015}
{Chang} C. {et~al.}, 2015, Physical Review Letters, 115, 051301

\bibitem[{{Chisari} {et~al}\mbox{.}(2015){Chisari}, {Dunkley}, {Miller}, \&
  {Allison}}]{CDM+15}
{Chisari} N.~E., {Dunkley} J., {Miller} L., {Allison} R., 2015, ArXiv e-prints

\bibitem[{{Chisari} {et~al}\mbox{.}(2014){Chisari}, {Mandelbaum}, {Strauss},
  {Huff}, \& {Bahcall}}]{CMS+14}
{Chisari} N.~E., {Mandelbaum} R., {Strauss} M.~A., {Huff} E.~M., {Bahcall}
  N.~A., 2014, \mnras, 445, 726

\bibitem[{{Clowe}, {De Lucia} \& {King}(2004){Clowe}, {De Lucia}, \&
  {King}}]{Cloweetal2004}
{Clowe} D., {De Lucia} G., {King} L., 2004, \mnras, 350, 1038

\bibitem[{{Dahle}(2006)}]{Dahle2006}
{Dahle} H., 2006, \apj, 653, 954

\bibitem[{{Desai} {et~al}\mbox{.}(2012){Desai}, {Armstrong}, {Mohr}, {Semler},
  {Liu}, {Bertin}, {Allam}, {Barkhouse}, {Bazin}, {Buckley-Geer}, {Cooper},
  {Hansen}, {High}, {Lin}, {Lin}, {Ngeow}, {Rest}, {Song}, {Tucker}, \&
  {Zenteno}}]{desai2012}
{Desai} S. {et~al.}, 2012, \apj, 757, 83

\bibitem[{Diamond \& Boyd(2015)}]{cvxpy}
Diamond S., Boyd S., 2015, Journal of Machine Learning Research

\bibitem[{Dietrich {et~al}\mbox{.}(2007)Dietrich, Erben, Lamer, Schneider,
  Schwope, Hartlap, \& Maturi}]{Dietrich2007}
Dietrich J.~P., Erben T., Lamer G., Schneider P., Schwope A., Hartlap J.,
  Maturi M., 2007, Astronomy and Astrophysics, 470, 821

\bibitem[{{Dietrich} \& {Hartlap}(2010)}]{DietrichHartlap2010}
{Dietrich} J.~P., {Hartlap} J., 2010, \mnras, 402, 1049

\bibitem[{Dietrich \& Hartlap(2010)}]{Dietrich2009}
Dietrich J.~P., Hartlap J., 2010, Monthly Notices of the Royal Astronomical
  Society, 402, 1049

\bibitem[{{Eifler}, {Schneider} \& {Hartlap}(2009){Eifler}, {Schneider}, \&
  {Hartlap}}]{Eifler2009}
{Eifler} T., {Schneider} P., {Hartlap} J., 2009, Astronomy \& Astrophysics,
  502, 721

\bibitem[{Eisenstein \& Hu(1998)}]{Eisenstein1998}
Eisenstein D.~J., Hu W., 1998, The Astrophysical Journal, 496, 605

\bibitem[{Flaugher {et~al}\mbox{.}(2015)Flaugher, Diehl, Honscheid, Abbott,
  Alvarez, Angstadt, Annis, Antonik, Ballester, Beaufore, Bernstein, Bernstein,
  Bigelow, Bonati, Boprie, Brooks, Buckley-Geer, Campa, Cardiel-Sas, Castander,
  Castilla, Cease, Cela-Ruiz, Chappa, Chi, Cooper, da~Costa, Dede, Derylo,
  DePoy, de~Vicente, Doel, Drlica-Wagner, Eiting, Elliott, Emes, Estrada,
  {Fausti Neto}, Finley, Flores, Frieman, Gerdes, Gladders, Gregory, Gutierrez,
  Hao, Holland, Holm, Huffman, Jackson, James, Jonas, Karcher, Karliner, Kent,
  Kessler, Kozlovsky, Kron, Kubik, Kuehn, Kuhlmann, Kuk, Lahav, Lathrop, Lee,
  Levi, Lewis, Li, Mandrichenko, Marshall, Martinez, Merritt, Miquel,
  Mu{\~{n}}oz, Neilsen, Nichol, Nord, Ogando, Olsen, Palaio, Patton, Peoples,
  Plazas, Rauch, Reil, Rheault, Roe, Rogers, Roodman, Sanchez, Scarpine,
  Schindler, Schmidt, Schmitt, Schubnell, Schultz, Schurter, Scott, Serrano,
  Shaw, Smith, Soares-Santos, Stefanik, Stuermer, Suchyta, Sypniewski, Tarle,
  Thaler, Tighe, Tran, Tucker, Walker, Wang, Watson, Weaverdyck, Wester, Woods,
  \& Yanny}]{Flaugher2015}
Flaugher B. {et~al.}, 2015, The Astronomical Journal, 150, 150

\bibitem[{{Graff} {et~al}\mbox{.}(2014){Graff}, {Feroz}, {Hobson}, \&
  {Lasenby}}]{Graff2013}
{Graff} P., {Feroz} F., {Hobson} M.~P., {Lasenby} A., 2014, \mnras, 441, 1741

\bibitem[{{Hamana} {et~al}\mbox{.}(2015){Hamana}, {Sakurai}, {Koike}, \&
  {Miller}}]{Hamanaetal2015}
{Hamana} T., {Sakurai} J., {Koike} M., {Miller} L., 2015, \pasj, 67, 34

\bibitem[{{Hamana}, {Takada} \& {Yoshida}(2004){Hamana}, {Takada}, \&
  {Yoshida}}]{Hamanaetal2004}
{Hamana} T., {Takada} M., {Yoshida} N., 2004, \mnras, 350, 893

\bibitem[{Hartlap, Simon \& Schneider(2007)Hartlap, Simon, \&
  Schneider}]{Hartlap2007}
Hartlap J., Simon P., Schneider P., 2007, Astronomy and Astrophysics, 464, 399

\bibitem[{{Heavens}, {Refregier} \& {Heymans}(2000){Heavens}, {Refregier}, \&
  {Heymans}}]{HRH2000}
{Heavens} A., {Refregier} A., {Heymans} C., 2000, \mnras, 319, 649

\bibitem[{{Hennawi} \& {Spergel}(2005)}]{HennawiSpergel2005}
{Hennawi} J.~F., {Spergel} D.~N., 2005, \apj, 624, 59

\bibitem[{Hetterscheidt {et~al}\mbox{.}(2005)Hetterscheidt, Erben, Schneider,
  Maoli, {Van Waerbeke}, \& Mellier}]{Hetterscheidt2005}
Hetterscheidt M., Erben T., Schneider P., Maoli R., {Van Waerbeke} L., Mellier
  Y., 2005, Astronomy and Astrophysics, 442, 43

\bibitem[{{Hetterscheidt} {et~al}\mbox{.}(2007){Hetterscheidt}, {Simon},
  {Schirmer}, {Hildebrandt}, {Schrabback}, {Erben}, \&
  {Schneider}}]{Hetterscheidtetal2007}
{Hetterscheidt} M., {Simon} P., {Schirmer} M., {Hildebrandt} H., {Schrabback}
  T., {Erben} T., {Schneider} P., 2007, \aap, 468, 859

\bibitem[{{Heymans} {et~al}\mbox{.}(2013{\natexlab{a}}){Heymans}, {Grocutt},
  {Heavens}, {Kilbinger}, {Kitching}, \& {etal.}}]{Heymansetal2013}
{Heymans} C., {Grocutt} E., {Heavens} A., {Kilbinger} M., {Kitching} T.~D.,
  {etal.}, 2013{\natexlab{a}}, \mnras, 432, 2433

\bibitem[{{Heymans} {et~al}\mbox{.}(2013{\natexlab{b}}){Heymans}, {Grocutt},
  {Heavens}, {Kilbinger}, {Kitching}, {Simpson}, {Benjamin}, {Erben},
  {Hildebrandt}, {Hoekstra}, {Mellier}, {Miller}, {Van Waerbeke}, {Brown},
  {Coupon}, {Fu}, {Harnois-D{\'e}raps}, {Hudson}, {Kuijken}, {Rowe},
  {Schrabback}, {Semboloni}, {Vafaei}, \& {Velander}}]{heymans13}
{Heymans} C. {et~al.}, 2013{\natexlab{b}}, \mnras, 432, 2433

\bibitem[{Heymans {et~al}\mbox{.}(2012)Heymans, {Van Waerbeke}, Miller, Erben,
  Hildebrandt, Hoekstra, Kitching, Mellier, Simon, Bonnett, Coupon, Fu,
  Harnois-D{\'{e}}raps, Hudson, Kilbinger, Kuijken, Rowe, Schrabback,
  Semboloni, van Uitert, Vafaei, \& Velander}]{heymans2012}
Heymans C. {et~al.}, 2012, Monthly Notices of the Royal Astronomical Society,
  427, 146

\bibitem[{Hilbert {et~al}\mbox{.}(2009)Hilbert, Hartlap, White, \&
  Schneider}]{Hilbert2009}
Hilbert S., Hartlap J., White S. D.~M., Schneider P., 2009, Astronomy and
  Astrophysics, 499, 31

\bibitem[{{Hilbert} {et~al}\mbox{.}(2012){Hilbert}, {Marian}, {Smith}, \&
  {Desjacques}}]{Hilbertetal2012}
{Hilbert} S., {Marian} L., {Smith} R.~E., {Desjacques} V., 2012, \mnras, 426,
  2870

\bibitem[{{Hirata} \& {Seljak}(2004)}]{HS04}
{Hirata} C.~M., {Seljak} U., 2004, \prd, 70, 063526

\bibitem[{{Hoekstra} {et~al}\mbox{.}(2006){Hoekstra}, {Mellier}, {van
  Waerbeke}, {Semboloni}, {Fu}, {Hudson}, {Parker}, {Tereno}, \&
  {Benabed}}]{Hoekstraetal2006}
{Hoekstra} H. {et~al.}, 2006, \apj, 647, 116

\bibitem[{Hu \& Kravtsov(2003)}]{Hu2003}
Hu W., Kravtsov A.~V., 2003, The Astrophysical Journal, 584, 702

\bibitem[{Huff {et~al}\mbox{.}(2011)Huff, Eifler, Hirata, Mandelbaum, Schlegel,
  \& Seljak}]{Huff2011}
Huff E.~M., Eifler T., Hirata C.~M., Mandelbaum R., Schlegel D., Seljak U.,
  2011, 23

\bibitem[{{Jarvis} {et~al}\mbox{.}(2003){Jarvis}, {Bernstein}, {Fischer},
  {Smith}, {Jain}, {Tyson}, \& {Wittman}}]{Jarvisetal2003}
{Jarvis} M., {Bernstein} G.~M., {Fischer} P., {Smith} D., {Jain} B., {Tyson}
  J.~A., {Wittman} D., 2003, \aj, 125, 1014

\bibitem[{Jarvis {et~al}\mbox{.}(2015)Jarvis, Sheldon, Zuntz, Kacprzak, Bridle,
  Amara, Armstrong, Becker, Bernstein, Bonnett, Chang, Das, Dietrich,
  Drlica-Wagner, Eifler, Gangkofner, Gruen, Hirsch, Huff, Jain, Kent, Kirk,
  MacCrann, Melchior, Plazas, Refregier, Rowe, Rykoff, Samuroff, S{\'{a}}nchez,
  Suchyta, Troxel, Vikram, Abbott, Abdalla, Allam, Annis, Benoit-L{\'{e}}vy,
  Bertin, Brooks, Buckley-Geer, Burke, Capozzi, Rosell, Kind, Carretero,
  Castander, Crocce, Cunha, D'Andrea, da~Costa, DePoy, Desai, Diehl, Doel,
  Neto, Flaugher, Fosalba, Frieman, Gaztanaga, Gerdes, Gruendl, Gutierrez,
  Honscheid, James, Kuehn, Kuropatkin, Lahav, Li, Lima, March, Martini, Miquel,
  Mohr, Neilsen, Nord, Ogando, Reil, Romer, Roodman, Sako, Sanchez, Scarpine,
  Schubnell, Sevilla-Noarbe, Smith, Soares-Santos, Sobreira, Swanson, Tarle,
  Thaler, Thomas, Walker, \& Wechsler}]{Jarvis2015}
Jarvis M. {et~al.}, 2015, ArXiv Astrophysics e-prints

\bibitem[{{Joachimi} \& {Bridle}(2010)}]{JB10}
{Joachimi} B., {Bridle} S.~L., 2010, \aap, 523, A1

\bibitem[{{Jouvel} {et~al}\mbox{.}(2009){Jouvel}, {Kneib}, {Ilbert},
  {Bernstein}, {Arnouts}, {Dahlen}, {Ealet}, {Milliard}, {Aussel}, {Capak},
  {Koekemoer}, {Le Brun}, {McCracken}, {Salvato}, \& {Scoville}}]{jouvel2009}
{Jouvel} S. {et~al.}, 2009, \aap, 504, 359

\bibitem[{Kacprzak {et~al}\mbox{.}(2014)Kacprzak, Bridle, Rowe, Voigt, Zuntz,
  Hirsch, \& MacCrann}]{Kacprzak2014}
Kacprzak T., Bridle S., Rowe B., Voigt L., Zuntz J., Hirsch M., MacCrann N.,
  2014, Monthly Notices of the Royal Astronomical Society, 441, 2528

\bibitem[{Kacprzak {et~al}\mbox{.}(2012)Kacprzak, Zuntz, Rowe, Bridle,
  Refregier, Amara, Voigt, \& Hirsch}]{Kacprzak2012}
Kacprzak T., Zuntz J., Rowe B., Bridle S., Refregier A., Amara A., Voigt L.,
  Hirsch M., 2012, Monthly Notices of the Royal Astronomical Society, 427, 2711

\bibitem[{{Kaiser} \& {Squires}(1993)}]{KaiserSquires1993}
{Kaiser} N., {Squires} G., 1993, \apj, 404, 441

\bibitem[{{Kaufman}(1967)}]{Kaufman}
{Kaufman} G.~M., 1967, Report No. 6710, Center for Operations Research and
  Econometrics, Catholic University of Louvain, Heverlee, Belgium

\bibitem[{Kilbinger(2015)}]{Kilbinger2015}
Kilbinger M., 2015, Reports on progress in physics. Physical Society (Great
  Britain), 78, 086901

\bibitem[{Kilbinger {et~al}\mbox{.}(2013)Kilbinger, Fu, Heymans, Simpson,
  Benjamin, Erben, Harnois-Deraps, Hoekstra, Hildebrandt, Kitching, Mellier,
  Miller, {Van Waerbeke}, Benabed, Bonnett, Coupon, Hudson, Kuijken, Rowe,
  Schrabback, Semboloni, Vafaei, \& Velander}]{Kilbinger2013}
Kilbinger M. {et~al.}, 2013, Monthly Notices of the Royal Astronomical Society,
  430, 2200

\bibitem[{{Kilbinger} {et~al}\mbox{.}(2013){Kilbinger}, {Fu}, {Heymans},
  {Simpson}, {Benjamin}, \& {etal.}}]{Kilbingeretal2013}
{Kilbinger} M., {Fu} L., {Heymans} C., {Simpson} F., {Benjamin} J., {etal.},
  2013, \mnras, 430, 2200

\bibitem[{{Kirk} {et~al}\mbox{.}(2015){Kirk}, {Omori}, {Benoit-L{\'e}vy},
  {Cawthon}, {Chang}, {Larsen}, {Amara}, {Bacon}, {Crawford}, {Dodelson},
  {Fosalba}, {Giannantonio}, {Holder}, {Jain}, {Kacprzak}, {Lahav}, {MacCrann},
  {Nicola}, {Refregier}, {Sheldon}, {Story}, {Troxel}, {Vieira}, {Vikram},
  {Zuntz}, {Abbott}, {Abdalla}, {Becker}, {Benson}, {Bernstein}, {Bernstein},
  {Bleem}, {Bonnett}, {Bridle}, {Brooks}, {Buckley-Geer}, {Burke}, {Capozzi},
  {Carlstrom}, {Carnero Rosell}, {Carrasco Kind}, {Carretero}, {Crocce},
  {Cunha}, {D'Andrea}, {da Costa}, {Desai}, {Diehl}, {Dietrich}, {Doel},
  {Eifler}, {Evrard}, {Flaugher}, {Frieman}, {Gerdes}, {Goldstein}, {Gruen},
  {Gruendl}, {Honscheid}, {James}, {Jarvis}, {Kent}, {Kuehn}, {Kuropatkin},
  {Lima}, {March}, {Martini}, {Melchior}, {Miller}, {Miquel}, {Nichol},
  {Ogando}, {Plazas}, {Reichardt}, {Roodman}, {Rozo}, {Rykoff}, {Sako},
  {Sanchez}, {Scarpine}, {Schubnell}, {Sevilla-Noarbe}, {Simard}, {Smith},
  {Soares-Santos}, {Sobreira}, {Suchyta}, {Swanson}, {Tarle}, {Thomas},
  {Wechsler}, \& {Weller}}]{Kirketal2015}
{Kirk} D. {et~al.}, 2015, ArXiv e-prints

\bibitem[{{Kirk} {et~al}\mbox{.}(2012){Kirk}, {Rassat}, {Host}, \&
  {Bridle}}]{KRH+12}
{Kirk} D., {Rassat} A., {Host} O., {Bridle} S., 2012, \mnras, 424, 1647

\bibitem[{{Kratochvil}, {Haiman} \& {May}(2010){Kratochvil}, {Haiman}, \&
  {May}}]{Kratochviletal2010}
{Kratochvil} J.~M., {Haiman} Z., {May} M., 2010, \prd, 81, 043519

\bibitem[{{Kratochvil} {et~al}\mbox{.}(2012){Kratochvil}, {Lim}, {Wang},
  {Haiman}, {May}, \& {Huffenberger}}]{KLW+12}
{Kratochvil} J.~M., {Lim} E.~A., {Wang} S., {Haiman} Z., {May} M.,
  {Huffenberger} K., 2012, \prd, 85, 103513

\bibitem[{{Laureijs et al.}(2011)}]{Euclid2011}
{Laureijs et al.}, 2011, arXiv:astro-ph/1110.3193

\bibitem[{{Li} {et~al}\mbox{.}(2013){Li}, {Wang}, {Yang}, {Chen}, {Xie}, \&
  {Wang}}]{LWY+13}
{Li} Z., {Wang} Y., {Yang} X., {Chen} X., {Xie} L., {Wang} X., 2013, \apj, 768,
  20

\bibitem[{{Lima} {et~al}\mbox{.}(2008){Lima}, {Cunha}, {Oyaizu}, {Frieman},
  {Lin}, \& {Sheldon}}]{lima2008}
{Lima} M., {Cunha} C.~E., {Oyaizu} H., {Frieman} J., {Lin} H., {Sheldon} E.~S.,
  2008, \mnras, 390, 118

\bibitem[{Lin \& Kilbinger(2015)}]{Lin2015}
Lin C.-A., Kilbinger M., 2015, Astronomy {\&} Astrophysics, 583, A70

\bibitem[{{Liu} {et~al}\mbox{.}(2015{\natexlab{a}}){Liu}, {Petri}, {Haiman},
  {Hui}, {Kratochvil}, \& {May}}]{Liuetal2015Z}
{Liu} J., {Petri} A., {Haiman} Z., {Hui} L., {Kratochvil} J.~M., {May} M.,
  2015{\natexlab{a}}, \prd, 91, 063507

\bibitem[{{Liu} {et~al}\mbox{.}(2015{\natexlab{b}}){Liu}, {Pan}, {Li}, {Shan},
  {Wang}, {Fu}, {Fan}, {Kneib}, {Leauthaud}, {Van Waerbeke}, {Makler},
  {Moraes}, {Erben}, \& {Charbonnier}}]{Liuetal2015W}
{Liu} X. {et~al.}, 2015{\natexlab{b}}, \mnras, 450, 2888

\bibitem[{{Liu J.} {et~al}\mbox{.}(2015){Liu J.}, Petri, Haiman, Hui,
  Kratochvil, \& May}]{Liu2014}
{Liu J.} J., Petri A., Haiman Z., Hui L., Kratochvil J.~M., May M., 2015,
  Physical Review D, 91, 063507

\bibitem[{{Liu X.} {et~al}\mbox{.}(2015){Liu X.}, Pan, Li, Shan, Wang, Fu, Fan,
  Kneib, Leauthaud, {Van Waerbeke}, Makler, Moraes, Erben, \&
  Charbonnier}]{Liu2014a}
{Liu X.} X. {et~al.}, 2015, Monthly Notices of the Royal Astronomical Society,
  450, 2888

\bibitem[{{LSST Science Collaborations} {et~al}\mbox{.}(2009){LSST Science
  Collaborations}, {Abell}, {Allison}, {Anderson}, {Andrew}, {Angel}, {Armus},
  {Arnett}, {Asztalos}, {Axelrod}, \& et~al.}]{lsst2009}
{LSST Science Collaborations} {et~al.}, 2009, arXiv:0912.0201

\bibitem[{Mandelbaum {et~al}\mbox{.}(2014)Mandelbaum, Rowe, Armstrong, Bard,
  Bertin, Bosch, Boutigny, Courbin, Dawson, Donnarumma, Conti, Gavazzi,
  Gentile, Gill, Hogg, Huff, Jee, Kacprzak, Kilbinger, Kuntzer, Lang, Luo,
  March, Marshall, Meyers, Miller, Miyatake, Nakajima, Mboula, Nurbaeva, Okura,
  Paulin-Henriksson, Rhodes, Schneider, Shan, Sheldon, Simet, Starck, Sureau,
  Tewes, Adami, Zhang, \& Zuntz}]{Mandelbaum2014}
Mandelbaum R. {et~al.}, 2014, Monthly Notices of the Royal Astronomical
  Society, 450, 2963

\bibitem[{{Mandelbaum} {et~al}\mbox{.}(2014){Mandelbaum}, {Rowe}, {Bosch},
  {Chang}, {Courbin}, {Gill}, {Jarvis}, {Kannawadi}, {Kacprzak}, {Lackner},
  {Leauthaud}, {Miyatake}, {Nakajima}, {Rhodes}, {Simet}, {Zuntz}, {Armstrong},
  {Bridle}, {Coupon}, {Dietrich}, {Gentile}, {Heymans}, {Jurling}, {Kent},
  {Kirkby}, {Margala}, {Massey}, {Melchior}, {Peterson}, {Roodman}, \&
  {Schrabback}}]{great3}
{Mandelbaum} R. {et~al.}, 2014, \apjs, 212, 5

\bibitem[{{Marian} \& {Bernstein}(2006)}]{MarianBernstein2006}
{Marian} L., {Bernstein} G.~M., 2006, \prd, 73, 123525

\bibitem[{{Marian} {et~al}\mbox{.}(2011){Marian}, {Hilbert}, {Smith},
  {Schneider}, \& {Desjacques}}]{Marianetal2011}
{Marian} L., {Hilbert} S., {Smith} R.~E., {Schneider} P., {Desjacques} V.,
  2011, \apjl, 728, L13+

\bibitem[{{Marian}, {Smith} \& {Bernstein}(2009){Marian}, {Smith}, \&
  {Bernstein}}]{Marianetal2009}
{Marian} L., {Smith} R.~E., {Bernstein} G.~M., 2009, \apjl, 698, L33

\bibitem[{Marian, Smith \& Bernstein(2009)Marian, Smith, \&
  Bernstein}]{Marian2009}
Marian L., Smith R.~E., Bernstein G.~M., 2009, The Astrophysical Journal, 709,
  286

\bibitem[{{Marian}, {Smith} \& {Bernstein}(2010){Marian}, {Smith}, \&
  {Bernstein}}]{Marianetal2010}
{Marian} L., {Smith} R.~E., {Bernstein} G.~M., 2010, \apj, 709, 286

\bibitem[{{Marian} {et~al}\mbox{.}(2012{\natexlab{a}}){Marian}, {Smith},
  {Hilbert}, \& {Schneider}}]{Marianetal2012}
{Marian} L., {Smith} R.~E., {Hilbert} S., {Schneider} P., 2012{\natexlab{a}},
  \mnras, 2969

\bibitem[{{Marian} {et~al}\mbox{.}(2012{\natexlab{b}}){Marian}, {Smith},
  {Hilbert}, \& {Schneider}}]{Marian2012}
{Marian} L., {Smith} R.~E., {Hilbert} S., {Schneider} P., 2012{\natexlab{b}},
  \mnras, 423, 1711

\bibitem[{{Marian} {et~al}\mbox{.}(2013){Marian}, {Smith}, {Hilbert}, \&
  {Schneider}}]{Marianetal2013}
{Marian} L., {Smith} R.~E., {Hilbert} S., {Schneider} P., 2013, \mnras, 432,
  1338

\bibitem[{{Maturi} {et~al}\mbox{.}(2010){Maturi}, {Angrick}, {Pace}, \&
  {Bartelmann}}]{Maturietal2010}
{Maturi} M., {Angrick} C., {Pace} F., {Bartelmann} M., 2010, \aap, 519, A23+

\bibitem[{{Maturi}, {Fedeli} \& {Moscardini}(2011){Maturi}, {Fedeli}, \&
  {Moscardini}}]{Maturietal2011}
{Maturi} M., {Fedeli} C., {Moscardini} L., 2011, \mnras, 416, 2527

\bibitem[{{Maturi} {et~al}\mbox{.}(2005){Maturi}, {Meneghetti}, {Bartelmann},
  {Dolag}, \& {Moscardini}}]{Maturietal2005}
{Maturi} M., {Meneghetti} M., {Bartelmann} M., {Dolag} K., {Moscardini} L.,
  2005, \aap, 442, 851

\bibitem[{{Maturi} {et~al}\mbox{.}(2007){Maturi}, {Schirmer}, {Meneghetti},
  {Bartelmann}, \& {Moscardini}}]{Maturietal2007}
{Maturi} M., {Schirmer} M., {Meneghetti} M., {Bartelmann} M., {Moscardini} L.,
  2007, \aap, 462, 473

\bibitem[{Melchior {et~al}\mbox{.}(2014)Melchior, Suchyta, Huff, Hirsch,
  Kacprzak, Rykoff, Gruen, Armstrong, Bacon, Bechtol, Bernstein, Bridle,
  Clampitt, Honscheid, Jain, Jouvel, Krause, Lin, MacCrann, Patton, Plazas,
  Rowe, Vikram, Wilcox, Young, Zuntz, Abbott, Abdalla, Allam, Banerji,
  Bernstein, Bernstein, Bertin, Buckley-Geer, Burke, Castander, da~Costa,
  Cunha, Depoy, Desai, Diehl, Doel, Estrada, Evrard, Neto, Fernandez, Finley,
  Flaugher, Frieman, Gaztanaga, Gerdes, Gruendl, Gutierrez, Jarvis, Karliner,
  Kent, Kuehn, Kuropatkin, Lahav, Maia, Makler, Marriner, Marshall, Merritt,
  Miller, Miquel, Mohr, Neilsen, Nichol, Nord, Reil, Roe, Roodman, Sako,
  Sanchez, Santiago, Schindler, Schubnell, Sevilla-Noarbe, Sheldon, Smith,
  Soares-Santos, Swanson, Sypniewski, Tarle, Thaler, Thomas, Tucker, Walker,
  Wechsler, Weller, \& Wester}]{Melchior2014}
Melchior P. {et~al.}, 2014, Monthly Notices of the Royal Astronomical Society,
  19, 19

\bibitem[{Melchior \& Viola(2012)}]{Melchior2012}
Melchior P., Viola M., 2012, Monthly Notices of the Royal Astronomical Society,
  424, 2757

\bibitem[{{Miller} {et~al}\mbox{.}(2013){Miller}, {Heymans}, {Kitching}, {van
  Waerbeke}, \& {etal}}]{Milleretal2013}
{Miller} L., {Heymans} C., {Kitching} T.~D., {van Waerbeke} L., {etal}, 2013,
  \mnras, 429, 2858

\bibitem[{Miralda-Escude(1991)}]{Miralda-Escude1991}
Miralda-Escude J., 1991, The Astrophysical Journal, 370, 1

\bibitem[{{Miyazaki} {et~al}\mbox{.}(2002){Miyazaki}, {Komiyama}, {Sekiguchi},
  {Okamura}, {Doi}, {Furusawa}, {Hamabe}, {Imi}, {Kimura}, {Nakata}, {Okada},
  {Ouchi}, {Shimasaku}, {Yagi}, \& {Yasuda}}]{Miyazakietal2002}
{Miyazaki} S. {et~al.}, 2002, \pasj, 54, 833

\bibitem[{Mohr {et~al}\mbox{.}(2012)Mohr, Armstrong, Bertin, Daues, Desai,
  Gower, Gruendl, Hanlon, Kuropatkin, Lin, Marriner, Petravic, Sevilla,
  Swanson, Tomashek, Tucker, \& Yanny}]{mohr2012}
Mohr J.~J. {et~al.}, 2012, in Software and Cyberinfrastructure for Astronomy
  II. Proceedings of the SPIE, Radziwill N.~M., Chiozzi G., eds., Vol. 8451, p.
  84510D

\bibitem[{Navarro, Frenk \& White(1997)Navarro, Frenk, \& White}]{Navarro1997}
Navarro J.~F., Frenk C.~S., White S. D.~M., 1997, The Astrophysical Journal,
  490, 493

\bibitem[{Osato, Shirasaki \& Yoshida(2015)Osato, Shirasaki, \&
  Yoshida}]{Osato2015a}
Osato K., Shirasaki M., Yoshida N., 2015, The Astrophysical Journal, 806, 186

\bibitem[{Paulin-Henriksson, Refregier \& Amara(2009)Paulin-Henriksson,
  Refregier, \& Amara}]{Paulin-Henriksson2009}
Paulin-Henriksson S., Refregier A., Amara A., 2009, Astronomy and Astrophysics,
  500, 647

\bibitem[{Petri {et~al}\mbox{.}(2014)Petri, May, Haiman, \&
  Kratochvil}]{Petri2014}
Petri A., May M., Haiman Z., Kratochvil J.~M., 2014, Physical Review D, 90

\bibitem[{Pires, Leonard \& Starck(2012)Pires, Leonard, \& Starck}]{Pires2012}
Pires S., Leonard A., Starck J.-L., 2012, Monthly Notices of the Royal
  Astronomical Society, 423, 983

\bibitem[{{Reblinsky} {et~al}\mbox{.}(1999){Reblinsky}, {Kruse}, {Jain}, \&
  {Schneider}}]{Reblinskyetal1999}
{Reblinsky} K., {Kruse} G., {Jain} B., {Schneider} P., 1999, \aap, 351, 815

\bibitem[{Refregier {et~al}\mbox{.}(2010)Refregier, Amara, Kitching, Rassat,
  Scaramella, Weller, \& Consortium}]{Refregier2010}
Refregier A., Amara A., Kitching T.~D., Rassat A., Scaramella R., Weller J.,
  Consortium f. t. E.~I., 2010, ArXiv e-prints

\bibitem[{Refregier {et~al}\mbox{.}(2012)Refregier, Kacprzak, Amara, Bridle, \&
  Rowe}]{Refregier2012}
Refregier A., Kacprzak T., Amara A., Bridle S., Rowe B., 2012, Monthly Notices
  of the Royal Astronomical Society, 425, 1951

\bibitem[{Reischke, Maturi \& Bartelmann(2015)Reischke, Maturi, \&
  Bartelmann}]{Reischke2015}
Reischke R., Maturi M., Bartelmann M., 2015, ArXiv e-prints, 14

\bibitem[{Rozo {et~al}\mbox{.}(2010)Rozo, Wechsler, Rykoff, Annis, Becker,
  Evrard, Frieman, Hansen, Hao, Johnston, Koester, McKay, Sheldon, \&
  Weinberg}]{Rozo2010}
Rozo E. {et~al.}, 2010, The Astrophysical Journal, 708, 645

\bibitem[{{S{\'a}nchez} {et~al}\mbox{.}(2014){S{\'a}nchez}, {Carrasco Kind},
  {Lin}, {Miquel}, {Abdalla}, {Amara}, {Banerji}, {Bonnett}, {Brunner},
  {Capozzi}, {Carnero}, {Castander}, {da Costa}, {Cunha}, {Fausti}, {Gerdes},
  {Greisel}, {Gschwend}, {Hartley}, {Jouvel}, {Lahav}, {Lima}, {Maia},
  {Mart{\'{\i}}}, {Ogando}, {Ostrovski}, {Pellegrini}, {Rau}, {Sadeh}, {Seitz},
  {Sevilla-Noarbe}, {Sypniewski}, {de Vicente}, {Abbot}, {Allam}, {Atlee},
  {Bernstein}, {Bernstein}, {Buckley-Geer}, {Burke}, {Childress}, {Davis},
  {DePoy}, {Dey}, {Desai}, {Diehl}, {Doel}, {Estrada}, {Evrard},
  {Fern{\'a}ndez}, {Finley}, {Flaugher}, {Frieman}, {Gaztanaga}, {Glazebrook},
  {Honscheid}, {Kim}, {Kuehn}, {Kuropatkin}, {Lidman}, {Makler}, {Marshall},
  {Nichol}, {Roodman}, {S{\'a}nchez}, {Santiago}, {Sako}, {Scalzo}, {Smith},
  {Swanson}, {Tarle}, {Thomas}, {Tucker}, {Uddin}, {Vald{\'e}s}, {Walker},
  {Yuan}, \& {Zuntz}}]{sanchez2014}
{S{\'a}nchez} C. {et~al.}, 2014, \mnras, 445, 1482

\bibitem[{Schirmer {et~al}\mbox{.}(2007)Schirmer, Erben, Hetterscheidt, \&
  Schneider}]{Schirmer2007}
Schirmer M., Erben T., Hetterscheidt M., Schneider P., 2007, Astronomy and
  Astrophysics, 462, 875

\bibitem[{{Schirmer} {et~al}\mbox{.}(2007){Schirmer}, {Erben}, {Hetterscheidt},
  \& {Schneider}}]{Schirmeretal2007}
{Schirmer} M., {Erben} T., {Hetterscheidt} M., {Schneider} P., 2007, \aap, 462,
  875

\bibitem[{Schneider \& Bridle(2010)}]{Schneider2010}
Schneider M.~D., Bridle S., 2010, Monthly Notices of the Royal Astronomical
  Society, 402, 2127

\bibitem[{{Schneider} \& {Bridle}(2010)}]{SB10}
{Schneider} M.~D., {Bridle} S., 2010, \mnras, 402, 2127

\bibitem[{Schneider(1996)}]{Schneider1996}
Schneider P., 1996, Monthly Notices of the Royal Astronomical Society, 283, 837

\bibitem[{Schrabback {et~al}\mbox{.}(2010)Schrabback, Hartlap, Joachimi,
  Kilbinger, Simon, Benabed, Brada{\v{c}}, Eifler, Erben, Fassnacht, High,
  Hilbert, Hildebrandt, Hoekstra, Kuijken, Marshall, Mellier, Morganson,
  Schneider, Semboloni, {Van Waerbeke}, \& Velander}]{Schrabback2010}
Schrabback T. {et~al.}, 2010, Astronomy and Astrophysics, 516, A63

\bibitem[{{Semboloni} {et~al}\mbox{.}(2006){Semboloni}, {Mellier}, {van
  Waerbeke}, {Hoekstra}, {Tereno}, {Benabed}, {Gwyn}, {Fu}, {Hudson}, {Maoli},
  \& {Parker}}]{Sembolonietal2006}
{Semboloni} E. {et~al.}, 2006, \aap, 452, 51

\bibitem[{Sheldon(2014)}]{Sheldon2014}
Sheldon E.~S., 2014, Monthly Notices of the Royal Astronomical Society:
  Letters, 444, L25

\bibitem[{Sheldon {et~al}\mbox{.}(2009)Sheldon, Johnston, Scranton, Koester,
  McKay, Oyaizu, Cunha, Lima, Lin, Frieman, Wechsler, Annis, Mandelbaum,
  Bahcall, \& Fukugita}]{Sheldon2009}
Sheldon E.~S. {et~al.}, 2009, The Astrophysical Journal, 703, 2217

\bibitem[{{Shi}, {Joachimi} \& {Schneider}(2010){Shi}, {Joachimi}, \&
  {Schneider}}]{SJS10}
{Shi} X., {Joachimi} B., {Schneider} P., 2010, \aap, 523, A60

\bibitem[{{Sif{\'o}n} {et~al}\mbox{.}(2015){Sif{\'o}n}, {Hoekstra}, {Cacciato},
  {Viola}, {K{\"o}hlinger}, {van der Burg}, {Sand}, \& {Graham}}]{SHC+15}
{Sif{\'o}n} C., {Hoekstra} H., {Cacciato} M., {Viola} M., {K{\"o}hlinger} F.,
  {van der Burg} R.~F.~J., {Sand} D.~J., {Graham} M.~L., 2015, \aap, 575, A48

\bibitem[{{Singh}, {Mandelbaum} \& {More}(2014){Singh}, {Mandelbaum}, \&
  {More}}]{SMM14}
{Singh} S., {Mandelbaum} R., {More} S., 2014, ArXiv e-prints

\bibitem[{Springel(2005)}]{Springel2005}
Springel V., 2005, Monthly Notices of the Royal Astronomical Society, 364, 1105

\bibitem[{Suchyta {et~al}\mbox{.}(2015)Suchyta, Huff, Aleksi{\'{c}}, Melchior,
  Jouvel, MacCrann, Crocce, Gaztanaga, Honscheid, Leistedt, Peiris, Ross,
  Rykoff, Sheldon, Abbott, Abdalla, Allam, Banerji, Benoit-L{\'{e}}vy, Bertin,
  Brooks, Burke, Rosell, Kind, Carretero, Cunha, D'Andrea, da~Costa, DePoy,
  Desai, Diehl, Dietrich, Doel, Eifler, Estrada, Evrard, Flaugher, Fosalba,
  Frieman, Gerdes, Gruen, Gruendl, James, Jarvis, Kuehn, Kuropatkin, Lahav,
  Lima, Maia, March, Marshall, Miller, Miquel, Neilsen, Nichol, Nord, Ogando,
  Percival, Reil, Roodman, Sako, Sanchez, Scarpine, Sevilla-Noarbe, Smith,
  Soares-Santos, Sobreira, Swanson, Tarle, Thaler, Thomas, Vikram, Walker,
  Wechsler, \& Zhang}]{Suchyta2015}
Suchyta E. {et~al.}, 2015, ArXiv e-prints, 24

\bibitem[{{Tang} \& {Fan}(2005)}]{TangFan2005}
{Tang} J.~Y., {Fan} Z.~H., 2005, \apj, 635, 60

\bibitem[{Taylor \& Joachimi(2014)}]{Taylor2014}
Taylor A., Joachimi B., 2014, Monthly Notices of the Royal Astronomical
  Society, 442, 2728

\bibitem[{{Taylor}, {Joachimi} \& {Kitching}(2013){Taylor}, {Joachimi}, \&
  {Kitching}}]{Taylor2013}
{Taylor} A., {Joachimi} B., {Kitching} T., 2013, Monthly Notices of the Royal
  Astronomical Society, 432, 1928

\bibitem[{{Tenneti}, {Mandelbaum} \& {Di Matteo}(2015){Tenneti}, {Mandelbaum},
  \& {Di Matteo}}]{TMM15}
{Tenneti} A., {Mandelbaum} R., {Di Matteo} T., 2015, ArXiv e-prints

\bibitem[{{The Dark Energy Survey
  Collaboration}(2005)}]{TheDarkEnergySurveyCollaboration2005}
{The Dark Energy Survey Collaboration}, 2005, ArXiv Astrophysics e-prints, 42

\bibitem[{{The Dark Energy Survey Collaboration}(2015)}]{DESCS2015}
{The Dark Energy Survey Collaboration}, 2015, ArXiv e-prints

\bibitem[{{The Dark Energy Survey Collaboration} {et~al}\mbox{.}(2015){The Dark
  Energy Survey Collaboration}, Abbott, Abdalla, Allam, Amara, Annis,
  Armstrong, Bacon, Banerji, Bauer, Baxter, Becker, Benoit-L{\'{e}}vy,
  Bernstein, Bernstein, Bertin, Blazek, Bonnett, Bridle, Brooks, Bruderer,
  Buckley-Geer, Burke, Busha, Capozzi, Rosell, Kind, Carretero, Castander,
  Chang, Clampitt, Crocce, Cunha, D'Andrea, da~Costa, Das, DePoy, Desai, Diehl,
  Dietrich, Dodelson, Doel, Drlica-Wagner, Efstathiou, Eifler, Erickson,
  Estrada, Evrard, Neto, Fernandez, Finley, Flaugher, Fosalba, Friedrich,
  Frieman, Gangkofner, Garcia-Bellido, Gaztanaga, Gerdes, Gruen, Gruendl,
  Gutierrez, Hartley, Hirsch, Honscheid, Huff, Jain, James, Jarvis, Kacprzak,
  Kent, Kirk, Krause, Kravtsov, Kuehn, Kuropatkin, Kwan, Lahav, Leistedt, Li,
  Lima, Lin, MacCrann, March, Marshall, Martini, McMahon, Melchior, Miller,
  Miquel, Mohr, Neilsen, Nichol, Nicola, Nord, Ogando, Palmese, Peiris, Plazas,
  Refregier, Roe, Romer, Roodman, Rowe, Rykoff, Sabiu, Sadeh, Sako, Samuroff,
  S{\'{a}}nchez, Sanchez, Seo, Sevilla-Noarbe, Sheldon, Smith, Soares-Santos,
  Sobreira, Suchyta, Swanson, Tarle, Thaler, Thomas, Troxel, Vikram, Walker,
  Wechsler, Weller, Zhang, \& Zuntz}]{TheDarkEnergySurveyCollaboration2015}
{The Dark Energy Survey Collaboration} {et~al.}, 2015, ArXiv e-prints, 20, 20

\bibitem[{{Troxel} \& {Ishak}(2012{\natexlab{a}})}]{TI12b}
{Troxel} M.~A., {Ishak} M., 2012{\natexlab{a}}, \mnras, 423, 1663

\bibitem[{{Troxel} \& {Ishak}(2012{\natexlab{b}})}]{TI12a}
{Troxel} M.~A., {Ishak} M., 2012{\natexlab{b}}, \mnras, 419, 1804

\bibitem[{Tyson, Wenk \& Valdes(1990)Tyson, Wenk, \& Valdes}]{Tyson1990}
Tyson J.~A., Wenk R.~A., Valdes F., 1990, The Astrophysical Journal, 349, L1

\bibitem[{{Vikram} {et~al}\mbox{.}(2015){Vikram}, {Chang}, {Jain}, {Bacon}, \&
  {etal.}}]{Vikrametal2015}
{Vikram} V., {Chang} C., {Jain} B., {Bacon} D., {etal.}, 2015, \prd, 92, 022006

\bibitem[{Voigt \& Bridle(2010)}]{Voigt2010}
Voigt L.~M., Bridle S.~L., 2010, Monthly Notices of the Royal Astronomical
  Society, 404, 458

\bibitem[{{Wang} {et~al}\mbox{.}(2004){Wang}, {Khoury}, {Haiman}, \&
  {May}}]{Wangetal2004}
{Wang} S., {Khoury} J., {Haiman} Z., {May} M., 2004, \prd, 70, 123008

\bibitem[{Yang {et~al}\mbox{.}(2012)Yang, Kratochvil, Huffenberger, Haiman, \&
  May}]{Yang2012}
Yang X., Kratochvil J.~M., Huffenberger K., Haiman Z., May M., 2012, Physical
  Review D, 87

\bibitem[{{Yang} {et~al}\mbox{.}(2011){Yang}, {Kratochvil}, {Wang}, {Lim},
  {Haiman}, \& {May}}]{YKW+11}
{Yang} X., {Kratochvil} J.~M., {Wang} S., {Lim} E.~A., {Haiman} Z., {May} M.,
  2011, \prd, 84, 043529

\bibitem[{Yang {et~al}\mbox{.}(2011)Yang, Kratochvil, Wang, Lim, Haiman, \&
  May}]{Yang2011a}
Yang X., Kratochvil J.~M., Wang S., Lim E.~A., Haiman Z., May M., 2011,
  Physical Review D, 84, 043529

\bibitem[{YuanShan {et~al}\mbox{.}(2014)YuanShan, Kneib, Comparat, Jullo,
  Charbonnier, Erben, Makler, Moraes, {Van Waerbeke}, Courbin, Meylan, Tao, \&
  Taylor}]{Shan2013}
YuanShan H. {et~al.}, 2014, Monthly Notices of the Royal Astronomical Society,
  442, 2534

\bibitem[{Zuntz {et~al}\mbox{.}(2013)Zuntz, Kacprzak, Voigt, Hirsch, Rowe, \&
  Bridle}]{Zuntz2013}
Zuntz J., Kacprzak T., Voigt L., Hirsch M., Rowe B., Bridle S., 2013, Monthly
  Notices of the Royal Astronomical Society, 434, 1604

\end{thebibliography}

~

\noindent
$[1]$ Department of Physics, ETH Zurich, Wolfgang-Pauli- Strasse 16, CH-8093 Zurich, Switzerland \\
$[2]$ Astrophysics Group, Department of Physics and Astronomy, University College London, 132 Hampstead Road, London, NW1 2PS, United Kingdom \\
$[3]$ University Observatory Munich, Scheinerstrasse 1, 81679 Munich, Germany \\
$[4]$ Max Planck Institute for Extraterrestrial Physics, Giessenbachstrasse, 85748 Garching, Germany \\
$[5]$ Department of Physics and Astronomy, University of Sussex, Brighton BN1 9QH, UK \\
$[6]$ Universit\"ats-Sternwarte, Fakult\"at f\"ur Physik, Ludwig-Maximilians Universit\"at M\"unchen, Scheinerstr. 1, 81679 M\"unchen, Germany\\
$[7]$ Excellence Cluster Universe, Boltzmannstr. 2, D-85748 Garching bei M\"unchen, Germany \\
$[8]$ Department of Physics and Astronomy, University of Pennsylvania, Philadelphia, PA 19104, USA \\
$[9]$ Institut de F\'isica d'Altes Energies (IFAE), The Barcelona Institute of Science and Technology, Campus UAB, 08193 Bellaterra (Barcelona) Spain \\
$[10]$ Institute of Cosmology \& Gravitation, University of Portsmouth, Portsmouth, PO1 3FX, UK \\
$[11]$ Department of Physics, Stanford University, 382 Via Pueblo Mall, Stanford, CA 94305, USA \\
$[12]$ Kavli Institute for Particle Astrophysics \& Cosmology, P. O. Box 2450, Stanford University, Stanford, CA 94305, USA \\
$[13]$ Jodrell Bank Center for Astrophysics, School of Physics and Astronomy, University of Manchester, Oxford Road, Manchester, M13 9PL, UK \\
$[14]$ Jet Propulsion Laboratory, California Institute of Technology, 4800 Oak Grove Dr., Pasadena, CA 91109, USA \\
$[15]$ Department of Physics, The Ohio State University, Columbus, OH 43210, USA \\
$[16]$ Center for Cosmology and Astro-Particle Physics, The Ohio State University, Columbus, OH 43210, USA \\
$[17]$ Department of Astrophysical Sciences, Princeton University, Peyton Hall, Princeton, NJ 08544, USA \\
$[18]$  Cerro Tololo Inter-American Observatory, National Optical Astronomy Observatory, Casilla 603, La Serena, Chile \\
$[19]$  Department of Physics \& Astronomy, University College London, Gower Street, London, WC1E 6BT, UK \\
$[20]$  Department of Physics and Electronics, Rhodes University, PO Box 94, Grahamstown, 6140, South Africa \\
$[21]$  Department of Astrophysical Sciences, Princeton University, Peyton Hall, Princeton, NJ 08544, USA \\
$[22]$  CNRS, UMR 7095, Institut d'Astrophysique de Paris, F-75014, Paris, France \\
$[23]$  Sorbonne Universit\'es, UPMC Univ Paris 06, UMR 7095, Institut d'Astrophysique de Paris, F-75014, Paris, France \\
$[24]$  Carnegie Observatories, 813 Santa Barbara St., Pasadena, CA 91101, USA \\
$[25]$  Kavli Institute for Particle Astrophysics \& Cosmology, P. O. Box 2450, Stanford University, Stanford, CA 94305, USA \\
$[26]$  SLAC National Accelerator Laboratory, Menlo Park, CA 94025, USA \\
$[27]$ Laborat\'orio Interinstitucional de e-Astronomia - LIneA, Rua Gal. Jos\'e Cristino 77, Rio de Janeiro, RJ - 20921-400, Brazil \\
$[28]$ Observat\'orio Nacional, Rua Gal. Jos\'e Cristino 77, Rio de Janeiro, RJ - 20921-400, Brazil \\
$[29]$ Department of Astronomy, University of Illinois, 1002 W. Green Street, Urbana, IL 61801, USA \\
$[30]$ National Center for Supercomputing Applications, 1205 West Clark St., Urbana, IL 61801, USA \\
$[31]$ Institut de Ci\`encies de l'Espai, IEEC-CSIC, Campus UAB, Carrer de Can Magrans, s/n,  08193 Bellaterra, Barcelona, Spain \\
$[32]$ Institut de F\'{\i}sica d'Altes Energies (IFAE), The Barcelona Institute of Science and Technology, Campus UAB, 08193 Bellaterra (Barcelona) Spain \\
$[33]$ Institute of Cosmology \& Gravitation, University of Portsmouth, Portsmouth, PO1 3FX, UK \\
$[34]$ School of Physics and Astronomy, University of Southampton,  Southampton, SO17 1BJ, UK \\
$[35]$ Excellence Cluster Universe, Boltzmannstr.\ 2, 85748 Garching, Germany \\
$[36]$ Faculty of Physics, Ludwig-Maximilians-Universit\"at, Scheinerstr. 1, 81679 Munich, Germany \\
$[37]$ Fermi National Accelerator Laboratory, P. O. Box 500, Batavia, IL 60510, USA \\
$[38]$ Department of Astronomy, University of Michigan, Ann Arbor, MI 48109, USA \\
$[39]$ Department of Physics, University of Michigan, Ann Arbor, MI 48109, USA \\
$[40]$ Kavli Institute for Cosmological Physics, University of Chicago, Chicago, IL 60637, USA \\
$[41]$ Department of Astronomy, University of California, Berkeley,  501 Campbell Hall, Berkeley, CA 94720, USA \\
$[42]$ Lawrence Berkeley National Laboratory, 1 Cyclotron Road, Berkeley, CA 94720, USA \\
$[43]$ Center for Cosmology and Astro-Particle Physics, The Ohio State University, Columbus, OH 43210, USA \\
$[44]$ Department of Physics, The Ohio State University, Columbus, OH 43210, USA \\
$[45]$ Australian Astronomical Observatory, North Ryde, NSW 2113, Australia \\
$[46]$ Departamento de F\'{\i}sica Matem\'atica,  Instituto de F\'{\i}sica, Universidade de S\~ao Paulo,  CP 66318, CEP 05314-970, S\~ao Paulo, SP,  Brazil \\
$[47]$ Department of Physics and Astronomy, University of Pennsylvania, Philadelphia, PA 19104, USA \\
$[48]$ George P. and Cynthia Woods Mitchell Institute for Fundamental Physics and Astronomy, and Department of Physics and Astronomy, Texas A\&M University, College Station, TX 77843,  USA \\
$[49]$ Department of Astronomy, The Ohio State University, Columbus, OH 43210, USA \\
$[50]$ Instituci\'o Catalana de Recerca i Estudis Avan\c{c}ats, E-08010 Barcelona, Spain \\
$[51]$ Max Planck Institute for Extraterrestrial Physics, Giessenbachstrasse, 85748 Garching, Germany \\
$[52]$ Jet Propulsion Laboratory, California Institute of Technology, 4800 Oak Grove Dr., Pasadena, CA 91109, USA \\
$[53]$ Department of Physics and Astronomy, Pevensey Building, University of Sussex, Brighton, BN1 9QH, UK \\
$[54]$ Centro de Investigaciones Energ\'eticas, Medioambientales y Tecnol\'ogicas (CIEMAT), Madrid, Spain \\
$[55]$ Argonne National Laboratory, 9700 South Cass Avenue, Lemont, IL 60439, USA \\

\end{document}